\documentclass[a4paper,fleqn,usenatbib]{mnras}

\usepackage{newtxtext,newtxmath}
\usepackage{pdflscape}

\usepackage[T1]{fontenc}
\usepackage{ae,aecompl}
\usepackage{booktabs,threeparttable}

\usepackage{graphicx}	
\usepackage{subcaption}	
\usepackage{amsmath}	
\usepackage{amssymb}	

 \title[POLAMI III: Total flux density and polarisation variability]
        {POLAMI: Polarimetric Monitoring of Active Galactic Nuclei at Millimetre Wavelengths\\
        {\LARGE III. Characterisation of total flux density and polarisation variability of relativistic jets}}                          

\author[I. Agudo et al.]{
            Iv\'an Agudo$^{1}$\thanks{E-mail: iagudo@iaa.es (IA)},
            Clemens Thum$^{2}\thanks{E-mail: thum@iram.es (CT)}$,
            Venkatessh Ramakrishnan$^{3,4}$,
            Sol N. Molina$^{1}$,
             \newauthor
            Carolina Casadio$^{5,1}$ 
            and Jos\'e L. G\'omez$^{1}$  
\\
$^{1}$Instituto de Astrof\'{\i}sica de Andaluc\'{\i}a (CSIC),  
                 Apartado 3004, E--18080 Granada, Spain\\
$^{2}$Instituto de Radio Astronom\'ia Millim\'etrica, 
                Avenida Divina Pastora, 7, Local 20, E--18012 Granada, Spain\\
$^{3}$Aalto University Mets\"ahovi Radio Observatory, 
                Mets\"ahovintie 114, 02540, Kylm\"al\"a, Finland\\
$^{4}$Universidad de Concepci\'on,  Departamento de Astronom\'ia, 
               Casilla 160--C, Concepci\'on, Chile\\
$^{5}$Max--Planck--Institut f\"ur Radioastronomie, 
                Auf dem H\"ugel, 69, D--53121, Bonn, Germany
               }

\date{Accepted XXX. Received YYY; in original form ZZZ}

\pubyear{2017}

\begin{document}
\label{firstpage}
\pagerange{\pageref{firstpage}--\pageref{lastpage}}
\maketitle

\begin{abstract}
We report on the first results of the POLAMI program, a simultaneous 3.5 and 1.3\,mm full--Stokes--polarisation monitoring of a sample of 36 of the brightest active galactic nuclei in the northern sky with the IRAM 30\,m Telescope.
Through a systematic statistical study of data taken from October 2006 (from December 2009 for the case of the 1.3\,mm observations) to August 2014, we characterise the variability of the total flux density and linear polarisation.
We find that all sources in the sample are highly variable in total flux density at both 3.5 and 1.3\,mm, as well as in spectral index, that (except in particularly prominent flares) is found to be optically thin between these two wavelengths.
The total flux--density variability at 1.3\,mm is found, in general, to be faster, and to have larger fractional amplitude and flatter power--spectral--density slopes than 3.5\,mm.
The polarisation degree is on average larger at 1.3\,mm than at 3.5\,mm, by a factor of 2.6.
The variability of linear polarisation degree is faster and has higher fractional amplitude than for total flux density, with the typical time scales during prominent polarisation peaks being significantly faster at 1.3\,mm than at 3.5\,mm.
The polarisation angle at both 3.5 and 1.3\,mm is highly variable.
Most of the sources show one or two excursions of $>180^{\circ}$ on time scales from a few weeks to about a year during the course of our observations. 
The 3.5 and 1.3\,mm polarisation angle evolution follow rather well each other, although the 1.3\,mm data show a clear preference to more prominent variability on the short time scales, i.e. weeks. 
The data are compatible with multi--zone models of conical jets involving smaller emission regions for the shortest--wavelength emitting sites.
Such smaller emitting regions should also be more efficient in energising particle populations, as implied by the coherent evolution of the spectral index and the total flux density during flaring activity of strong enough sources. 
The data also favours the integrated emission at 1.3\,mm to have better ordered magnetic fields than the one at 3.5\,mm. 
\end{abstract}




\section{Introduction}
\label{Intr}

Radio--loud active galactic nuclei (AGN) are among the most powerful emitters at spectral ranges from radio to $\gamma$--rays.
Rapid and strong variability is one of the most salient properties of radio--loud AGN, which is inherent to the relativistic nature of their powerful jets \citep[e.g.][]{Marscher:2008p15675,Abdo:2010p11811,Aleksic:2015p23952}.
The millimetre--range jet emission is produced by the synchrotron process in the presence of magnetic fields \citep[that are also an essential ingredient for the jet--formation mechanism, e.g.][]{Tchekhovskoy:2015p25829}, which makes the emission significantly polarised \citep[e.g.][]{Marscher:2010p11374,Agudo:2011p15946}.

Recent polarisation variability studies of AGN have demonstrated to be a powerful tool to deepen our understanding of the relativistic jet phenomenon. 
Polarisation observations carry information about the magnetic fields responsible for the AGN jet emission, and allow us to constrain the possible physical conditions of the emitting plasma by eliminating some degrees of freedom inherent to non--polarimetric observations.
They also allow us to identify individual events (and even emission regions) along the spectrum by comparing similar polarimetric properties and time--dependent behaviour \citep[e.g.][]{Marscher:2010p11374,Abdo:2010p11811}.
In particular, short millimetre observations have proven to be instrumental in these tasks (specially when combined with multi--spectral--range monitoring) when well--sampled time--evolution tracks of the polarimetric properties of radio loud AGN are compiled \citep[e.g.][]{Jorstad:2010p11830,Jorstad:2013p21321,Agudo:2011p14707,Agudo:2011p15946}.
This is mainly because short mm wavelengths access the innermost regions of relativistic jets where their synchrotron emission is optically thin \citep{2007AJ.134.799J, Agudo:2014p22485}.
At millimetre wavelengths, typical Faraday rotation measures are in the range of $10^{2}$ to $10^{5}$\,rad~m$^{-2}$  \citep[e.g.,][]{Zavala:2004p138} implying that the associated small Faraday depths only mildly modify the intrinsic polarisation angles.

In the current study (the third of a series of three, Paper III hereafter), we focus on the properties of the total flux density (total flux hereafter) and linear polarisation variability for a sample of 36 of the brightest radio--loud AGN monitored within the POLAMI program (Polarimetric Monitoring of AGN with Millimetre Wavelengths, see \href{url}{http://polami.iaa.es}) at 3.5\,mm and 1.3\,mm\footnote{For easier reading, hereafter we will use 3\,mm and 1\,mm instead of 3.5\,mm and 1.3\,mm, respectively.} with the IRAM 30m Telescope.
The 3\,mm observations presented here were performed from October 2006 to August 2014, with a median sampling of 22 days, and slightly faster, 19 days, after 2010.
The simultaneous 1\,mm observations were performed from December 2009 only, therefore not covering the first three years of 3\,mm observations.
Despite this partial mismatch, we also study in this work the interrelation of the variability at the two observing bands.
This paper complements the POLAMI results that are presented in \citet[][Paper I hereafter]{PaperI}
where we provide the basic information regarding this long--term monitoring program (including sample selection, observing strategy, and data reduction and calibration), and  \citet[][Paper II hereafter]{PaperII}
where we focus on the circular polarisation properties of the sample.
Further publications on the results of the comparison of the POLAMI data with the $\gamma$--ray, optical polarimetric, and 7\,mm VLBI polarimetric behaviour of the sample of sources is already in preparation, together with more detailed studies of particularly interesting sources or events.

Here, in Paper III we present our results regarding long--term variability of the sample of targets (Section~\ref{Res}), and we discuss the implication of those observing results (Section~\ref{Dis}), first in terms of total flux (Section~\ref{tfluxvar}), and then in linear polarisation degree (Section~\ref{pvar}), and linear polarisation angle (Section~\ref{chivar}).
The inter--relation of the total flux variability with the linear polarisation is discussed in Section~\ref{SmLchicorrel}, whereas the summary and main conclusions of our work are presented in Section~\ref{sumcon}.

\section{Results}
\label{Res}

The end product of our observations, that were acquired and calibrated as discussed in detail in Paper I, 
is presented in Fig.~\ref{timevol}.
This figure shows the time evolution of the daily averages of the 3 and 1\,mm fully calibrated measurements of total flux ($S$), linear polarisation degree ($m_{\rm{L}}$), and linear polarisation angle ($\chi$), for the 36 variable sources in the POLAMI sample\footnote{Note that 1328+307 (3C~286) is a standard total flux and linear polarisation calibrator \citep[see][and references therein]{Agudo:2012p17464} 
and has therefore not been included in this variability study.}
Figure~\ref{timevol} also includes the comparison of $S$, $m_{\rm{L}}$, and $\chi$ at the two observing wavelengths, by providing measurements of spectral index ($\alpha$), ratio of 1\,mm to 3\,mm linear polarisation degree ($m_{\rm{L,1}}/m_{\rm{L,3}}$), and rotation measure (RM). 

For defining the 3\,mm polarisation angle ($\chi_{3}$) curves, we took care of the ${n}\pi$--ambiguity of the polarisation vector.
Although we tested several different methods \citep[mainly following the prescriptions by][]{Kiehlmann:2016p25237}, we found that there is no optimum method for all modes of $\chi_{3}$ variability and uneven time sampling on our data set.
We therefore chose the simplest method among all tested ones that still gave point--to--point smooth variations of the polarisation angle.
For this, we adopted a procedure for rotation of $\chi_{3}$ measurements such that every third point was rotated by $\pm{n}\times\pi$ if the angle difference with the weighted mean of the two preceding points was larger than $\pi/2$, with $n$ being a natural number selected to minimise such angle difference. 
For the 1\,mm polarisation angle measurements ($\chi_{1}$), we applied rotations by $\pm{m}\times\pi$ so that $|\chi_{3}-\chi_{1}|<\pi/2$ for the final representation of every independent $\chi_{1}$ measurements and its corresponding $\chi_{3}$.

\section{Discussion}
\label{Dis}

The methods and definitions employed for the characterisation and the analysis of the total flux and polarisation variability are described in Appendix~\ref{appendix}.
These include a standard $\chi^2$ test for variability, a power spectral density (PSD) analysis for the estimation of the PSD slopes, a definition of the fractional variability amplitude ($F$), and a method for computing the discrete correlation function based on a Monte Carlo simulation scheme for the assessment of the statistical significance of correlations peaks. 
The following subsections describe the results obtained from implementing these analysis tools and definitions on the data sets, as well as their astrophysical implications.

\subsection{Total flux variability}
\label{tfluxvar}

Figure~\ref{timevol} shows that all sources in the sample are strongly variable in total flux within the time range of our observations, both at 3 and 1\,mm.
Representative cases are 0829+046 and 1055+018, with maximum to minimum ratios of total flux by factors $\sim4$ at both wavelengths (see Table~\ref{tab:I}).
These amplitudes are moderate compared to more extreme cases like 2251+158, with max/min ratio up to $\approx42$ at 1\,mm ($\approx17$ at 3\,mm), and 1406$-$076, with max/min$\approx2$ at both 3 and 1\,mm.

Table~\ref{tab:I} shows that all 36 sources show a probability to be variable of more than $99.73$\,\% at both observing wavelengths.
However, different variability modes are identified for different sources in the sample.
This is clearly reflected in the formal variability analysis that we have performed for the total flux for every source in the sample (see Table~\ref{tab:I} and subsections below).

\begin{figure}
   \centering
   \includegraphics[width=\columnwidth]{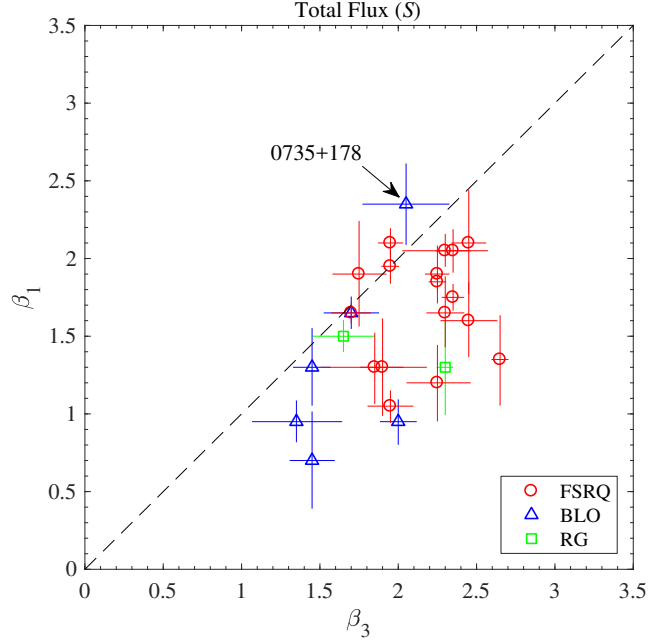}
      \caption{PSD slope ($\beta$, see Table~\ref{tab:I}) computed from the 1\,mm total flux light curve of every source with regard to the corresponding one at 3\,mm. Horizontal and vertical lines correspond to confidence intervals. The label FSRQ stands for flat spectrum radio quasars, where BLO and RG indicate BL~Lac type objects and radio galaxies, respectively. Outliers of the distribution of points in this plot are labeled by their corresponding source name.}
       \label{I_beta_3_1mm}
\end{figure}

The power spectral density (PSD) slope for 3\,mm total flux ($\beta_3$), computed for the time spanned between December 2009 and August 2014; i.e. the same one as the 1\,mm data, shows a median $\sim1.9$ (Table~\ref{tab:I}).
A small fraction of cases shows more prominent variability on the short time scales $\lesssim1$\,month (with $\beta_3\lesssim1.5$), e.g. 0716+714 and 2200+420. 
Others show a much smoother mode of variability on the short time scales, with $\beta_3\gtrsim2.5$, e.g. 0316+413, 0336$-$019, and 0836+710.
The maximum and minimum of $\beta_3$ are $\sim2.9$ for 0219+428 and $\sim0.6$ for 1219+285, respectively, but these extreme cases are affected by large uncertainties ($\sim2.9$) in the computation of $\beta_3$.
The PSD slope at 1\,mm ($\beta_1$), with median $\sim1.6$, shows, in general, significantly smaller PSD slopes (i.e. more prominent variability on the short time--scales with regard to the long ones) as for the total flux variability at 3\,mm.
This is clearly seen in Figure~\ref{I_beta_3_1mm} which shows that the entire distribution of 3\,mm PSD slopes is systematically shifted towards higher values as compared to those at 1\,mm.
Our results thus imply that, in general, radio--loud AGN show a tendency to display more prominent short time--scales of variability as compared to that at longer time--scales when observing wavelength decreases from 3 to 1\,mm.

\begin{figure}
   \centering
   \includegraphics[width=\columnwidth]{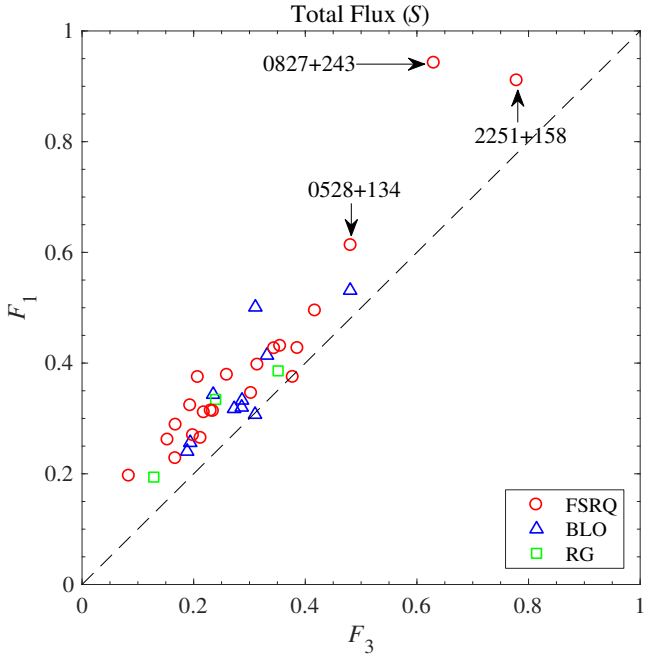}
      \caption{Fractional variability ($F$, see Table~\ref{tab:I}) of 1\,mm flux versus those at 3\,mm for every source in the variable sample. The 3\,mm values of $F$ shown here are those computed for the common time period of the 3 and 1\,mm data, i.e. December 2009 to August 2014. Outliers of the distribution of points in this plot are labeled by their corresponding source name.}
       \label{I_Fvar_3_1mm}
\end{figure}

The 3\,mm fractional variability amplitude ($F_3$, also shown in Table~\ref{tab:I}) reaches maximum and minimum levels of $\sim$0.7 (for 2251+158, a well known extremely variable source from 2008 to 2011), and $\sim0.1$ (for 1406$-$076 that did not show extreme fractional amplitude during our observations), respectively, but the typical (median) $\tilde F_3\sim$0.3, where the tilde denotes the median of $F_3$. 
The 1\,mm fractional variability amplitude goes from $\sim0.9$ (for 0827+243 and 2251+158) to $\sim0.2$ (for the cases of 0316+413 and 1406$-$076), with typical (median) $\tilde F_1=0.4$. 
A comparison of $F_3$, computed by using the 3\,mm flux densities in the same time span covered by the 1\,mm data, and $F_1$ also shows a general trend of $F_1$ to display larger values than the corresponding $F_3$ for a given source, see Fig.~\ref{I_Fvar_3_1mm}.
Therefore the data demonstrate that, in general, the total flux variability of radio loud AGN at 1\,mm have higher fractional amplitude than those at 3\,mm.

\begin{figure}
   \centering
   \includegraphics[width=\columnwidth]{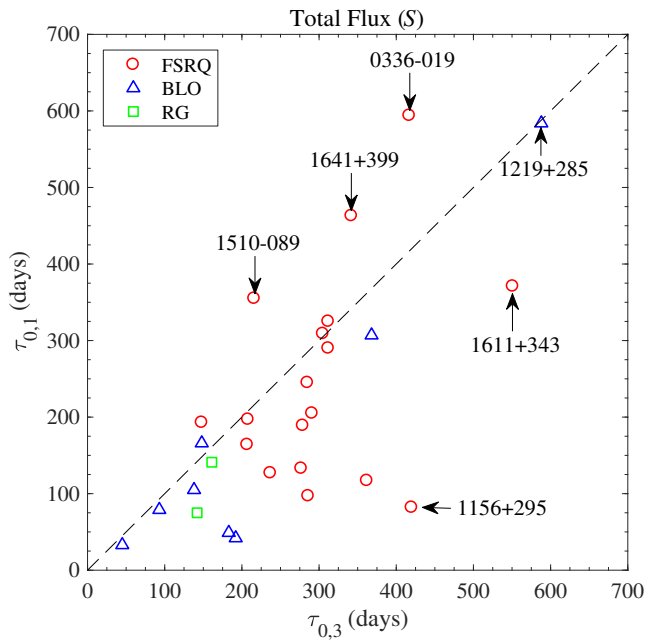}
      \caption{Time scale of variability computed from the autocorrelation function ($\tau_0$, see Table~\ref{tab:I}) of 1\,mm light curves of every source in the sample versus the corresponding 3\,mm value of $\tau_0$. The 3\,mm values of $\tau_0$ shown here are those computed for the common time period of the 3 and 1\,mm data, i.e. December 2009 to August 2014. Outliers of the distribution of points in this plot are labeled by their corresponding source name.}
       \label{I_tACF_3_1mm}
\end{figure}

To characterise the time scales of total flux variability for each one of the sources we use the zero--crossing time of the main lobe of the auto--correlation function (ACF), i.e. $\tau_0$, see Appendix~\ref{appendix}.
Note that, by definition, the discrete correlation function (DCF, and therefore the ACF) are primarily sensitive to the most prominent peaks in the data trains, and therefore $\tau_0$ mainly provides a representation of the time scales of the most prominent outburst in the total--flux light curves.
At 3\,mm, the median of $\tau_0$ for the entire sample is 318 days.
However, time scales of prominent variability as large as 805 days (for 0735+178), and as small as 45 days (for 1101+384), are found in the data set.
The time scale of total flux variability computed from $\tau_0$ is also smaller at 1\,mm (with a median of 197 days) as compared to that at 3\,mm. 
The general trend for larger $\tau_0$ in the 3\,mm light curves as compared with the 1\,mm ones is shown in Fig.~\ref{I_tACF_3_1mm}, where $\tau_{0,3}$ was computed from data covering the same time period as the one for the 1\,mm data, see also Table~\ref{tab:I}.

To summarise, we find that, in general, the variability of total flux millimetre emission in radio--loud AGN is more prominent on short time--scales as compared to that on the longer time--scales when going from 1\,mm to 3\,mm observing wavelengths.
Additionally, the variability also shows larger fractional variability amplitudes and shorter time--scales at 1\,mm as compared to 3\,mm. 
These observations are consistent with the assumption that the shorter wavelength emission originates at smaller and more violently variable regions.
This is in agreement with models that attribute the jet variability to turbulent processes with cell sizes becoming smaller towards shorter wavelengths \citep[e.g.,][]{Marscher:2014p22506}.

\subsection{Spectral index}

As it can be seen from the light curves in Fig.~\ref{timevol}, the 3--to--1\,mm spectral indices ($\alpha$)\footnote{We define the spectral index between 3 and 1\,mm as $\alpha=ln(S_{3}/S_{1})/ln(86/229)$, where $S_{3}$ and $S_{1}$ are simultaneous 3 and 1\,mm (86 and 229\,GHz) total flux measurements, respectively.} are almost constantly negative.
This contrasts with what is found at longer centimetre wavelengths, where the spectral index is typically found close to zero; i.e. flat spectra; or positive; corresponding to optically thick spectra \citep{Angelakis:2012p27463,Fuhrmann:2016p27459}.
This confirms that the spectrum in the short millimetre range is in general  optically thin \citep[see also][]{Agudo:2014p22485}. 
Only for a small fraction of sources (namely 0415+379, 0716+714, 0735+178, 0827+243, 0829+046, 1219+285, and 2200+420) the spectral index raises to positive values but (a) never larger than $\sim$0.2 and (b) only during brief periods in the proximity of bright total--flux flares.
The median spectral index is $\tilde\alpha\approx-0.6$ for the source sample (see Table~\ref{tab:alpha}), whereas the standard deviation is $\rm{rms}_{\alpha}\approx0.2$).
This implies an average energy index of the emitting particle distribution for the source sample in the range $\sim2.2$ to $\sim2.4$ \citep{Rybicki:1979p6159}, provided such distribution follows a power law.

The general optically thin spectrum of sources is also consistent with the time delays between the 3 and 1\,mm light curves being compatible with zero for all strongly flaring sources (within the time resolution given by our DCF, not shown here).
The latter agrees with standard shock-in-jet models for the most prominent flares in the data \citet{1985ApJ...298..114M,Valtaoja:1992p27721}.
A study of our total flux and polarization data in the framework of these and other models will be presented in further publications including also the analysis of millimetre VLBI polarization data.

The variability analysis of $\alpha$ also shows clear evidence of strong variability (Table~\ref{tab:alpha}).
Indeed, the $\chi^2$ test shows a probability $>99.73\,\%$ for most sources to be variable in spectral index.
Only 0316+413 shows a probability slightly smaller than that, but this is 98.5\,\%, therefore also showing high probability to be variable\footnote{In this paper, we claim variability from a data train whenever its $\chi^2$ test gives a probability $\geq95$\,\%.}.

The PSD slopes of $\alpha$ for each source range from 2.9 (for 1253$-$055) to $\sim0.5$ (for 1730$-$130), while the entire source sample shows a median PSD value $=1.1$, hence showing in general flatter PSD (i.e. more prominent short time--scale of variability as compared to the long time--scale one) than both the 3 and 1\,mm total--flux light curves.
Indeed, $\tau_0$ is actually smaller in general for $\alpha$ than for $S_{3}$ and $S_{1}$, respectively (with the median of ${\tau_{0,\alpha}}=45$\,days), whereas the fractional variability amplitude of $\alpha$ shows similar values in general than those for $S_{3}$ (i.e. median of ${F_\alpha}=0.27$).

The fact that we find a larger fractional amplitude of variability at 1\,mm as compared to the 3\,mm (see previous section) should affect the spectral index if the variability at both wavelengths is coherent, as it seems to be the case at first sight from Fig.~\ref{timevol}.
Therefore, for a larger fractional variability amplitude at 1\,mm, we expect that for intense flares, the spectral index (and hence the energy index of the emitting particle distribution) should raise to levels higher than those at quiescence, even while staying optically thin (i.e. on negative values).
This is exactly the observed effect that is shown for those sources where sufficiently isolated, bright, and well monitored flares were observed at 3 and 1\,mm during the course of our monitoring, e.g. 0528+134, 0735+178, 0827+243, 1226+023, 1611+343, 2230+114, 2251+158.
This spectral hardening in bright flares indicates that the spectrum of the newly injected electron distribution responsible for the emission is harder than the quiescent one, and such that the spectral changes are visible in the short mm range, mostly unaffected by opacity effects.

\subsection{Variability of the linear polarisation degree}
\label{pvar}

Strong variability of the linear polarisation degree ($m_{\rm{L}}$) with time is also quite noticeable from Fig.~\ref{timevol} for almost all sources.
Inspection of these plots shows that the properties of $m_{\rm{L}}$ variability are notably different from those of total flux, and also that the variability amplitude of $m_{\rm{L}}$ is higher at 1\,mm than at 3\,mm. 
This is formally demonstrated from the variability analysis of $m_{\rm{L}}$, see Table~\ref{tab:pL} and paragraphs below.

At 3\,mm, the values of $m_{\rm{L}}$ range from  $\sim0$\,\% (as observed for many sources, see Fig.~\ref{timevol}) up to $\sim15$\,\% (for 0954+658 and 1055+018).
The median $m_{\rm{L}}$ for the entire source sample $\tilde m_{\rm{L}}\approx3$\,\%.
At 1\,mm, $m_{\rm{L}}$ ranges from $\sim0$\,\% to  $16$\,\% (for 1222+216 and 1253$-$055) and the entire source sample has a median $\tilde m_{\rm{L}}\approx6$\,\%.
The comparison of these statistics at both wavelengths suggests a general higher degree of polarisation at 1\,mm as compared to the one at 3\,mm.
This result is not biased by the fact that the 3\,mm and 1\,mm data trains cover different time periods. 
Indeed, the same effect was reported before for a much larger source sample  measuring $m_{\rm{L}}$ simultaneously at 3\,mm and 1\,mm at a single epoch) \citep{Agudo:2014p22485}.
This is confirmed again by the new data set presented here through the analysis of the ratio of 1 to 3\,mm degree of polarization made for every single simultaneous observation, see next section.

At 3\,mm, almost all sources are variable in $m_{\rm{L}}$ according to the $\chi^2$ test with probability $>99.73$\,\% ($P=95\,\%$ for the case of 0430+052), see Table~\ref{tab:pL}.
Only 0316+413 is an exception and cannot be claimed to be variable in $m_{\rm{L}}$ with the current data set.
At 1\,mm  most of the sources (i.e. 24) are variable in $m_{\rm{L}}$ at 1\,mm with $P>95\,\%$.
The remaining 1\,mm sources may well also be variable in $m_{\rm{L}}$, but a formal test does not give significant results due to fewer data points and the generally higher measurement errors at 1\,mm.

\begin{figure}
   \centering
   \includegraphics[width=\columnwidth]{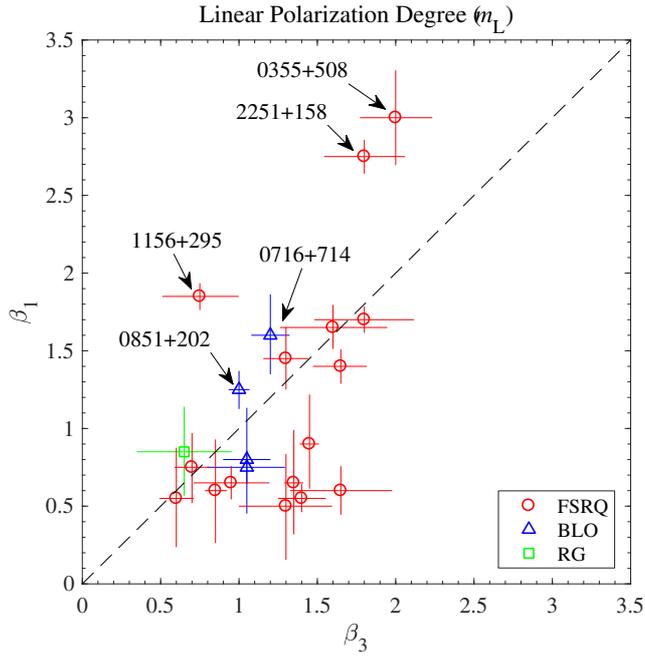}
      \caption{Same as Fig.~\ref{I_beta_3_1mm} but for $m_{\rm{L}}$.}
       \label{p_beta_3_1mm}
\end{figure}

The typical (median) PSD slopes of $m_{\rm{L}}$ data trains at 3 and 1\,mm are $1.3$ and $0.8$, respectively, which are significantly smaller values as compared to those for the total flux ones.
This points out a more prominent short time--scale of $m_{\rm{L}}$ variability than the long time--scale one as compared to the total flux variability, which applies for both the 3 and the 1\,mm data.
However, unlike for the total flux behaviour, for $m_{\rm{L}}$ the $\beta_1$ values only show a rather weak trend to cluster at significantly smaller values as compared to the $\beta_3$ ones, Fig.~\ref{p_beta_3_1mm}.
The main reason for the weakening of this trend for $m_{\rm{L}}$ is the difference in time-range and time sampling for some sources that appear as outliers in Fig.~\ref{p_beta_3_1mm}.

The fractional variability amplitude of $m_{\rm{L}}$ at 3\,mm, with median $\tilde{F_3}=0.53$, show typical values larger than those for the total flux in general. 
However, at 1\,mm, the fractional variability amplitudes of $m_{\rm{L}}$, although with a similar median ($\tilde{F_1}=0.52$), are systematically smaller as compared to the 3\,mm ones (Fig.~\ref{p_Fvar_3_1mm}).
Our definition of $F$ (see Appendix~\ref{appendix}) accounts for both the mean squared error of every data train, which is subtracted from the variance of such data train.
The 1\,mm $m_{\rm{L}}$ errors are rather large (Fig.~\ref{timevol}), therefore probably decreasing the values of $F$.
This may lead to a systematic underestimation of the $F$ values resulting from the measurement errors being systematically higher at 1\,mm than at 3\,mm. 
Although this is a hypothesis that cannot be tested with the current data, we speculate that in analogy with the spectral behaviour of the total flux, the fractional variability amplitude a 1\,mm should show larger values in general than at 3\,mm.
However, the current data actually show the opposite effect, with $F$ at 1\,mm showing systematically smaller values than those at 3\,mm, see Fig.~\ref{p_Fvar_3_1mm}.
Again, we attribute this behaviour to the larger relative $m_{\rm{L}}$ errors at 1\,mm, and not to the intrinsic properties of the $m_{\rm{L}}$ variability at 1\,mm.
 
\begin{figure}
   \centering
   \includegraphics[width=\columnwidth]{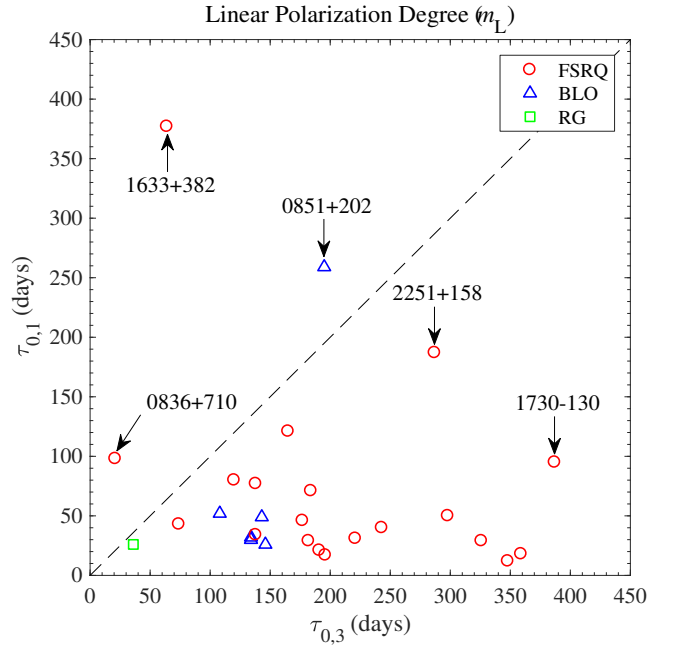}
      \caption{Same as Fig.~\ref{I_tACF_3_1mm} but for $m_{\rm{L}}$. Two outliers, 1730$-$130 and 2200+420, are excluded from the selected scale of the plot, see Table~\ref{tab:pL}.}
       \label{p_tACF_3_1mm}
\end{figure}

The time scales of variability of prominent peaks of $m_{\rm{L}}$, characterised by $\tau_0$, are  longer for the 3\,mm measurements ($\tilde{\tau_0}=171$ days) as compared to the 1\,mm ones ($\tilde{\tau_0}=45$ days).
These numbers are smaller than those characterising the time scale of total flux variability, therefore providing an additional piece of information in support of the faster time scale of linear polarisation variability as compared to the total flux, both for 3 and for 1\,mm.
The effect of more rapid variability at the shortest wavelength detected by $\tau_0$ (also observed for the total flux variability behaviour), is also well reproduced in Fig.~\ref{p_tACF_3_1mm} , where we compare $\tau_0$ at 3 and 1\,mm for every given source with enough $m_{\rm{L}}$ data for the variability analysis.

In summary, we have found that $m_{\rm{L}}$ measured at 1\,mm is larger in general than at 3\,mm.
The variability of linear polarisation degree is significantly faster in prominent flares at 1\,mm as compared to 3\,mm, and it is also more prominent on the shorter time scales studied in this work than on the longer ones as compared to the total flux.
These results are again consistent with a jet model where the bulk of the millimetre emission is produced by different independent inhomogeneity (or turbulence) cells evolving in the jet with different configuration of the magnetic field.
In this way, more rapid variability would be observed in $m_{\rm{L}}$ than in total flux, since the total flux emission is not affected by emission cancelation of orthogonal polarisation components. 
Moreover, the fact that the 1\,mm polarisation is larger than at 3\,mm implies that the average magnetic field is better ordered in the 1\,mm emission regions, which because of their smaller size they produce more rapid variability than at 3\,mm.

\begin{figure}
   \centering
   \includegraphics[width=\columnwidth]{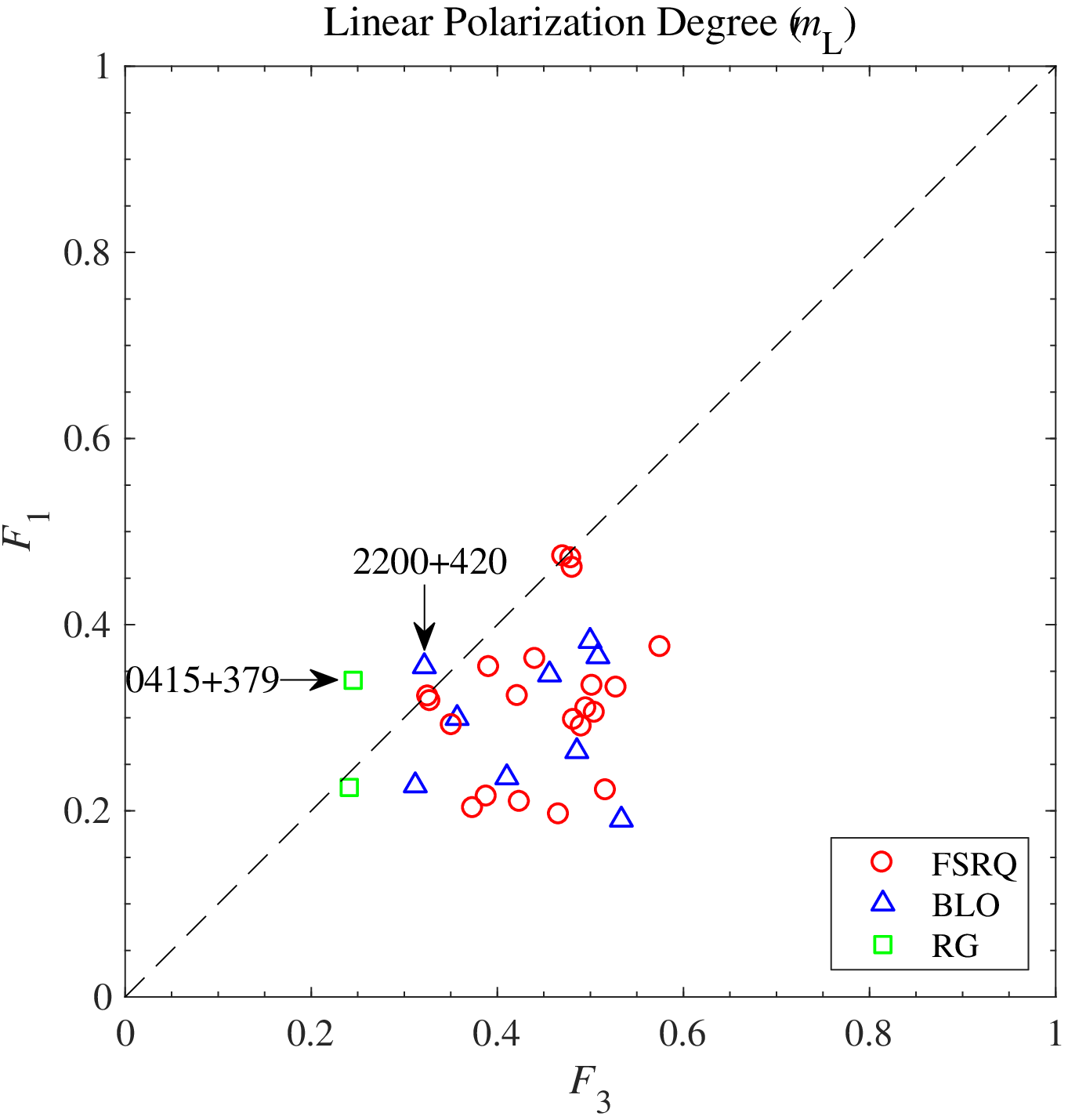}
      \caption{Same as Fig.~\ref{I_Fvar_3_1mm} but for $m_{\rm{L}}$.}
       \label{p_Fvar_3_1mm}
\end{figure}

\subsection{Ratio of 1 to 3\,mm degree of polarisation}

\citet{Agudo:2010p12104} found that the linear polarisation degree measured at a single epoch at 3\,mm was, in median, a factor of $\sim2$ larger than the one measured at 2\,cm for a large sample of 71 AGN (dominated by blazars). 
A similar result was found by \citet{Agudo:2014p22485}, who reported the single epoch 1\,mm linear polarisation degree of a sample of 22 sources to be a factor $\sim1.7$ larger than that at 3\,mm for a smaller subsample of 22 sources for which the $m_{\rm{L,1}}/m_{\rm{L,3}}$ coefficient was estimated from strictly simultaneous data at both wavelengths. 
This evidence was used in support of claims that the magnetic field is progressively better ordered in the blazar jet regions responsible for the emission at progressively smaller millimetre wavelengths \citep{Agudo:2014p22485}.

The new database that we present in this work offer the possibility of a similar study, but this time using a large number of measurements per source and exploring the time variability of the $m_{\rm{L,1}}/m_{\rm{L,3}}$ ratio, see Fig.~\ref{timevol}.
Table~\ref{tab:polfrac} shows the statistical variability analysis of $m_{\rm{L,1}}/m_{\rm{L,3}}$.
The 36 sources showed maxima of $m_{\rm{L,1}}/m_{\rm{L,3}}$ ratios ranging from 1.7 (for 2200+420) and 7.4 (for 1219+285), and minima ranging from 0.3 (for 0954+658) and 2.3 (for 0430+052).
The median values of $m_{\rm{L,1}}/m_{\rm{L,3}}$ for every source range from $1.0$ and $3.9$, with typical (median of the median) values around 2.6. 
The latter number confirms that in general, the 1\,mm linear polarisation degree is higher than the 3\,mm one for AGN jetted sources.
Once again, we note that these are numbers that characterise the relatively small population of sources for which we could perform the variability study.
However, every one of these sources show different properties with regard to their $m_{\rm{L,1}}/m_{\rm{L,3}}$ behaviour, see Fig.~\ref{timevol} and Table~\ref{tab:polfrac}.

The variability study concentrates on 29 out of the 36 sample sources for which we have more than 9 $m_{\rm{L}}$ measurements with SNR$>1$ at both 3 and 1\,mm (Table~\ref{tab:polfrac}), a condition for us to make the variability study.
Among these 29 sources, 10 of them show $>95\,\%$ probability to be variable according to the $\chi^2$ test.
Once again, the large $m_{\rm{L}}$ uncertainties at 1\,mm do not allow to ensure variability of the $m_{\rm{L,1}}/m_{\rm{L,3}}$ fraction for a larger fraction of the sources, although that variability is probably a property of all AGN radio loud jets as well.

\subsection{Variability of linear polarisation angle}
\label{chivar}

The linear polarisation angle ($\chi$) is also found to vary very prominently at both 3 and 1\,mm in all sources of the sample, see Fig.~\ref{timevol}.
This Figure also shows the good (general) correspondence of the $\chi$ evolution curves at both wavelengths on the long time--scales (of months/years), although on the short time--scales (of weeks), such correspondence is somehow weakened.
Indeed, Table~\ref{tab:chi} shows that all sources for which we had enough measurements to perform the $\chi^2$ test, are variable at the $P>99.73$\,\% confidence level.

\begin{figure}
   \centering
   \includegraphics[width=\columnwidth]{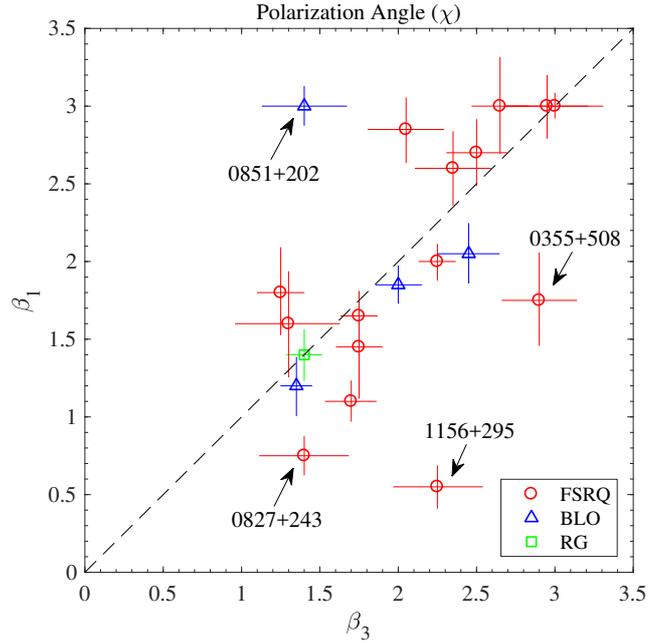}
      \caption{Same as Fig.~\ref{I_beta_3_1mm} but for the linear polarisation angle.}
       \label{chi_beta_3_1mm}
\end{figure}

The $\chi$ variability found in the source sample shows PSD slopes as flat as $\beta_3=0.5$ at 3\,mm (for 0735+178), and as steep as $\beta\approx3.0$ (e.g. for 2230+114 and 2251+158) at both wavelengths.
The median values of $\beta$ of the $\chi$ data trains are 2.1 at 3\,mm, and 1.8 at 1\,mm. 
These similar ranges of values of the PSD slopes at both wavelengths for the source population are also reflected in the one to one comparison for every given source in Fig.~\ref{chi_beta_3_1mm}.

\begin{figure}
   \centering
   \includegraphics[width=\columnwidth]{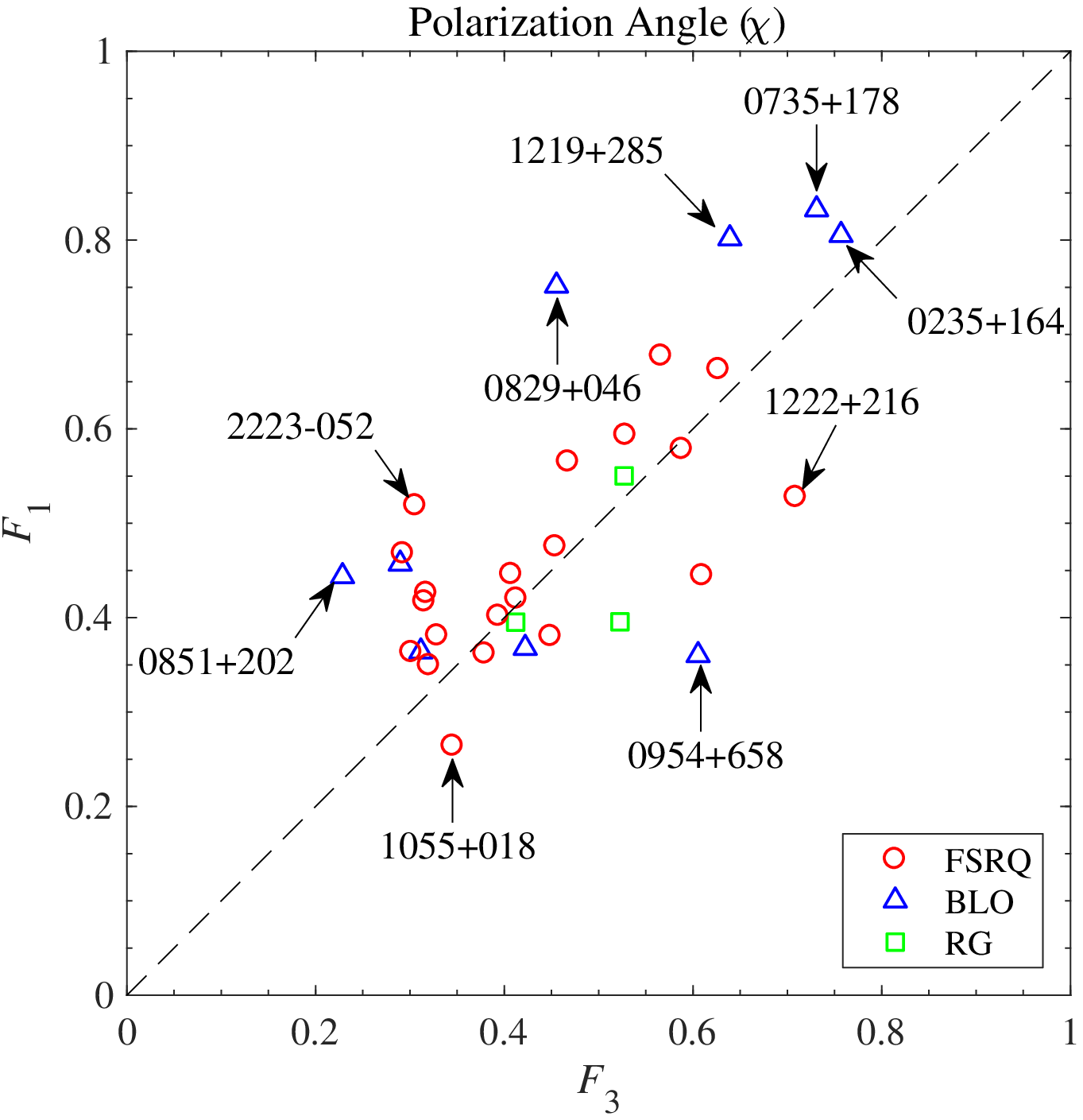}
      \caption{Same as Fig.~\ref{I_Fvar_3_1mm} but for the linear polarisation angle.}
       \label{chi_Fvar_3_1mm}
\end{figure}

The fractional variability amplitude of $\chi$ is very similar in general at both 3\,mm and 1\,mm, with median $\sim0.5$ at both wavelengths.
Such a good correspondence, also illustrated by the limited spread of the cloud of points on Fig.~\ref{chi_Fvar_3_1mm},  is not surprising taking into account the good match of the 3\,mm and 1\,mm $\chi$ evolutions on the long time scales shown in Fig.~\ref{timevol}.
Even if there is more (either intrinsic or artificial) short time scale of variability on the 1\,mm data, the definition of the fractional variability amplitude (Section~\ref{Dis}) makes the scatter on the short time scales to be compensated by the larger mean measurement uncertainties.

\begin{figure}
   \centering
   \includegraphics[width=\columnwidth]{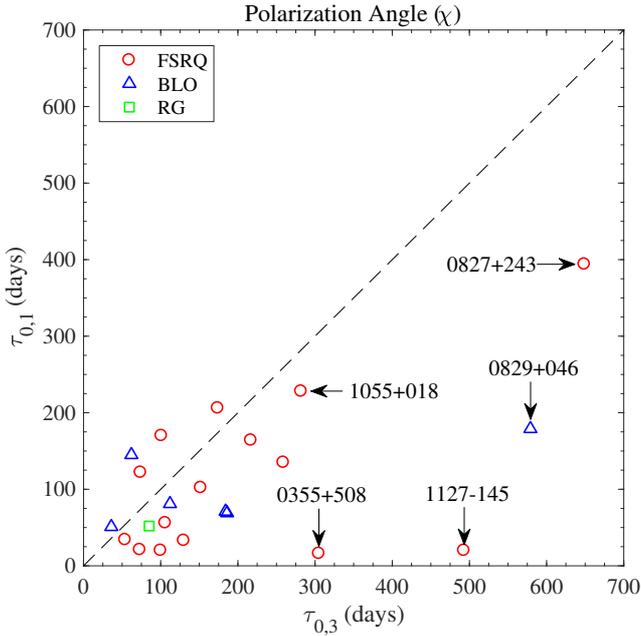}
      \caption{Same as Fig.~\ref{I_tACF_3_1mm} but for the linear polarisation angle. Two outliers, 0235+164 and 0716+714, are excluded from the selected scale of the plot, see Table~\ref{tab:chi}.}
       \label{chi_tACF_3_1mm}
\end{figure}

The auto--correlation function is however more sensitive to changes of the time--scale of variability at different wavebands, irrespective of the measurement error. 
Therefore, Table~\ref{tab:chi} shows significantly shorter $\tau_0$ time scales of variability of $\chi$ for the 1\,mm (with median $=102$\,days) as compared to those at 3\,mm (with median $=281$\,days), see also Fig.~\ref{chi_tACF_3_1mm}.
Together with $\tau_0$, we also computed the minimum time for the linear polarisation angle to rotate in every source by $180^{\circ}$ and $90^{\circ}$, see Table~\ref{tab:chi}.
For 3\,mm, we find that 33 out of 36 sources showed at least one $\chi$ rotation by $\geq90^{\circ}$, whereas 21 of the 36 sources showed at least one $\geq180^{\circ}$ rotation.
For 1\,mm, the number of sources showing large rotations are similar. 
At this wavelength, 32 sources showed at least one $\chi$ rotation by $\geq90^{\circ}$, and 25 sources showed at least a $\geq180^{\circ}$ rotation, which seems to suggest that large $\chi$ rotations is a phenomenon probably inherent to a majority of radio--loud AGN jets.
The typical time scales of the $\geq180^{\circ}$ swings range from medians of 81 days at 1\,mm, to 147 days at 3\,mm, although very different time scales can be found in every particular source (Table~\ref{tab:chi}).
Indeed, the standard deviations on the distribution of time scales of $\geq180^{\circ}$ rotations are 208 and 368 days at 1 and 3\,mm respectively.
This large scatter accounts both for cases of long time scales of $\geq180^{\circ}$ rotations as large as a year or more (e.g. 0829+046 or 1611+343), or rotations in time scales as short as several weeks (e.g. 2251+158), see Fig.~\ref{timevol} and Table~\ref{tab:chi}.
Much faster swings are routinely observed in the optical range \citep[e.g.,][]{Blinov:2016p27469,Blinov:2016p27467}, where sources may rotate by $>180^{\circ}$ in time scales of a few days \citep[e.g.,][]{Marscher:2008p15675,Abdo:2010p11811,Blinov:2016p27469,Blinov:2016p27467}.
However, our time sampling does not permit to check if those fast swings also happen in the millimeter bands.

\subsection{Misalignment between linear polarisation and jet position angles}
\label{misal}
The prominent polarisation angle variability that characterizes most of our sources causes $\chi$ to be, in general, rarely closely aligned with the position angle of the inner jet (see Fig.~\ref{timevol}).
This explains why previous large single--epoch millimetre--polarimetric surveys \citep{Agudo:2010p12104,Agudo:2014p22485} did not find clear signs of $\chi$ aligning parallel (or perpendicular) to the jet position angle, as predicted by axial--symmetric two--dimensional jet--models with the magnetic fields oriented predominantly perpendicular (or parallel, respectively) to the jet \citep[e.g,][]{Lyutikov:2005p321,Cawthorne:2006p409}.
Rather, the misalignment between the linear polarisation angle and the jet position angle was shown to take essentially any possible value within the first quarter of the circle for the entire source sample, quasars and BL~Lac objects.
This is a clear sign of polarisation emission in AGN jets having a remarkable departure from two--dimensional structure. 

The new 3 and 1\,mm data sets of $\chi$ presented here are an order of magnitude larger than previous similar millimeter studies \citep{Agudo:2014p22485,Agudo:2010p12104}.
For calculating the distribution of the misalignment angle $\lvert\chi-\phi_{\rm{jet}}\rvert$, in most of the cases we take as a measurement of the jet position angle ($\phi_{\rm{jet}}$) the averaged values given by \citet{Jorstad:2017VLBApaper} from sequences of 7\,mm VLBI images taken from 2007 to 2014 by the VLBA--BU--BLAZAR Monitoring Program\footnote{\href{url}{http://www.bu.edu/blazars/VLBAproject.html}}. 
However, the 0235+164 value is not given by \citet{Jorstad:2017VLBApaper} and we measured it from the 7\,mm VLBI image on the 23th of October 2016 provided by the VLBA--BU--BLAZAR program. 
For 0355+508, a source not included in \citet{Jorstad:2017VLBApaper}, $\phi_{\rm{jet}}$ was taken from the 3\,mm VLBI image obtained in May 2010 as reported by \citet{Molina:2014p22482}.

\begin{figure}
   \centering
   \includegraphics[width=\columnwidth]{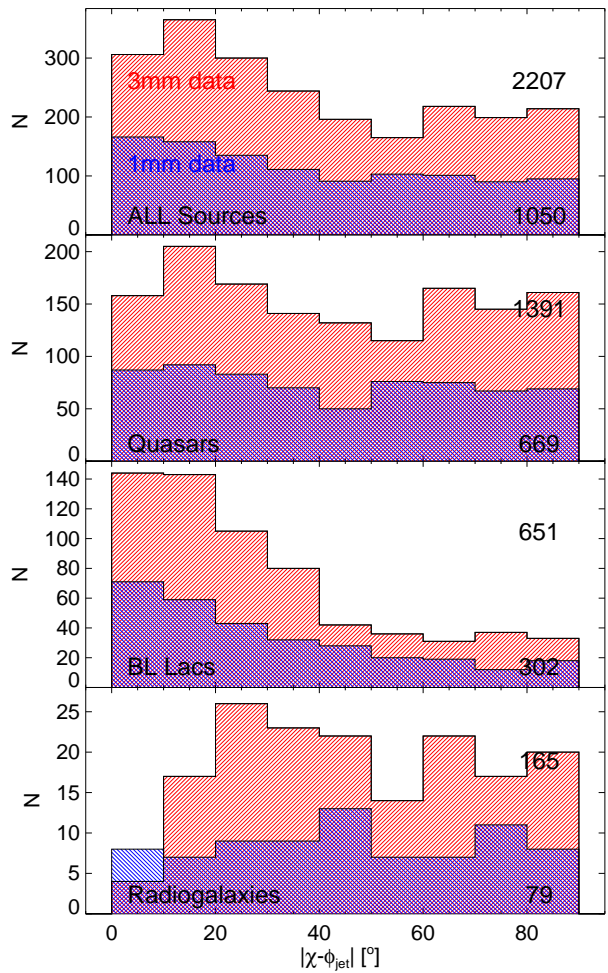}
      \caption{Histograms of the misalignment between the jet position angle ($\phi_{\rm{jet}}$) and all linear polarisation angle measurements for the entire source sample, quasars, BL~Lac objects, and radiogalaxies at both 3 and 1\,mm. Numbers displayed to the right indicate the number of measurements for every distribution. The $\phi_{\rm{jet}}$ measurement is taken in most of the cases from \citet{Jorstad:2017VLBApaper}. Only the measurements of 0235+164 and 0355+508 were taken from a different source, see text.}
       \label{misalign}
\end{figure}

Our new histograms of $\lvert\chi-\phi_{\rm{jet}}\rvert$ (Fig.~\ref{misalign}) confirm the previous results by \citet{Agudo:2010p12104,Agudo:2014p22485} only for quasars.
However, BL~Lac objects show a remarkable tendency to cluster at small misalignment angles, i.e. $\chi$ closely aligned with $\phi_{\rm{jet}}$  \citep[like expected e.g. for jet emission dominated by plane perpendicular shock waves or by toroidal homogeneous magnetic fields, e.g.,][]{Wardle:1994p27464,2000MNRAS.319.1109G,Lister:2001p994}.
Radio galaxies, show a flatter distribution similar to the one of quasars, but with a remarkable deficit of 3\,mm data at small misalignment angles that is not present on the 1\,mm data.
We speculate that this behaviour is produced by the small number of radio galaxies in the sample (only three), that by chance, do not seem to cover the small misalignment region during the time span of our observations.
In further studies covering significantly broader time ranges we will test this hypothesis.

Previous work studying the polarisation properties of the cores of VLBI images report contradicting results on which regard to the alignment of the linear polarisation with the jet position angle.
In particular, \citet{2000MNRAS.319.1109G} observed the polarisation angle of the 6\,cm VLBI cores of a sample of BL Lac objets to align predominantly either parallel or perpendicular to the jet position angle.
\cite{Lister:2001p994} found the cores of both quasars and BL~Lac objects on a sample of 32 blazars observed with 7\,mm VLBI to show a strong tendency to align their polarisation angles parallel to the jet position angle.
Using a larger sample of 177 sources observed with 6\,cm VLBI, \cite{2003ApJ...589..733P} reported a strong tendency of the polarisation angle of the 6 cm VLBI cores of a large sample of 177 AGN to align perpendicular to the jet position angle in short jets. 
No preference was found for BL~Lac objects for the alignment of their polarisation angle with regard to the jet axis.
Later, \cite{Lister:2005p261}, through their 2 cm VLBI observations of a big sample of 133 MOJAVE sources, found a remarkable tendency for BL~Lac objects to cluster at small misalignment angles, whereas quasars showed a flatter distribution of misalignment angles.
The apparent contradiction of the results from these studies and ours, that cannot be explained in its entirety by opacity effects, is again consistent with the remarkable variability of the polarisation angle shown in Fig.~\ref{timevol} and Table~\ref{tab:chi}.

The behaviour of BL~Lac objects shown by Fig.~\ref{misalign} is dominated by 7 of the 11 BL~Lac objects in the sample (i.e. 0219+428, 0829+046, 0851+202, 0954+658, 1101+384, 1219+285, and 2200+420), that seem to prefer to orient $\chi$ at small angles with regard to the 7\,mm jet position angle, whereas the remaining 4 sources do not have a clear preference for the orientation of $\chi$.
The fact that this did not appear on our previous \citet{Agudo:2010p12104,Agudo:2014p22485} surveys, with a far larger number of sources measured at a single epoch, may mean that these 7 BL~Lac objects may be special with regard to the entire BL Lac class, or that we caught them in particularly quiescent states with regard to $\chi$ variability.
The latter hypothesis will be testable by the analysis of the data on a longer time baseline, and by direct comparison with the polarimetric 7\,mm VLBI image sequences accumulated by the VLBA--BU--BLAZAR Monitoring Program \citep{Jorstad:2016p25971,Jorstad:2017VLBApaper}\footnote{\href{url}{https://www.bu.edu/blazars/VLBAproject.html}}.
Note that something similar than found for BL~Lac objects is also observed for a small number of quasars, e.g. 0336$-$019, 0836+710, 1222+216, 1226+023 (in this case with $\chi$ preferentially oriented at $\sim90^{\circ}$ of $\phi_{\rm{jet}}$), and 1641+399, which may be indicating that there are periods when relativistic jets in AGN show their two dimensional character more prominently, perhaps when they are not much affected with dynamic inhomogeneities in their flow.

The polarisation variability behaviour that we report in this paper suggests that there are still some other ingredients, apart from non--axisymmetry, producing a non-alignment of $\chi$ and $\phi_{\rm{jet}}$ that we report for a majority of sources (23, out of the total 36).
In particular, the faster variability of linear polarisation degree with regard to the total flux one reported above indicates that there is in general more than one single dynamical polarisation component driving the observed emission of the sources.
This is evident from the sequences of 7\,mm VLBI polarimetric images of the VLBA--BU-BLAZAR Program. 

The presence of some level of structure in polarisation is known since long ago \citep[e.g.][]{1966MNRAS.133...67B}, since it was needed to explain the low polarisation degrees of relativistic jets in AGN as compared to the theoretical expectation ($\sim70$\,\% for optically thin synchrotron emission of AGN jet plasmas with ordered magnetic field), see \citet[][]{1966MNRAS.133...67B,Rybicki:1979p6159,Marscher:2014p22506}.
To decrease $m_{\rm{L}}$ from $\sim70$\,\% to the average $\sim6$\,\% at 1\,mm that we report in this paper, and under the assumption of a certain level of turbulence in the plasma, $\sim150$ turbulent cells are needed \citep[][]{Marscher:2014p22506}, which would produce a random behaviour of the resultant polarisation angle.
The $\chi$ variability that we report in Fig.~\ref{timevol} does not appear to be totally random in general, which suggests that there is a significantly smaller number of competing dynamical polarisation components governing the main polarisation behaviour on the top of the lower layer of plasma turbulence.
Indeed a small number of such independent dynamical polarisation--components \citep[i.e. around $2$ or $3$, as usually seen on ultra--high resolution VLBI images of AGN jets, e.g.][]{Attridge:2005p220,MartiVidal:2012p26864, Molina:2014p22482,Hada:2016p26899} could reproduce any $\chi$ behaviour as those shown in Fig.~\ref{timevol}.
The fact that we do not see an integrated polarisation degree much larger than $\sim15$\,\% suggests that this is about the maximum degree of polarisation of every one of these most prominent polarisation components (see references above).
This supports the idea of an additional scale of turbulence smaller than the size of the emission components usually see on jet VLBI images.
 
This is again compatible with a model of jets that is mainly driven by a combination of turbulence, perhaps on different scales, and evolving shocked plasma regions.
Turbulent magnetic fields are an ideal scenario for magnetic reconnection \citep[e.g.][and references therein]{Sironi:2015p26776}, which we speculate that should also play a role on the variability properties of the total flux and polarisation of AGN jets, even perhaps with some level of feedback with regard to the production of turbulence.

\subsection{Rotation measure between 3 and 1\,mm}

For estimating the observer's--frame rotation measures (RM), we used, the data from every single epoch for which we had simultaneous measurements of $\chi$ both at 3 and 1\,mm, see Fig.~\ref{timevol}.
The errors in the determination of RM are dominated by the uncertainties on the measurements of $\chi$ at 1\,mm which are $\sim10^{\circ}$, therefore implying errors $\sim2\times10^4\,\rm{rad}\,\rm{m}^{-2}$ in RM.
Indeed, this is the median error of all our computations of RM (for the entire source sample) shown in Fig.~\ref{timevol}, although many particular $\chi$ measurements at 1\,mm affected by larger non systematic uncertainties give much larger RM errors.
Because of the large uncertainties of the RM measurements, we do not provide a detailed variability analysis of this variable, although we can provide constraints about its typical values.

Based on the statistics shown in Table~\ref{tab:RM}, the estimates of RM variations that we report lie $\lesssim10^5\,\rm{rad}\,\rm{m}^{-2}$, which is about an order of magnitude larger than the typical RM detected by \citet{Hovatta:2012p18733} in the large MOJAVE sample \citep[see also][for studies over smaller source samples]{Gabuzda:2015p25146,Kravchenko:2017p26927}.
Although not frequent, values $\sim10^5\,\rm{rad}\,\rm{m}^{-2}$  for the RM in AGN jets are not unusual \citep{Hovatta:2012p18733}, and even seem to be typical in radio--galaxies \citep[e.g.,][]{Zavala:2004p138,Gomez:2008p30,Gomez:2011p16108}

However, the $\sim10^5\,\rm{rad}\,\rm{m}^{-2}$ upper limit is about two orders of magnitude smaller than the high RM reported by \citet{MartiVidal:2015p26904} in the gravitationally lensed AGN PKS~1830$-$211 through ALMA measurements from up to four sky wavelengths ranging from 3 to 1\,mm.
Because we only measured $\chi$ at two wavelengths, and therefore we cannot unambiguously solve for the $\pi$ ambiguity affecting the RM,  our data do not allow us to give a certified explanation about the apparent mismatch between \citet{MartiVidal:2015p26904} observations and ours.
Such mismatch would be solved if a rotation of a multiple of $\sim180^\circ$ between the 3 and 1\,mm measurements of $\chi$ would be applied, which would be allowed to overcome a possible $\pi$ ambiguity.
In this case, the corresponding RM would be a multiple of $\approx3\times10^6\,\rm{rad}\,\rm{m}^{-2}$, which would be more consistent with the measurements by \citet{MartiVidal:2015p26904}.
However, the good general match of $\chi_3$ and $\chi_1$ shown in Fig.~\ref{timevol} suggests that it would be unlikely that either all or a fraction of sources need such a rotation.

An ongoing dedicated program to measure high precision RM with  6 millimetre wavelengths at the IRAM 30\,m Telescope includes some of the sources.
We hope this program will provide robust answers regarding this topic.  
If the low ($\lesssim10^5\,\rm{rad}\,\rm{m}^{-2}$) RM regime is confirmed for the sample, this may support the possibility of the particular character of PKS~1830$-$211 with regard to its large Faraday rotation will be evidenced.
Although it is certain that PKS~1830$-$211 is a high redshift object ($z=2.5$), which makes the mm wavelengths observed  by \citet{MartiVidal:2015p26904} to be shifted to the short sub--millimeter range, this does not explain the apparent mismatch in RM with all sources.
The POLAMI sample also includes high--redshift sources, i.e. 0528+134 (at $z=2.1$) and 0836+710 (at $z=2.2$), but all their RM measurements are still consistent with values $\lesssim10^5\,\rm{rad}\,\rm{m}^{-2}$.
Alternatively, the high angular resolution observations of PKS~1830$-$211 made with ALMA may also play a role on removing some polarisation smearing that may affect the POLAMI single dish observations.

\subsection{Relation between total flux, linear polarisation degree, and polarisation angle variability}
\label{SmLchicorrel}

Inspection of Fig.~\ref{timevol} makes it very difficult to find any noticeable relation of the variability in total flux with that of the linear polarisation degree and polarisation angle both at 3\,mm and 1\,mm for any of the sources in the sample.
Each of these three variables (i.e. $S$, $m_{\rm{L}}$, and $\chi$) seems to follow a complex and independent evolution from each other.
To be sure about this, we made a formal cross--correlation analysis by using the DCF defined in Appendix~\ref{appendix}.
The result of the cross--correlations of all $S$, $m_{\rm{L}}$, and $\chi$ variables of all sources, at both observing wavelengths, did not show any single correlation peak with a significance larger than 99.73, therefore confirming that correlation of $S$, $m_{\rm{L}}$, and $\chi$ is definitely not a general property of the millimetre wave emission of relativistic jets in AGN.

This result may appear surprising if one takes into account that, in origin, the polarised emission of every single emission zone should be directly tied to that of the total flux through the magnetic field in such zone.
For a single zone emission system, that should lead to a coherent time evolution in $S$, $m_{\rm{L}}$, and $\chi$ all along the spectrum. 
However, the fact that we actually observe a radically different behaviour implies that, in general, the millimetre emission in AGN relativistic jets is driven by at least two (probably more) emission components.

However, there are particular sources for which particular time ranges may reveal a coherent behaviour on $S$, $m_{\rm{L}}$, and $\chi$.
This is e.g. the case of 3C454.3 that in the time between mid 2011 and mid 2012 showed a period of exceptionally weak total flux, high polarisation fraction, and stable polarisation angle at a value very close to the jet position angle, see Fig.~\ref{timevol}. 
In rare cases like this, the emission may be dominated by a single region, and therefore perhaps could be well reproduced by a simple one--zone model.
This seems to be the case for 3C454.3, that during the intriguing event mentioned above seems to be dominated by the highly polarised emission of a single traveling jet component detected in the 7\,mm VLBI images of the VLBA--BU-BLAZAR Monitoring Program \citep[see][]{Jorstad:2013p21321,Jorstad:2017VLBApaper}.

We also looked for signs of periodic behaviour in the $S$, $m_{\rm{L}}$, and $\chi$ time evolutions at both 3 and 1\,mm for all sources by analyzing their ACF.
No clear sign of such periodicity was revealed by the data on any of the variables during the time spanned by our observations.

\section{Summary and conclusions}
\label{sumcon}

We find that all sources in the sample are highly variable in total flux at both 3 and 1\,mm.
The same is found for the spectral index, that is found to be optically thin between these two wavelengths, except in some specially prominent flares (not all of them), and for short time-scales as compared to the duration of the flares.
Therefore, although a small portion of the millimetre emission of the source sample might be affected by opacity effects even in non flaring states, we can safely assume that such opacity effects will not be the most dominant cause of the behaviour of our sources in general.
The total flux variability at 1\,mm is found, in general, to be faster, and to have larger fractional amplitude than at 3\,mm, which in turn should lead to flatter PSD slopes at 1\,mm.
This finding is consistent with models of the inner jets where the shortest--wavelength millimetre emission is produced by smaller regions. 
This fits perfectly in the framework of new models for jet emission variability produced by turbulent processes with turbulence cell sizes becoming smaller as emitting wavelength decreases \citep[see][]{Marscher:2014p22506}.
Moreover, the larger fractional amplitude of variability at 1\,mm as compared to the 3\,mm, together with the observed flattening of the spectrum during total flux increases, evidences that the spectrum of the electron distribution injected in flares is harder than the quiescent one. 
The injected electron distribution is such that the spectral changes are visible in the short mm range, and mostly unaffected by opacity effects.

Confirming previous results by \citet{Agudo:2014p22485}, the 1\,mm polarisation degree is also found higher in general than the 3\,mm one by a median factor of 2.6. 
As we rule out strong opacity effects, this implies that the magnetic field is better ordered on the shorter wavelength regions as compared to the longer wavelength ones.
The linear polarisation degree is also highly variable in general, with the time scales of variability in big flares being significantly faster at 1\,mm as compared to the 3\,mm one.
This is compatible with the total flux emission at the short wavelength being also faster than the longer wavelength one if both the polarised and total flux emission is produced in the same sets of regions.
Under the internal turbulence scenario proposed by \citet{Marscher:2014p22506}, such turbulence would distribute the resultant linear polarisation in a different angle for every one of the turbulence cells, which would result not only in a linear polarisation degree far below the theoretical maximum (as it is the case in our observations), but also in faster variability of the linear polarisation with regard to the total flux, as the data shows from the $\tau_0$ time--scales.

The polarisation angle at both 3 and 1\,mm has also been shown to be highly variable, with most of the sources showing at least one excursion of $>180^{\circ}$ on time scales from a few weeks to a year or more during the course of the observations. 
This strong variability is a likely explanation for the general lack of relation found between the jet position angle and the polarisation angle of the monitored sources, although there is a minority of targets that show a remarkable tendency to align the polarisation angle either parallel to the jet (for 7 of the 11 BL~Lac objects observed plus the quasar 1641+399), or perpendicular  (for the case of 4 quasars).
The reason for the preferential alignment in these sources remains to be investigated through the analysis of longer monitoring data and their combination with polarimetric VLBI image sequences.

The 3 and 1\,mm polarisation angle evolution follow rather well each other, although the 1\,mm data shows a clear preference to vary slightly more prominently on the short time scales (of weeks).
This might be produced by a mix of intrinsic short--time--scale variability and scatter on the more noisy 1\,mm data, though.  
In both cases, the high amplitude variations of the polarisation angle do not seem to have any direct correlation with the total flux and polarisation degree variability in general.
On contrary, the polarisation angle variability seems to follow a rather complex evolution as compared to the total flux, the polarisation degree, and even the jet position angle for each particular source.
The latter is, by itself, a sign of non--axisymmetry of the millimeter emission zones of AGN jets.
Moreover, the variability on the polarisation degree also do not seem to be directly related to the one of the total flux.
All this results together automatically imply that, apart from exceptional events of particular sources (like 3C454.3, pointed out in Section~\ref{SmLchicorrel}), the variability of the linearly polarised emission cannot be explained by the time evolution of a single emission component.
This is, any single zone model will fail to explain the polarisation behaviour of radio--loud AGN like those represented in the sample.
Therefore, the number of emission zones in a model capable of explaining together the total flux and polarisation emission of radio--loud AGN should then be larger than one (probably larger than two in some cases) in order to reproduce the apparently erratic behaviour of  $S$, $m_L$ and $\chi$, and their apparent lack of direct relation.

Therefore, in summary, the data rule out single--zone jet models for the standard state of radio loud AGN, and are compatible with general multi--zone non-axisymmetric jet models that involve smaller emission regions for the short wavelength emitting sites, which in turn should also be more efficient in energising particle populations than the long wavelength ones. 
The data also favours the short wavelength emitting regions to have better ordered integrated magnetic fields in general, with different magnetic field orientation for the different emitting regions dominating the emission at every given observing wavelength. 

\section*{Acknowledgements}
     The authors acknowledge the anonymous referee for his/her constructive revision of this paper.
     We gratefully acknowledge Emmanouil Angelakis (MPIfR, Germany) for his careful revision and useful comments to improve this manuscript.
     This paper is based on observations carried out with the IRAM 30~m Telescope. 
     IRAM is supported by INSU/CNRS (France), MPG (Germany) and IGN (Spain).
     IA acknowledges support by a Ram\'on y Cajal grant of the Ministerio de Econom\'ia, Industria y Competitividad (MINECO) of Spain.
     The research at the IAA--CSIC was supported in part by the MINECO through grants AYA2016--80889--P, AYA2013--40825--P, and AYA2010--14844, and by the regional government of Andaluc\'{i}a through grant P09--FQM--4784.
     This research has made use of the VLBA--BU-BLAZAR Program database of 7\,mm VLBA polarimetric images  \citep{Jorstad:2016p25971,Jorstad:2017VLBApaper}, and the NASA/IPAC Extragalactic Database.


\begin{landscape}
\begin{table}
	\centering
	\caption{Variability analysis for total flux ($S$).}
	\label{tab:I}
	\scriptsize
	\begin{threeparttable}
	\begin{tabular}{lcccccccccccccccccccccc}
	\toprule
	& \multicolumn{12}{c}{3~mm} & \multicolumn{10}{c}{1.3~mm} \\
	\cmidrule(lr){2-13} \cmidrule(lr){14-23} 
	Source & $N_{\rm obs}$ & min & max & ratio & $\tilde{S}$ & ${\sigma}_{S}$ & Prob. & $\beta$ & $F_{\rm var}$ & $F_{\rm var}^*$ & $\tau_0$ & $\tau_0^*$ & $N_{\rm obs}$ & min & max & ratio & $\tilde{S}$ & ${\sigma}_{S}$ & Prob. & $\beta$ & $F_{\rm var}$ & $\tau_0$ \\
	& & (Jy) & (Jy) & & (Jy) & (Jy) & & & & & (days) & (days) & & (Jy) & (Jy) &  & (Jy) & (Jy) & & & & (days) \\
	 (0) & (1) & (2) & (3) & (4) & (5) & (6) & (7) & (8) & (9) & (10) & (11) & (12) & (13) & (14) & (15) & (16) & (17) & (18) & (19) & (20) & (21) & (22)  \\
	\midrule
	0219+428 & 31 & 0.30 & 0.83 & 2.77 & 0.54 & 0.15 & $>$99.73 &2.85\,[2.51,\,3.18] &0.26 & 0.26 & 543 & 543 &  7 & 0.30 & 0.68 & 2.27 & 0.49 & 0.14 & $^a$ &$^a$ & $^a$ & $^a$ \\
	0235+164 & 50 & 0.69 & 4.68 & 6.78 & 1.61 & 1.27 & $>$99.73 &1.55\,[1.31,\,1.78] &0.57 & 0.19 & 172 & 454 & 15 & 0.49 & 1.35 & 2.76 & 0.91 & 0.24 & $>$99.73 &$^b$ & 0.26 & $^c$ \\
	0316+413 & 59 & 6.82 & 21.60 & 3.17&18.19 & 2.73 & $>$99.73 &2.30\,[2.25,\,2.35] &0.19 & 0.13 & $^c$&$^c$ & 29 & 5.27 &10.31 & 1.96 & 8.88 & 1.52 & $>$99.73 &1.30\,[0.99,\,1.60] & 0.19 & 490 \\
	0336$-$019 & 37 & 1.03 & 3.38 & 3.28 & 2.25 & 0.57 & $>$99.73 &2.10\,[1.98,\,2.22] &0.28 & 0.24 & 530 & 417 & 19 & 0.59 & 2.21 & 3.75 & 1.31 & 0.40 & $>$99.73 &$^b$ & 0.31 & 594 \\
	0355+508 & 76 & 2.28 & 7.40 & 3.25 & 4.62 & 1.02 & $>$99.73 &1.75\,[1.58,\,1.92] &0.19 & 0.17 & 427 & 291 & 35 & 1.40 & 3.30 & 2.36 & 1.91 & 0.52 & $>$99.73 &1.90\,[1.56,\,2.24] & 0.23 & 205 \\
	0415+379 & 91 & 1.66 & 13.46 & 8.11& 3.70 & 3.20 & $>$99.73 &1.65\,[1.46,\,1.85] &0.54 & 0.24 & 681 & 161 & 47 & 1.04 & 4.12 & 3.96 & 2.44 & 0.81 & $>$99.73 &1.50\,[1.40,\,1.60] & 0.33 & 141 \\
	0420$-$014 & 46 & 1.79 & 7.91 & 4.42 & 4.21 & 1.16 & $>$99.73 &1.90\,[1.76,\,2.03] &0.24 & 0.21 & 318 & 286 & 29 & 0.90 & 3.29 & 3.66 & 2.28 & 0.57 & $>$99.73 &1.30\,[0.99,\,1.61] & 0.26 & 97 \\
	0430+052 & 26 & 0.83 & 5.26 & 6.34 & 2.21 & 1.10 & $>$99.73 &$^b$                &0.43 & 0.35 & 127 & 142 & 14 & 0.80 & 2.80 & 3.50 & 1.37 & 0.60 & $>$99.73 &$^b$ & 0.39 & 75 \\
	0528+134 & 68 & 0.63 & 5.57 & 8.84 & 2.76 & 1.24 & $>$99.73 &2.25\,[2.17,\,2.33] &0.55 & 0.48 & 368 & 285 & 26 & 0.26 & 2.18 & 8.38 & 1.60 & 0.60 & $>$99.73 &1.90\,[1.71,\,2.08] & 0.61 & 245 \\
	0716+714 & 108 & 0.66 & 8.69 &13.17& 3.78 & 1.67 & $>$99.73 &1.45\,[1.33,\,1.57] &0.51 & 0.48 & 301 &  93 & 64 & 0.52 & 7.66 &14.73 & 2.87 & 1.56 & $>$99.73 &1.30\,[1.05,\,1.55] & 0.53 & 79 \\
	0735+178 & 52 & 0.31 & 1.08 & 3.48 & 0.87 & 0.21 & $>$99.73 &2.05\,[1.77,\,2.33] &0.31 & 0.29 & 805 & 769 & 31 & 0.21 & 0.90 & 4.29 & 0.61 & 0.18 & $>$99.73 &2.35\,[2.09,\,2.61] & 0.32 & $^c$ \\
	0827+243 & 73 & 0.50 & 4.02 & 8.04 & 1.52 & 1.01 & $>$99.73 &1.70\,[1.57,\,1.83] &0.60 & 0.63 & 189 & 207 & 42 & 0.26 & 3.45 &13.27 & 1.93 & 1.14 & $>$99.73 &1.65\,[1.58,\,1.72] & 0.94 & 164 \\
	0829+046 & 52 & 0.36 & 1.59 & 4.42 & 0.70 & 0.29 & $>$99.73 &1.45\,[1.31,\,1.59] &0.34 & 0.33 & 236 & 192 & 30 & 0.22 & 1.07 & 4.86 & 0.53 & 0.23 & $>$99.73 &0.70\,[0.39,\,1.02] & 0.41 & 42 \\
	0836+710 & 91 & 0.70 & 3.21 & 4.59 & 1.48 & 0.57 & $>$99.73 &2.65\,[2.59,\,2.70] &0.34 & 0.31 & 743 &$^a$ & 48 & 0.25 & 1.54 & 6.16 & 0.65 & 0.29 & $>$99.73 &1.35\,[1.05,\,1.64] & 0.40 & 294 \\
	0851+202 & 92 & 1.26 & 9.16 & 7.27 & 5.52 & 1.58 & $>$99.73 &2.00\,[1.88,\,2.12] &0.33 & 0.27 & 150 & 138 & 57 & 1.16 & 5.94 & 5.12 & 3.63 & 1.04 & $>$99.73 &0.95\,[0.80,\,1.09] & 0.32 & 105 \\
	0954+658 & 79 & 0.75 & 2.04 & 2.72 & 1.36 & 0.29 & $>$99.73 &$^b$                &0.21 & 0.19 & 111 & 183 & 42 & 0.52 & 1.37 & 2.63 & 0.97 & 0.23 & $>$99.73 &0.60\,[0.41,\,0.79] & 0.24 & 49 \\
	1055+018 & 61 & 1.40 & 5.64 & 4.03 & 4.10 & 1.08 & $>$99.73 &2.35\,[2.24,\,2.46] &0.30 & 0.30 & 280 & 312 & 38 & 0.92 & 4.23 & 4.60 & 2.30 & 0.81 & $>$99.73 &2.05\,[1.91,\,2.19] & 0.35 & 325 \\
	1101+384 & 25 & 0.30 & 0.85 & 2.83 & 0.50 & 0.13 & $>$99.73 &$^b$                &0.23 & 0.23 & 45  &  45 & 18 & 0.10 & 0.58 & 5.80 & 0.32 & 0.11 & $>$99.73 &$^b$ & 0.34 & 33 \\
	1127$-$145 & 56 & 0.86 & 2.85 & 3.31 & 1.64 & 0.41 & $>$99.73 &2.30\,[2.18,\,2.42] &0.24 & 0.22 & 256 & 237 & 28 & 0.43 & 1.28 & 2.98 & 0.72 & 0.25 & $>$99.73 &1.65\,[1.43,\,1.89] & 0.31 & 127 \\
	1156+295 & 59 & 0.62 & 3.07 & 4.95 & 1.37 & 0.63 & $>$99.73 &1.85\,[1.51,\,2.18] &0.42 & 0.34 & 328 & 420 & 32 & 0.34 & 1.54 & 4.53 & 0.82 & 0.35 & $>$99.73 &1.30\,[1.06,\,1.52] & 0.43 & 82 \\
	1219+285 & 33 & 0.27 & 0.72 & 2.67 & 0.44 & 0.15 & $>$99.73 &0.55\,[0.29,\,0.81] &0.31 & 0.31 & 603 & 588 & 17 & 0.10 & 0.72 & 7.20 & 0.34 & 0.16 & $>$99.73 &$^b$ & 0.50 & 584 \\
	1222+216 & 60 & 0.56 & 2.58 & 4.61 & 1.65 & 0.37 & $>$99.73 &1.95\,[1.80,\,2.10] &0.24 & 0.21 & 278 & 279 & 45 & 0.29 & 1.82 & 6.28 & 1.01 & 0.34 & $>$99.73 &1.05\,[0.94,\,1.15] & 0.37 & 189 \\
	1226+023 & 83 & 5.65 & 28.14 & 4.98&11.61 & 5.23 & $>$99.73 &2.45\,[2.34,\,2.56] &0.39 & 0.26 & $^c$&$^c$ & 53 & 1.87 & 8.68 & 4.64 & 4.08 & 1.62 & $>$99.73 &2.10\,[1.77,\,2.44] & 0.38 & 416 \\
	1253$-$055 & 77 & 11.10 & 32.81 &2.96&20.71 & 4.72 & $>$99.73 &$^b$                &0.22 & 0.20 & 418 & 305 & 37 & 6.73 &19.11 & 2.84 &12.35 & 3.20 & $>$99.73 &2.05\,[1.92,\,2.18] & 0.27 & 309 \\
	1308+326 & 54 & 1.11 & 2.33 & 2.10 & 1.66 & 0.28 & $>$99.73 &2.35\,[2.28,\,2.42] &0.16 & 0.17 & 288 & 277 & 35 & 0.36 & 1.44 & 4.00 & 0.97 & 0.25 & $>$99.73 &1.75\,[1.66,\,1.84] & 0.29 & 133 \\                                                                                                                                                                                              
	1406$-$076 & 32 & 0.55 & 1.00 & 1.82 & 0.75 & 0.08 & $>$99.73 &1.15\,[0.80,\,1.47] &0.10 & 0.08 & 464 & 148 & 14 & 0.23 & 0.49 & 2.13 & 0.36 & 0.07 & $>$99.73 &$^b$ & 0.20 & 193 \\
	1510$-$089 & 53 & 1.44 & 5.81 & 4.03 & 2.75 & 1.20 & $>$99.73 &$^b$                &0.38 & 0.38 & 160 & 216 & 27 & 0.80 & 3.20 & 4.00 & 1.89 & 0.67 & $>$99.73 &1.65\,[1.56,\,1.74] & 0.37 & 355 \\
	1611+343 & 69 & 1.02 & 2.70 & 2.65 & 2.18 & 0.43 & $>$99.73 &2.25\,[2.05,\,2.46] &0.24 & 0.19 & $^c$&$^c$ & 43 & 0.40 & 1.52 & 3.80 & 1.13 & 0.28 & $>$99.73 &1.20\,[0.95,\,1.44] & 0.32 & 371 \\
	1633+382 & 76 & 1.92 & 8.03 & 4.18 & 4.11 & 1.64 & $>$99.73 &1.95\,[1.89,\,2.00] &0.38 & 0.36 & 254 & 208 & 48 & 0.99 & 5.50 & 5.56 & 2.85 & 1.16 & $>$99.73 &1.95\,[1.84,\,2.06] & 0.43 & 197 \\
	1641+399 & 80 & 2.65 & 6.37 & 2.40 & 3.65 & 0.94 & $>$99.73 &2.45\,[2.27,\,2.63] &0.21 & 0.15 & 395 & 342 & 52 & 1.09 & 3.69 & 3.39 & 1.76 & 0.59 & $>$99.73 &1.60\,[1.37,\,1.85] & 0.26 & 463 \\
	1730$-$130 & 48 & 2.05 & 5.97 & 2.91 & 3.13 & 0.82 & $>$99.73 &2.30\,[2.03,\,2.57] &0.22 & 0.23 & 668 &$^a$ & 26 & 0.95 & 2.92 & 3.07 & 1.50 & 0.56 & $>$99.73 &2.05\,[1.95,\,2.16] & 0.31 & 709 \\
	1749+096 & 62 & 1.77 & 5.45 & 3.08 & 3.22 & 1.00 & $>$99.73 &1.70\,[1.52,\,1.88] &0.28 & 0.29 & 152 & 148 & 44 & 0.80 & 3.36 & 4.20 & 1.70 & 0.65 & $>$99.73 &1.65\,[1.55,\,1.76] & 0.33 & 166 \\
	2200+420 & 70 & 1.77 & 13.41 & 7.58& 6.98 & 2.64 & $>$99.73 &1.35\,[1.07,\,1.64] &0.45 & 0.31 & $^c$&$^c$ & 44 & 2.85 &10.94 & 3.84 & 5.59 & 1.88 & $>$99.73 &0.95\,[0.82,\,1.09] & 0.31 & 307 \\
	2223$-$052 & 60 & 0.91 & 6.49 & 7.13 & 4.58 & 1.75 & $>$99.73 &$^b$                &0.58 & 0.42 & $^c$&$^c$ & 29 & 0.40 & 2.01 & 5.03 & 1.21 & 0.49 & $>$99.73 &2.10\,[1.97,\,2.23] & 0.49 & 654 \\
	2230+114 & 67 & 0.99 & 5.23 & 5.28 & 3.43 & 1.08 & $>$99.73 &1.95\,[1.87,\,2.03] &0.37 & 0.39 & 472 & 362 & 30 & 0.62 & 2.92 & 4.71 & 1.34 & 0.66 & $>$99.73 &2.10\,[2.01,\,2.20] & 0.43 & 117 \\
	2251+158 & 80 & 2.68 & 45.19 &16.86&24.20 &11.53 & $>$99.73 &2.25\,[2.19,\,2.31] &0.68 & 0.78 & 458 & 312 & 39 & 0.98 &41.37 &42.21 &24.22 & 11.56 & $>$99.73 &1.85\,[1.79,\,1.91] & 0.91 & 290 \\
	\midrule
	min & 25 & 0.27 & 0.72 & 1.82 & 0.44 & 0.08 & -- &0.55\,[0.29,\,0.81] &0.10 & 0.08 & 45 & 45 & 7 & 0.10 & 0.49 & 1.96 & 0.32 & 0.07 & -- &0.60\,[0.41,\,0.79] & 0.19 & 33 \\
	max & 108 & 11.10 & 45.19 & 16.86 & 24.20 & 11.53 & -- &2.85\,[2.51,\,3.18] &0.68 & 0.78 & 805 & 769 & 64 & 6.73 & 41.37 & 42.21 & 24.22 & 11.56 & -- &2.35\,[2.09,\,2.61] & 0.94 & 709 \\
	median & 60 & 2.28 & 13.41 & 4.98 & 6.98 & 2.64 & -- &1.90\,[1.70,\,2.05] &0.28 & 0.24 & 318 & 285 & 33 & 1.40 & 8.68 & 4.64 & 12.35 & 1.56 & -- &1.60\,[1.30,\,1.90] & 0.33 & 197 \\
	std. dev. & 20 & 3.74 & 14.99 & 4.87 & 9.36 & 3.56 & -- &0.54 &0.16 & 0.19 & 198 & 160 & 13 & 2.39 & 15.39 & 16.27 & 10.53 & 4.69 & -- &0.47 & 0.22 & 187 \\
	\bottomrule
	\end{tabular}
	\begin{tablenotes}[para,flushleft]
	Columns are as follows: (0)~Source names following the IAU convention as defined in Paper I, where we also give other common source names; (1)~number of observations; (2)~minimum flux; (3)~maximum flux; (4)~ratio of maximum to minimum flux; (5)~median flux; (6)~standard deviation of flux; (7)~probability of the source being variable according to $\chi^2$ test; (8)~PSD slope with power-law $\propto f^{-\beta}$, {\bf for the time period from December 2009 to August 2014}, along with 68\% confidence intervals (in square brackets); (9)~fractional variability amplitude {\bf for the entire time range spanned by every data train}; {\bf (10)~same as (9) but for the time period from December 2009 to August 2014; (11)~}Zero-crossing time scale from autocorrelation function; {\bf (12) same as (11) but for the time period from December 2009 to August 2014; (13)--(20)} same as {\bf (1)--(8)} but for 1\,mm; {\bf (21) and (22) are same as (10) and (12) but for 1\,mm, respectively.}\\
	$^{a}$ Not enough data. Number of data points $<$10.\\
	$^{b}$ Probability of PSD slope $<$0.05.\\
	$^{c}$ No zero-crossing points in ACF.\\
	\end{tablenotes}
	\end{threeparttable}
\end{table}
\end{landscape}

\begin{landscape}
\begin{table}
	\centering
	\caption{Variability analysis for the Spectral Index  $\alpha$. Columns are same as in Table~\ref{tab:I}}.
	\label{tab:alpha}
	\scriptsize
	\begin{tabular}{lccccccccc}
	\toprule
	Source & $N_{\rm obs}$ & min & max & $\tilde{\alpha}$ & ${\sigma}_{\alpha}$ & Prob. & $\beta$ & $F_{\rm var}$ & $\tau_0$ \\
	&  &  &  &  &  &  &  &  & (days) \\
	\midrule
	0219+428 & 7 & -0.66 & -0.18 & -0.37 & 0.17 & $^a$ &$^a$ & $^a$ & $^a$ \\
	0235+164 & 15 & -0.80 & -0.35 & -0.59 & 0.15 & $>$99.73 &$^b$ & 0.25 & 86 \\
	0316+413 & 29 & -0.97 & -0.53 & -0.76 & 0.10 & 98.58 &0.60\,[0.46,\,0.74] & 0.08 & 20 \\
	0336$-$019 & 19 & -1.08 & -0.41 & -0.62 & 0.18 & $>$99.73 &$^b$ & 0.25 & 45 \\
	0355+508 & 35 & -1.17 & -0.62 & -0.89 & 0.14 & $>$99.73 &1.40\,[1.09,\,1.69] & 0.12 & 196 \\
	0415+379 & 47 & -1.07 & 0.21 & -0.46 & 0.31 & $>$99.73 &1.15\,[1.05,\,1.25] & 0.68 & 169 \\
	0420$-$014 & 29 & -1.40 & -0.50 & -0.72 & 0.20 & $>$99.73 &1.55\,[1.44,\,1.66] & 0.25 & 97 \\
	0430+052 & 14 & -0.76 & -0.28 & -0.42 & 0.15 & $>$99.73 &$^b$ & 0.27 & 26 \\
	0528+134 & 26 & -1.27 & -0.38 & -0.85 & 0.27 & $>$99.73 &1.55\,[1.40,\,1.70] & 0.30 & 314 \\
	0716+714 & 64 & -1.10 & 0.07 & -0.27 & 0.20 & $>$99.73 &0.60\,[0.26,\,0.93] & 0.60 & 22 \\
	0735+178 & 31 & -0.71 & 0.00 & -0.36 & 0.16 & $>$99.73 &2.30\,[2.13,\,2.47] & 0.35 & 93 \\
	0827+243 & 42 & -1.40 & 0.23 & -0.76 & 0.42 & $>$99.73 &1.70\,[1.46,\,1.94] & 0.62 & 164 \\
	0829+046 & 30 & -1.16 & 0.09 & -0.48 & 0.31 & $>$99.73 &0.95\,[0.85,\,1.06] & 0.59 & 35 \\
	0836+710 & 48 & -1.42 & -0.04 & -0.79 & 0.29 & $>$99.73 &1.00\,[0.87,\,1.14] & 0.35 & 40 \\
	0851+202 & 57 & -1.05 & -0.15 & -0.47 & 0.17 & $>$99.73 &0.90\,[0.72,\,1.07] & 0.32 & 22 \\
	0954+658 & 42 & -0.81 & -0.04 & -0.41 & 0.18 & $>$99.73 &0.90\,[0.77,\,1.03] & 0.41 & 41 \\
	1055+018 & 38 & -1.02 & -0.26 & -0.57 & 0.16 & $>$99.73 &$^b$ & 0.19 & 20 \\
	1101+384 & 18 & -1.49 & -0.13 & -0.58 & 0.34 & $>$99.73 &$^b$ & 0.60 & 54 \\
	1127$-$145 & 28 & -1.33 & -0.61 & -0.95 & 0.19 & $>$99.73 &0.70\,[0.47,\,0.92] & 0.17 & 21 \\
	1156+295 & 32 & -1.23 & -0.15 & -0.49 & 0.23 & $>$99.73 &1.40\,[1.30,\,1.50] & 0.39 & 36 \\
	1219+285 & 17 & -1.25 & 0.06 & -0.41 & 0.32 & $>$99.73 &$^b$ & 0.68 & 40 \\
	1222+216 & 45 & -1.34 & -0.34 & -0.61 & 0.25 & $>$99.73 &0.65\,[0.35,\,0.95] & 0.35 & 119 \\
	1226+023 & 53 & -1.61 & -0.54 & -0.95 & 0.22 & $>$99.73 &1.80\,[1.65,\,1.93] & 0.18 & 108 \\
	1253$-$055 & 37 & -0.91 & -0.38 & -0.66 & 0.14 & $>$99.73 &2.90\,[2.78,\,3.02] & 0.19 & 412 \\
	1308+326 & 35 & -1.15 & -0.33 & -0.64 & 0.19 & $>$99.73 &2.85\,[2.65,\,3.06] & 0.25 & 32 \\
	1406$-$076 & 14 & -1.22 & -0.52 & -0.72 & 0.16 & $>$99.73 &$^b$ & 0.20 & 111 \\
	1510$-$089 & 27 & -1.23 & -0.35 & -0.66 & 0.21 & $>$99.73 &1.35\,[1.16,\,1.54] & 0.30 & 28 \\
	1611+343 & 43 & -1.21 & -0.52 & -0.78 & 0.19 & $>$99.73 &0.70\,[0.47,\,0.93] & 0.20 & 191 \\
	1633+382 & 48 & -1.07 & -0.33 & -0.54 & 0.15 & $>$99.73 &1.10\,[0.83,\,1.36] & 0.22 & 46 \\
	1641+399 & 52 & -1.20 & -0.49 & -0.77 & 0.14 & $>$99.73 &1.10\,[1.04,\,1.16] & 0.16 & 54 \\
	1730$-$130 & 26 & -1.01 & -0.48 & -0.76 & 0.15 & $>$99.73 &0.50\,[0.30,\,0.70] & 0.16 & 457 \\
	1749+096 & 44 & -0.93 & -0.27 & -0.60 & 0.16 & $>$99.73 &1.50\,[1.43,\,1.57] & 0.22 & 35 \\
	2200+420 & 44 & -0.59 & -0.00 & -0.21 & 0.13 & $>$99.73 &0.60\,[0.40,\,0.81] & 0.47 & 36 \\
	2223$-$052 & 29 & -1.52 & -0.63 & -0.88 & 0.23 & $>$99.73 &1.10\,[0.83,\,1.38] & 0.24 & 36 \\
	2230+114 & 30 & -0.94 & -0.28 & -0.63 & 0.14 & $>$99.73 &0.75\,[0.47,\,1.04] & 0.19 & 30 \\
	2251+158 & 39 & -1.40 & -0.03 & -0.52 & 0.29 & $>$99.73 &1.35\,[1.15,\,1.56] & 0.53 & 349 \\
	\midrule
	min & 7 & -1.61 & -0.63 & -0.95 & 0.10 & -- &0.50\,[0.30,\,0.70] & 0.08 & 20 \\
	max & 64 & -0.59 & 0.23 & -0.21 & 0.42 & -- &2.90\,[2.78,\,3.02] & 0.68 & 457 \\
	median & 33 & -1.16 & -0.33 & -0.59 & 0.21 & -- &1.10\,[0.90,\,1.40] & 0.27 & 45 \\
	std. dev. & 13 & 0.25 & 0.25 & 0.18 & 0.08 & -- &0.61 & 0.18 & 115 \\
	\bottomrule
	\end{tabular}
\end{table}
\end{landscape}

\begin{landscape}
\begin{table}
	\centering
	\caption{Variability analysis for linear polarization degree($m_{\rm L}$). Columns are same as in Table~\ref{tab:I}.}
	\label{tab:pL}
	\scriptsize
	\begin{tabular}{lcccccccccccccccccccccc}
	\toprule
	& \multicolumn{12}{c}{3~mm} & \multicolumn{10}{c}{1.3~mm} \\
	\cmidrule(lr){2-13} \cmidrule(lr){14-23} 
	Source & $N_{\rm obs}$ & min & max & diff & $\tilde{m_{\rm L}}$ & ${\sigma}_{m_{\rm L}}$ & Prob. & $\beta$ & $F_{\rm var}$ & $F_{\rm var}^*$ & $\tau_0$ & $\tau_0^*$ & $N_{\rm obs}$ & min & max & diff & $\tilde{m_{\rm L}}$ & ${\sigma}_{m_{\rm L}}$ & Prob. & $\beta$ & $F_{\rm var}$ & $\tau_0$ \\
	& & (\%) & (\%) &  & (\%) & (\%) & & & & & (days) & (days) & & (\%) & (\%) &  & (\%) & (\%) & & & & (days) \\
	\midrule
	0219+428 & 30 & 1.0 & 6.5 & 5.5 & 2.3 & 1.3 & $>$99.73 &0.55\,[0.26,\,0.84] &0.41 & 0.41 & 33 & 33 & 4 & 4.7 & 10.6 & 5.9 & 8.9 & 2.5 & $^a$ &$^a$ & $^a$ & $^a$ \\
	0235+164 & 45 & 0.6 & 6.0 & 5.4 & 2.5 & 1.2 & $>$99.73 &0.50\,[0.41,\,0.59] &0.42 & 0.41 & 229 & 422 & 10 & 1.9 & 9.2 & 7.3 & 4.8 & 2.2 & 87.52 &$^b$ & 0.24 & $^c$ \\
	0316+413 & 39 & 0.4 & 2.1 & 1.7 & 0.9 & 0.4 & 42.36 &$^b$ &0.18 & 0.17 & 52 & 34 & 9 & 1.3 & 5.0 & 3.7 & 2.6 & 1.3 & $^a$ &2.85\,[2.59,\,3.11] & $^a$ & $^a$ \\
	0336$-$019 & 36 & 0.7 & 5.6 & 4.9 & 3.0 & 1.2 & $>$99.73 &2.10\,[1.97,\,2.24] &0.41 & 0.37 & 213 & 184 & 12 & 4.0 & 8.4 & 4.4 & 5.7 & 1.5 & 11.34 &$^b$ & 0.20 & 71 \\
	0355+508 & 67 & 0.6 & 7.5 & 6.9 & 2.3 & 1.5 & $>$99.73 &2.00\,[1.77,\,2.23] &0.53 & 0.47 & 195 & 221 & 28 & 1.5 & 6.8 & 5.3 & 4.2 & 1.4 & 53.11 &3.00\,[2.70,\,3.31] & 0.20 & 31 \\
	0415+379 & 66 & 0.5 & 3.2 & 2.7 & 1.4 & 0.6 & $>$99.73 &0.65\,[0.35,\,0.95] &0.25 & 0.25 & 41 & 36 & 25 & 1.4 & 8.7 & 7.3 & 3.8 & 2.2 & 98.70 &0.85\,[0.57,\,1.14] & 0.34 & 26 \\
	0420$-$014 & 45 & 0.5 & 5.8 & 5.3 & 2.5 & 1.4 & $>$99.73 &0.70\,[0.59,\,0.81] &0.49 & 0.51 & 193 & 196 & 23 & 2.2 & 11.7 & 9.5 & 5.1 & 2.3 & $>$99.73 &0.75\,[0.52,\,0.97] & 0.30 & 17 \\
	0430+052 & 19 & 0.9 & 3.0 & 2.1 & 1.8 & 0.7 & 94.87 &$^b$ &0.23 & 0.24 & 219 & 145 & 10 & 2.8 & 9.3 & 6.5 & 5.4 & 2.2 & 86.85 &$^b$ & 0.23 & $^c$ \\
	0528+134 & 62 & 0.5 & 6.8 & 6.3 & 2.1 & 1.2 & $>$99.73 &0.60\,[0.33,\,0.86] &0.46 & 0.42 & 279 & 120 & 18 & 2.2 & 10.2 & 8.0 & 4.8 & 2.4 & 97.56 &$^b$ & 0.32 & 80 \\
	0716+714 & 107 & 0.5 & 8.8 & 8.3 & 3.5 & 2.0 & $>$99.73 &1.20\,[1.08,\,1.32] &0.52 & 0.51 & 94 & 134 & 46 & 1.6 & 11.6 & 10.0 & 4.8 & 2.6 & $>$99.73 &1.60\,[1.35,\,1.86] & 0.37 & 30 \\
	0735+178 & 49 & 0.8 & 7.3 & 6.5 & 2.4 & 1.4 & $>$99.73 &0.90\,[0.64,\,1.16] &0.48 & 0.49 & 170 & 143 & 16 & 1.9 & 9.7 & 7.8 & 3.9 & 2.3 & 95.79 &$^b$ & 0.26 & 49 \\
	0827+243 & 70 & 0.6 & 6.5 & 5.9 & 2.4 & 1.3 & $>$99.73 &1.45\,[1.39,\,1.51] &0.47 & 0.48 & 227 & 191 & 26 & 1.7 & 11.6 & 9.9 & 5.6 & 2.6 & 99.60 &0.90\,[0.61,\,1.22] & 0.30 & 21 \\
	0829+046 & 50 & 1.1 & 9.7 & 8.6 & 3.6 & 2.3 & $>$99.73 &1.30\,[0.96,\,1.64] &0.54 & 0.53 & 160 & 134 & 16 & 4.4 & 14.9 & 10.5 & 8.0 & 2.7 & 87.91 &$^b$ & 0.19 & 32 \\
	0836+710 & 84 & 0.6 & 4.8 & 4.2 & 2.1 & 1.0 & $>$99.73 &0.85\,[0.78,\,0.92] &0.39 & 0.39 & 46 & 21 & 30 & 1.6 & 9.3 & 7.7 & 4.7 & 2.3 & 93.59 &0.60\,[0.26,\,0.93] & 0.21 & 98 \\
	0851+202 & 92 & 1.1 & 12.7 & 11.6 & 7.3 & 2.8 & $>$99.73 &1.00\,[0.93,\,1.07] &0.37 & 0.36 & 169 & 195 & 48 & 2.2 & 14.4 & 12.2 & 8.9 & 3.1 & $>$99.73 &1.25\,[1.13,\,1.37] & 0.30 & 259 \\
	0954+658 & 79 & 2.2 & 14.6 & 12.4 & 8.1 & 2.6 & $>$99.73 &1.05\,[0.90,\,1.20] &0.32 & 0.31 & 62 & 108 & 36 & 2.8 & 12.8 & 10.0 & 6.7 & 2.3 & $>$99.73 &0.80\,[0.45,\,1.13] & 0.23 & 52 \\
	1055+018 & 60 & 1.0 & 14.8 & 13.8 & 5.2 & 3.0 & $>$99.73 &1.35\,[1.29,\,1.41] &0.54 & 0.44 & 320 & 359 & 34 & 1.7 & 12.6 & 10.9 & 6.5 & 2.9 & $>$99.73 &0.65\,[0.32,\,0.99] & 0.36 & 18 \\
	1101+384 & 25 & 1.0 & 5.3 & 4.3 & 3.0 & 1.1 & $>$99.73 &$^b$ &0.27 & 0.27 & 36 & 36 & 5 & 5.0 & 6.1 & 1.1 & 5.3 & 0.4 & $^a$ &$^a$ & $^a$ & $^a$ \\
	1127$-$145 & 54 & 0.9 & 10.1 & 9.2 & 3.0 & 2.2 & $>$99.73 &1.65\,[1.36,\,1.95] &0.56 & 0.58 & 139 & 138 & 21 & 2.4 & 14.2 & 11.8 & 6.4 & 3.2 & $>$99.73 &$^b$ & 0.38 & 77 \\
	1156+295 & 53 & 0.6 & 7.4 & 6.8 & 3.1 & 1.7 & $>$99.73 &0.75\,[0.51,\,1.00] &0.50 & 0.52 & 277 & 177 & 27 & 1.7 & 11.0 & 9.3 & 5.7 & 2.3 & 96.50 &1.85\,[1.76,\,1.93] & 0.22 & 46 \\
	1219+285 & 31 & 0.6 & 6.7 & 6.1 & 3.7 & 1.7 & $>$99.73 &0.75\,[0.64,\,0.87] &0.44 & 0.46 & 62 & 162 & 11 & 3.5 & 13.5 & 10.0 & 5.8 & 3.0 & 99.67 &$^b$ & 0.35 & $^c$ \\
	1222+216 & 60 & 1.1 & 11.7 & 10.6 & 7.0 & 2.3 & $>$99.73 &1.65\,[1.33,\,1.98] &0.35 & 0.33 & 359 & 348 & 37 & 2.9 & 15.7 & 12.8 & 7.0 & 3.0 & $>$99.73 &0.60\,[0.44,\,0.76] & 0.32 & 12 \\
	1226+023 & 82 & 0.6 & 6.5 & 5.9 & 3.4 & 1.2 & $>$99.73 &1.30\,[1.00,\,1.59] &0.32 & 0.33 & 101 & 326 & 32 & 1.4 & 8.1 & 6.7 & 3.3 & 2.0 & 98.79 &0.50\,[0.16,\,0.83] & 0.32 & 29 \\
	1253$-$055 & 76 & 0.7 & 12.9 & 12.2 & 6.2 & 3.3 & $>$99.73 &1.60\,[1.26,\,1.95] &0.55 & 0.47 & $^c$ & $^c$ & 32 & 1.2 & 15.9 & 14.7 & 8.0 & 3.9 & $>$99.73 &1.65\,[1.51,\,1.80] & 0.47 & $^c$ \\
	1308+326 & 52 & 0.6 & 7.1 & 6.5 & 3.0 & 1.8 & $>$99.73 &1.30\,[1.16,\,1.44] &0.51 & 0.50 & 155 & 138 & 26 & 2.1 & 14.6 & 12.5 & 6.0 & 3.0 & 99.07 &1.45\,[1.25,\,1.65] & 0.33 & 34 \\
	1406$-$076 & 30 & 1.0 & 7.0 & 6.0 & 4.0 & 1.5 & $>$99.73 &1.75\,[1.48,\,2.02] &0.35 & 0.36 & 53 & 196 & 6 & 3.1 & 13.8 & 10.7 & 7.0 & 3.6 & $^a$ &$^a$ & $^a$ & $^a$ \\
	1510$-$089 & 53 & 0.6 & 6.5 & 5.9 & 2.6 & 1.4 & $>$99.73 &0.60\,[0.49,\,0.71] &0.46 & 0.48 & 43 & 74 & 21 & 2.1 & 12.5 & 10.4 & 5.6 & 3.3 & $>$99.73 &0.55\,[0.24,\,0.88] & 0.47 & 43 \\
	1611+343 & 65 & 0.5 & 8.1 & 7.6 & 3.6 & 1.9 & $>$99.73 &1.65\,[1.47,\,1.82] &0.53 & 0.49 & 188 & 165 & 34 & 1.9 & 13.6 & 11.7 & 5.4 & 2.5 & $>$99.73 &1.40\,[1.29,\,1.51] & 0.29 & 121 \\
	1633+382 & 71 & 0.4 & 5.6 & 5.2 & 2.5 & 1.2 & $>$99.73 &0.95\,[0.71,\,1.19] &0.43 & 0.42 & 85 & 64 & 32 & 1.9 & 9.3 & 7.4 & 4.6 & 1.9 & 84.02 &0.65\,[0.54,\,0.76] & 0.21 & 377 \\
	1641+399 & 79 & 0.5 & 8.6 & 8.1 & 3.5 & 1.6 & $>$99.73 &1.80\,[1.48,\,2.12] &0.43 & 0.39 & 363 & 182 & 44 & 1.6 & 12.4 & 10.8 & 4.3 & 2.5 & $>$99.73 &1.70\,[1.62,\,1.78] & 0.35 & 29 \\
	1730$-$130 & 48 & 0.6 & 6.5 & 5.9 & 2.4 & 1.5 & $>$99.73 &1.50\,[1.28,\,1.72] &0.50 & 0.50 & 538 & 387 & 13 & 1.9 & 9.2 & 7.3 & 3.0 & 2.1 & 82.76 &$^b$ & 0.31 & 95 \\
	1749+096 & 60 & 0.4 & 7.2 & 6.8 & 3.3 & 1.7 & $>$99.73 &1.05\,[0.80,\,1.29] &0.50 & 0.50 & 171 & 146 & 37 & 1.5 & 10.2 & 8.7 & 5.2 & 2.5 & $>$99.73 &0.75\,[0.64,\,0.86] & 0.38 & 26 \\
	2200+420 & 70 & 2.4 & 13.5 & 11.1 & 8.1 & 2.7 & $>$99.73 &1.95\,[1.63,\,2.27] &0.34 & 0.32 & 844 & 465 & 41 & 2.6 & 14.9 & 12.3 & 7.5 & 3.1 & $>$99.73 &$^b$ & 0.36 & 395 \\
	2223$-$052 & 59 & 0.8 & 6.9 & 6.1 & 2.7 & 1.5 & $>$99.73 &1.75\,[1.65,\,1.85] &0.48 & 0.48 & 328 & 243 & 19 & 1.5 & 10.3 & 8.8 & 4.0 & 2.8 & $>$99.73 &$^b$ & 0.46 & 40 \\
	2230+114 & 63 & 0.6 & 6.7 & 6.1 & 3.2 & 1.2 & $>$99.73 &1.40\,[1.25,\,1.55] &0.35 & 0.35 & 257 & 298 & 22 & 1.7 & 8.9 & 7.2 & 4.8 & 2.3 & 95.53 &0.55\,[0.46,\,0.64] & 0.29 & 50 \\
	2251+158 & 73 & 0.6 & 7.5 & 6.9 & 3.2 & 2.1 & $>$99.73 &1.80\,[1.54,\,2.06] &0.59 & 0.53 & 274 & 287 & 28 & 1.7 & 9.7 & 8.0 & 4.7 & 2.1 & $>$99.73 &2.75\,[2.64,\,2.86] & 0.33 & 187 \\
	\midrule
	min & 19 & 0.4 & 2.1 & 1.7 & 0.9 & 0.4 & -- &0.50\,[0.41,\,0.59] &0.18 & 0.17 & 33 & 21 & 4 & 1.2 & 5.0 & 1.1 & 2.6 & 0.4 & -- &0.50\,[0.16,\,0.83] & 0.19 & 12 \\
	max & 107 & 2.4 & 14.8 & 13.8 & 8.1 & 3.3 & -- &2.10\,[1.97,\,2.24] &0.59 & 0.58 & 844 & 465 & 48 & 5.0 & 15.9 & 14.7 & 8.9 & 3.9 & -- &3.00\,[2.70,\,3.31] & 0.47 & 395 \\
	median & 60 & 0.6 & 7.3 & 6.2 & 3.2 & 1.7 & -- &1.30\,[0.95,\,1.50] &0.35 & 0.36 & 171 & 165 & 25 & 2.1 & 11.0 & 9.1 & 5.6 & 2.5 & -- &0.80\,[0.65,\,0.90] & 0.30 & 45 \\
	std. dev. & 19 & 0.4 & 3.1 & 2.8 & 1.7 & 0.7 & -- &0.47 &0.13 & 0.13 & 162 & 115 & 12 & 1.0 & 2.7 & 2.8 & 1.6 & 0.6 & -- &0.80 & 0.08 & 101 \\
	\bottomrule
	\end{tabular}
\end{table}
\end{landscape}

\begin{landscape}
\begin{table}
	\centering
	\caption{Variability analysis for the polarization fraction  $m_{\rm L,1/3}$. Columns are same as in Table~\ref{tab:I}}.
	\label{tab:polfrac}
	\scriptsize
	\begin{tabular}{lccccccccc}
	\toprule
	Source & $N_{\rm obs}$ & min & max & $\tilde{m_{\rm L,1/3}}$ & ${\sigma}_{m_{\rm L,1/3}}$ & Prob. & $\beta$ & $F_{\rm var}$ & $\tau_0$ \\
	& &  &  &  &  &  &  &  & (days) \\
	\midrule
	0219+428 & 4 & 2.2 & 4.4 & 2.4 & 1.0 & $^a$ &$^a$ & $^a$ & $^a$ \\
	0235+164 & 9 & 0.7 & 2.7 & 2.2 & 0.5 & $^a$ &$^a$ & $^a$ & $^a$ \\
	0316+413 & 3 & 1.1 & 3.5 & 3.5 & 0.9 & $^a$ &$^a$ & $^a$ & $^a$ \\
	0336$-$019 & 12 & 1.3 & 3.6 & 2.4 & 0.8 & 19.87 &$^b$ & 0.19 & 125 \\
	0355+508 & 22 & 0.7 & 4.0 & 1.8 & 1.0 & 22.98 &$^b$ & 0.26 & 40 \\
	0415+379 & 13 & 0.9 & 3.4 & 2.0 & 1.0 & 16.60 &$^b$ & 0.56 & 88 \\
	0420$-$014 & 22 & 0.5 & 4.2 & 2.1 & 1.0 & $>$99.73 &$^b$ & 0.04 & 31 \\
	0430+052 & 6 & 2.3 & 6.1 & 3.9 & 1.4 & $^a$ &$^a$ & $^a$ & $^a$ \\
	0528+134 & 13 & 0.8 & 3.4 & 2.4 & 0.8 & 55.31 &$^b$ & 0.26 & 80 \\
	0716+714 & 44 & 0.5 & 3.8 & 1.5 & 0.9 & 85.66 &2.40\,[2.24,\,2.55] & 0.24 & 21 \\
	0735+178 & 16 & 0.9 & 4.8 & 2.7 & 1.2 & 59.12 &$^b$ & 0.04 & 26 \\
	0827+243 & 21 & 0.7 & 6.3 & 2.5 & 1.7 & 98.70 &$^b$ & 0.31 & 42 \\
	0829+046 & 15 & 0.9 & 5.0 & 2.8 & 1.5 & 97.22 &$^b$ & 0.49 & $^c$ \\
	0836+710 & 28 & 1.0 & 4.5 & 2.6 & 1.1 & 18.05 &0.50\,[0.31,\,0.70] & 0.35 & 25 \\
	0851+202 & 48 & 0.5 & 2.1 & 1.1 & 0.4 & 97.90 &2.95\,[2.82,\,3.08] & 0.08 & 17 \\
	0954+658 & 36 & 0.3 & 2.1 & 1.0 & 0.5 & $>$99.73 &0.70\,[0.45,\,0.95] & 0.25 & 45 \\
	1055+018 & 34 & 0.4 & 3.4 & 1.5 & 0.8 & 99.31 &0.60\,[0.30,\,0.90] & 0.26 & 24 \\
	1101+384 & 4 & 1.3 & 2.7 & 2.7 & 0.6 & $^a$ &$^a$ & $^a$ & $^a$ \\
	1127$-$145 & 20 & 0.6 & 3.8 & 1.9 & 1.0 & 63.08 &$^b$ & 0.11 & 29 \\
	1156+295 & 21 & 0.9 & 6.1 & 2.5 & 1.5 & 94.35 &$^b$ & 0.30 & 30 \\
	1219+285 & 9 & 0.6 & 7.4 & 3.4 & 2.4 & $^a$ &$^a$ & $^a$ & $^a$ \\
	1222+216 & 37 & 0.4 & 2.8 & 1.4 & 0.7 & $>$99.73 &0.60\,[0.38,\,0.81] & 0.37 & 107 \\
	1226+023 & 32 & 0.4 & 5.3 & 1.3 & 1.4 & 88.47 &0.80\,[0.56,\,1.04] & 0.56 & 26 \\
	1253$-$055 & 32 & 0.7 & 2.3 & 1.2 & 0.4 & 35.78 &2.95\,[2.89,\,3.01] & 0.15 & 39 \\
	1308+326 & 25 & 0.4 & 6.1 & 2.6 & 1.5 & 95.88 &$^b$ & 0.43 & 146 \\
	1406$-$076 & 6 & 0.5 & 5.8 & 3.0 & 2.0 & $^a$ &$^a$ & $^a$ & $^a$ \\
	1510$-$089 & 21 & 0.8 & 5.0 & 2.2 & 1.0 & 93.79 &$^b$ & 0.17 & 27 \\
	1611+343 & 30 & 0.7 & 4.2 & 2.0 & 0.9 & 94.73 &2.70\,[2.36,\,3.05] & 0.12 & 116 \\
	1633+382 & 30 & 0.5 & 3.1 & 1.8 & 0.6 & 29.99 &2.55\,[2.31,\,2.79] & 0.41 & 19 \\
	1641+399 & 43 & 0.4 & 5.7 & 1.6 & 1.4 & 87.56 &2.70\,[2.55,\,2.84] & 0.42 & 127 \\
	1730$-$130 & 13 & 0.4 & 3.4 & 1.8 & 0.9 & 81.77 &$^b$ & 0.14 & 82 \\
	1749+096 & 34 & 0.6 & 4.2 & 2.1 & 1.2 & 99.56 &0.70\,[0.59,\,0.81] & 0.33 & 116 \\
	2200+420 & 41 & 0.5 & 1.7 & 1.0 & 0.3 & 87.22 &2.80\,[2.75,\,2.85] & 0.12 & 33 \\
	2223$-$052 & 19 & 0.5 & 4.0 & 1.8 & 1.1 & 95.48 &$^b$ & 0.15 & 77 \\
	2230+114 & 22 & 0.6 & 2.8 & 1.4 & 0.6 & 45.02 &$^b$ & 0.15 & 38 \\
	2251+158 & 27 & 0.5 & 3.0 & 1.4 & 0.8 & 40.98 &2.65\,[2.51,\,2.79] & 0.32 & 28 \\
	\midrule
	min & 3 & 0.3 & 1.7 & 1.0 & 0.3 & -- &0.50\,[0.31,\,0.70] & 0.04 & 17 \\
	max & 48 & 2.3 & 7.4 & 3.9 & 2.4 & -- &2.95\,[2.82,\,3.08] & 0.56 & 146 \\
	median & 21 & 0.8 & 4.2 & 2.6 & 1.2 & -- &0.80\,[0.60,\,2.55] & 0.15 & 38 \\
	std. dev. & 12 & 0.5 & 1.3 & 0.7 & 0.5 & -- &1.02 & 0.15 & 40 \\
	\bottomrule
	\end{tabular}
\end{table}
\end{landscape}

\begin{landscape}
\begin{table}
	\centering
	\caption{Variability analysis for Polarization Angle ($\chi$). Columns are same as in Table~\ref{tab:I}. Also shown are the minimum number of days for $\chi$ to rotate by 180$^{\circ}$ and 90$^{\circ}$ (rotate) at 3 and 1.3~mm, respectively.}
	\label{tab:chi}
	\scriptsize
	\begin{tabular}{lcccccccccccccccccccccccc}
	\toprule
	& \multicolumn{13}{c}{3~mm} & \multicolumn{11}{c}{1.3~mm} \\
	\cmidrule(lr){2-14} \cmidrule(lr){15-25} 
	Source & $N_{\rm obs}$ & min & max & diff & $\tilde{\chi}$ & ${\sigma}_{\chi}$ & Prob. & $\beta$ & $F_{\rm var}$ & $F_{\rm var}^*$ & $\Delta(\chi)$ & $\tau_0$ & $\tau_0^*$ & $N_{\rm obs}$ & min & max & diff & $\tilde{\chi}$ & ${\sigma}_{\chi}$ & Prob. & $\beta$ & $F_{\rm var}$ & $\Delta(\chi)$ & $\tau_0$ \\
	& & ($^{\circ}$) & ($^{\circ}$) &  & ($^{\circ}$) & ($^{\circ}$) & & & & & (days) & (days) & (days) & & ($^{\circ}$) & ($^{\circ}$) &  & ($^{\circ}$) & ($^{\circ}$) & & & & (days) & (days) \\
	\midrule
	0219+428 & 30 & -106 & 51 & 157 & -5 & 27 & $>$99.73 &3.00\,[2.91,\,3.09] & 0.29 & 0.29 & $^a$~/~170 &40 & 40 & 4 & -42 & 37 & 79 & 34 & 35 & $^a$ &$^a$ & $^a$ & $^a$ & $^a$ \\
	0235+164 & 50 & -238 & 89 & 327 & -177 & 92 & $>$99.73 &2.75\,[2.58,\,2.92] & 0.64 & 0.76 & 244~/~96 &793 & 184 & 14 & -253 & -7 & 246 & -168 & 57 & $>$99.73 &$^b$ & 0.81 & 196~/~15 & 71 \\
	0316+413 & 54 & -169 & 59 & 228 & -42 & 51 & $>$99.73 &1.05\,[0.84,\,1.25] & 0.53 & 0.53 & 295~/~16 &86 & 85 & 24 & -163 & 126 & 289 & -10 & 73 & $>$99.73 &$^b$ & 0.55 & 30~/~21 & 52 \\
	0336$-$019 & 36 & 29 & 123 & 94 & 76 & 30 & $>$99.73 &1.00\,[0.70,\,1.32] & 0.62 & 0.63 & $^a$~/~868 &601 & 106 & 14 & 25 & 113 & 88 & 73 & 28 & $>$99.73 &$^b$ & 0.66 & $^a$ & 56 \\
	0355+508 & 76 & -49 & 165 & 214 & 88 & 50 & $>$99.73 &2.90\,[2.66,\,3.14] & 0.55 & 0.32 & 518~/~98 &426 & 305 & 34 & -24 & 210 & 234 & 119 & 55 & $>$99.73 &1.75\,[1.46,\,2.06] & 0.42 & 518~/~6 & 16 \\
	0415+379 & 87 & -276 & 276 & 552 & -24 & 161 & $>$99.73 &1.40\,[1.29,\,1.51] & 0.57 & 0.52 & 109~/~25 &$^c$ & $^c$ & 43 & -145 & 321 & 466 & -4 & 132 & $>$99.73 &1.40\,[1.23,\,1.56] & 0.40 & 78~/~12 & $^c$ \\
	0420$-$014 & 46 & -179 & 118 & 297 & -37 & 73 & $>$99.73 &2.35\,[2.11,\,2.59] & 0.44 & 0.45 & 239~/~52 &$^c$ & $^c$ & 29 & -216 & 51 & 267 & -16 & 62 & $>$99.73 &2.60\,[2.36,\,2.84] & 0.38 & 81~/~28 & 164 \\
	0430+052 & 24 & -84 & 38 & 122 & -20 & 36 & $>$99.73 &$^b$ & 0.40 & 0.41 & $^a$~/~19 &63 & 63 & 12 & -220 & 100 & 320 & 29 & 81 & $>$99.73 &$^b$ & 0.40 & 12~/~12 & $^c$ \\
	0528+134 & 66 & -221 & 43 & 264 & -65 & 63 & $>$99.73 &1.65\,[1.31,\,1.98] & 0.38 & 0.38 & 145~/~27 &130 & 152 & 21 & -250 & 28 & 278 & -42 & 71 & $>$99.73 &$^b$ & 0.36 & 44~/~44 & 102 \\
	0716+714 & 108 & -116 & 295 & 411 & 147 & 113 & $>$99.73 &2.45\,[2.25,\,2.65] & 0.43 & 0.31 & 95~/~29 &598 & $^a$ & 60 & -109 & 382 & 491 & 193 & 141 & $>$99.73 &2.05\,[1.86,\,2.25] & 0.36 & 37~/~0 & 743 \\
	0735+178 & 52 & -91 & 86 & 177 & -49 & 38 & $>$99.73 &0.50\,[0.37,\,0.63] & 0.74 & 0.73 & $^a$~/~49 &451 & 186 & 21 & -68 & 119 & 187 & -9 & 56 & $>$99.73 &$^b$ & 0.83 & 468~/~168 & 69 \\
	0827+243 & 71 & -268 & 106 & 374 & 2 & 97 & $>$99.73 &1.40\,[1.11,\,1.68] & 0.52 & 0.41 & 232~/~39 &$^c$ & $^c$ & 32 & -254 & 117 & 371 & 15 & 79 & $>$99.73 &0.75\,[0.62,\,0.88] & 0.45 & 148~/~14 & 394 \\
	0829+046 & 52 & -174 & 77 & 251 & -81 & 70 & $>$99.73 &2.65\,[2.48,\,2.82] & 0.53 & 0.46 & 396~/~140 &529 & 579 & 16 & -145 & 77 & 222 & -92 & 58 & $>$99.73 &$^b$ & 0.75 & 566~/~270 & 179 \\
	0836+710 & 91 & -272 & 219 & 491 & -108 & 129 & $>$99.73 &1.75\,[1.60,\,1.90] & 0.45 & 0.47 & 145~/~30 &$^c$ & $^c$ & 38 & -300 & 111 & 411 & -113 & 106 & $>$99.73 &1.45\,[1.12,\,1.77] & 0.57 & 144~/~9 & $^c$ \\
	0851+202 & 92 & -97 & 27 & 124 & -32 & 21 & $>$99.73 &1.40\,[1.13,\,1.67] & 0.24 & 0.23 & $^a$~/~485 &303 & 62 & 49 & -57 & 8 & 65 & -33 & 15 & $>$99.73 &3.00\,[2.87,\,3.13] & 0.44 & $^a$ & 145 \\
	0954+658 & 79 & -19 & 65 & 84 & 1 & 18 & $>$99.73 &$^b$ & 0.65 & 0.61 & $^a$ &36 & 36 & 36 & -28 & 29 & 57 & -1 & 14 & $>$99.73 &$^b$ & 0.36 & $^a$ & 51 \\
	1055+018 & 60 & -101 & 68 & 169 & -27 & 36 & $>$99.73 &2.05\,[1.80,\,2.29] & 0.34 & 0.35 & $^a$~/~132 &195 & 282 & 38 & -123 & 48 & 171 & -13 & 38 & $>$99.73 &2.85\,[2.63,\,3.06] & 0.26 & $^a$~/~29 & 228 \\
	1101+384 & 25 & -78 & 14 & 92 & -32 & 26 & $>$99.73 &$^b$ & 0.55 & 0.56 & $^a$~/~663 &45 & 45 & 7 & -63 & 35 & 98 & -13 & 29 & $^a$ &$^a$ & $^a$ & $^a$~/~138 & $^a$ \\
	1127$-$145 & 55 & -129 & 8 & 137 & -61 & 31 & $>$99.73 &1.30\,[0.96,\,1.63] & 0.32 & 0.29 & $^a$~/~171 &417 & 493 & 26 & -134 & 113 & 247 & -47 & 75 & $>$99.73 &1.60\,[1.25,\,1.94] & 0.47 & 278~/~13 & 20 \\
	1156+295 & 58 & -39 & 177 & 216 & 69 & 52 & $>$99.73 &2.25\,[1.97,\,2.54] & 0.54 & 0.59 & 116~/~42 &93 & 101 & 30 & -44 & 220 & 264 & 68 & 75 & $>$99.73 &0.55\,[0.41,\,0.69] & 0.58 & 33~/~13 & 170 \\
	1219+285 & 33 & -127 & 27 & 154 & -84 & 48 & $>$99.73 &0.95\,[0.77,\,1.13] & 0.63 & 0.64 & $^a$~/~30 &362 & 280 & 11 & -118 & 33 & 151 & -69 & 48 & $>$99.73 &$^b$ & 0.80 & $^a$~/~354 & $^c$ \\
	1222+216 & 60 & -22 & 93 & 115 & -8 & 28 & $>$99.73 &2.50\,[2.31,\,2.69] & 0.74 & 0.71 & $^a$~/~61 &180 & 100 & 37 & -46 & 101 & 147 & -8 & 22 & $>$99.73 &2.70\,[2.49,\,2.92] & 0.53 & $^a$~/~9 & 20 \\
	1226+023 & 82 & -74 & -14 & 60 & -44 & 13 & $>$99.73 &$^b$ & 0.35 & 0.32 & $^a$ &51 & 73 & 40 & -111 & 36 & 147 & -46 & 29 & $>$99.73 &0.50\,[0.26,\,0.75] & 0.35 & $^a$~/~15 & 21 \\
	1253$-$055 & 77 & -192 & 220 & 412 & -53 & 129 & $>$99.73 &2.65\,[2.47,\,2.83] & 0.54 & 0.57 & 168~/~35 &$^c$ & $^c$ & 34 & -118 & 223 & 341 & 5 & 129 & $>$99.73 &3.00\,[2.69,\,3.32] & 0.68 & 22~/~22 & $^c$ \\
	1308+326 & 54 & -183 & 13 & 196 & -83 & 46 & $>$99.73 &1.25\,[1.10,\,1.40] & 0.32 & 0.30 & 1750~/~55 &89 & 74 & 32 & -185 & -12 & 173 & -70 & 49 & $>$99.73 &1.80\,[1.53,\,2.09] & 0.36 & $^a$~/~42 & 122 \\
	1406$-$076 & 32 & -42 & 162 & 204 & 116 & 64 & $>$99.73 &$^b$ & 0.57 & 0.58 & 616~/~78 &345 & 339 & 7 & -45 & 144 & 189 & 124 & 54 & $^a$ &$^a$ & $^a$ & 309~/~276 & $^a$ \\
	1510$-$089 & 53 & -277 & 300 & 577 & -81 & 129 & $>$99.73 &2.25\,[2.13,\,2.37] & 0.53 & 0.53 & 101~/~31 &$^c$ & $^c$ & 27 & -238 & 128 & 366 & -70 & 97 & $>$99.73 &2.00\,[1.88,\,2.11] & 0.59 & 56~/~14 & 206 \\
	1611+343 & 68 & -138 & 271 & 409 & -67 & 114 & $>$99.73 &1.75\,[1.63,\,1.87] & 0.48 & 0.32 & 507~/~37 &$^c$ & $^c$ & 36 & -157 & 110 & 267 & -80 & 52 & $>$99.73 &1.65\,[1.49,\,1.81] & 0.43 & 763~/~26 & 135 \\
	1633+382 & 75 & -311 & 230 & 541 & 40 & 164 & $>$99.73 &1.65\,[1.52,\,1.78] & 0.36 & 0.41 & 133~/~26 &$^c$ & $^c$ & 41 & -332 & 181 & 513 & -21 & 166 & $>$99.73 &$^b$ & 0.42 & 58~/~9 & 805 \\
	1641+399 & 80 & -3 & 103 & 106 & 66 & 24 & $>$99.73 &1.70\,[1.53,\,1.86] & 0.33 & 0.33 & $^a$~/~22 &100 & 54 & 49 & -37 & 116 & 153 & 66 & 36 & $>$99.73 &1.10\,[0.97,\,1.23] & 0.38 & $^a$~/~2 & 34 \\
	1730$-$130 & 48 & -247 & 72 & 319 & -52 & 90 & $>$99.73 &2.35\,[2.06,\,2.64] & 0.64 & 0.61 & 147~/~45 &$^c$ & $^c$ & 18 & -132 & 87 & 219 & 14 & 74 & $>$99.73 &$^b$ & 0.44 & 42~/~42 & $^c$ \\
	1749+096 & 60 & -228 & 82 & 310 & -23 & 71 & $>$99.73 &2.00\,[1.85,\,2.15] & 0.32 & 0.29 & 101~/~56 &$^c$ & $^c$ & 42 & -171 & 167 & 338 & -31 & 91 & $>$99.73 &1.85\,[1.73,\,1.97] & 0.46 & 23~/~6 & 175 \\
	2200+420 & 70 & -9 & 62 & 71 & 16 & 12 & $>$99.73 &1.35\,[1.25,\,1.45] & 0.39 & 0.42 & $^a$ &281 & 112 & 42 & -11 & 46 & 57 & 9 & 14 & $>$99.73 &1.20\,[1.01,\,1.39] & 0.37 & $^a$ & 81 \\
	2223$-$052 & 59 & -2 & 119 & 121 & 86 & 28 & $>$99.73 &2.55\,[2.40,\,2.70] & 0.42 & 0.31 & $^a$~/~614 &597 & 130 & 21 & 34 & 154 & 120 & 103 & 32 & $>$99.73 &$^b$ & 0.52 & $^a$~/~75 & 33 \\
	2230+114 & 65 & -144 & 291 & 435 & 195 & 111 & $>$99.73 &3.00\,[2.69,\,3.31] & 0.33 & 0.39 & 62~/~23 &$^c$ & $^c$ & 27 & -87 & 292 & 379 & 193 & 93 & $>$99.73 &3.00\,[2.92,\,3.09] & 0.40 & 229~/~64 & $^c$ \\
	2251+158 & 79 & -206 & 257 & 463 & -52 & 113 & $>$99.73 &2.95\,[2.67,\,3.21] & 0.58 & 0.45 & 91~/~19 &769 & $^a$ & 38 & -228 & 219 & 447 & -84 & 134 & $>$99.73 &3.00\,[2.79,\,3.20] & 0.48 & 105~/~21 & $^c$ \\
	\midrule
	min & 24 & -311 & -14 & 60 & -177 & 12 & -- &0.50\,[0.37,\,0.63] & 0.24 & 0.23 & 62~/~16 &36 & 36 & 4 & -332 & -12 & 57 & -168 & 14 & -- &0.50\,[0.26,\,0.75] & 0.26 & 12~/~0 & 16 \\
	max & 108 & 29 & 300 & 577 & 195 & 164 & -- &3.00\,[2.91,\,3.09] & 0.74 & 0.76 & 1750~/~868 &793 & 798 & 60 & 34 & 382 & 513 & 193 & 166 & -- &3.00\,[2.87,\,3.13] & 0.83 & 763~/~354 & 805 \\
	median & 60 & -129 & 72 & 215 & -37 & 73 & -- &2.05\,[1.65,\,2.35] & 0.53 & 0.53 & 147~/~45 &281 & 141 & 31 & -132 & 119 & 240 & -4 & 73 & -- &1.75\,[1.45,\,2.00] & 0.52 & 81~/~21 & 102 \\
	std. dev. & 20 & 80 & 87 & 152 & 97 & 43 & -- &0.68 & 0.12 & 0.14 & 368~/~209 &241 & 213 & 13 & 85 & 95 & 128 & 89 & 35 & -- &0.78 & 0.17 & 208~/~89 & 203 \\
	\bottomrule
	\end{tabular}
	\begin{tablenotes}[para,flushleft]
	$^{d}$ No rotation $>180^{\circ}$.\\
	\end{tablenotes}
\end{table}
\end{landscape}


\begin{landscape}
\begin{table}
	\centering
	\caption{Variability analysis for the rotation measure $RM$. Columns are same as six first columns in Table~\ref{tab:I}}.
	\label{tab:RM}
	\scriptsize
	\begin{tabular}{lccccc}
	\toprule
	Source & $N_{\rm obs}$ & min & max & $\tilde{\rm RM}$ & ${\sigma}_{\rm RM}$ \\
	& & $[\times 10^3~rad/m^2]$ & $[\times 10^3~rad/m^2]$ & $[\times 10^3~rad/m^2]$ & $[\times 10^3~rad/m^2]$ \\
	\midrule
	0219+428 & 4  & -32  & 49  & -14 & 32 \\
	0235+164 & 14 & -225 & 57  & -45 & 101\\
	0316+413 & 21 & -190 & 80  & -31 & 71 \\
	0336$-$019 & 14 & -93  & 102 & -4  & 60 \\
	0355+508 & 34 & -179 & 114 & -21 & 63 \\
	0415+379 & 41 & -218 & 91  & -6  & 75 \\
	0420$-$014 & 29 & -122 & 73  & 15  & 47 \\
	0430+052 & 11 & -166 & 63  & -73 & 91 \\
	0528+134 & 21 & -156 & 106 & 7   & 78 \\
	0716+714 & 60 & -219 & 119 & -6  & 83 \\
	0735+178 & 21 & -157 & 36  & -56 & 63 \\
	0827+243 & 32 & -160 & 123 & 17  & 76 \\
	0829+046 & 16 & -101 & 27  & -16 & 31 \\
	0836+710 & 38 & -172 & 122 & 13  & 75 \\
	0851+202 & 49 & -108 & 19  & -4  & 23 \\
	0954+658 & 36 & -51  & 52  & -2  & 22 \\
	1055+018 & 38 & -138 & 77  & -7  & 51 \\
	1101+384 & 7  & -134 & 113 & -17 & 71 \\
	1127$-$145 & 25 & -205 & 107 & 1.1 & 91 \\
	1156+295 & 29 & -181 & 97  & -40 & 72 \\
	1219+285 & 11 & -155 & 82  & -13 & 58 \\
	1222+216 & 37 & -47  & 114 & -6  & 33 \\
	1226+023 & 40 & -161 & 85  & 11  & 49 \\
	1253$-$055 & 34 & -182 & 56  & 7   & 68 \\
	1308+326 & 32 & -103 & 74  & -12 & 44 \\
	1406$-$076 & 7  & -51  & 35  & -2  & 33 \\
	1510$-$089 & 27 & -138 & 87  & -3  & 57 \\
	1611+343 & 36 & -132 & 131 & 20  & 61 \\
	1633+382 & 41 & -191 & 88  & 7   & 77 \\
	1641+399 & 49 & -65  & 120 & 1.3 & 44 \\
	1730$-$130 & 18 & -142 & 36  & 9   & 50 \\
	1749+096 & 41 & -221 & 136 & 13  & 81 \\
	2200+420 & 42 & -12  & 29  & 6   & 10 \\
	2223$-$052 & 21 & -75  & 111 & -3  & 53 \\
	2230+114 & 27 & -168 & 116 & -15 & 56 \\
	2251+158 & 38 & -158 & 64  & -2  & 52 \\
	\midrule
	min & 4       & -225 & 19  & -73 & 10 \\
	max & 60      & -12  & 136 & 20  &101 \\
	median & 30   & -166 & 74  & -12 & 68 \\
	std. dev. & 13& 52   & 29  & 26  & 19 \\
	\bottomrule
	\end{tabular}
\end{table}
\end{landscape}

\begin{figure*}
	\begin{subfigure}[c]{0.49\textwidth}
		\includegraphics[width=\textwidth]{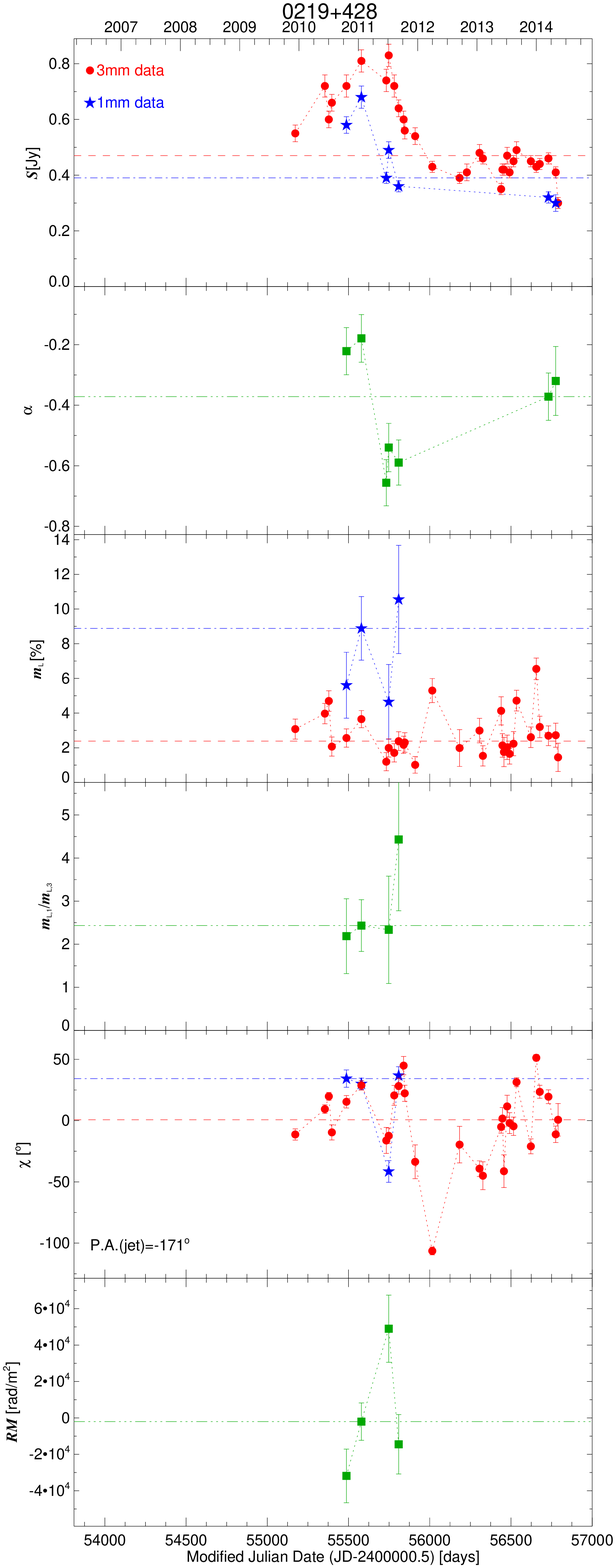}
	\end{subfigure}
	\begin{subfigure}[c]{0.49\textwidth}
		\includegraphics[width=\textwidth]{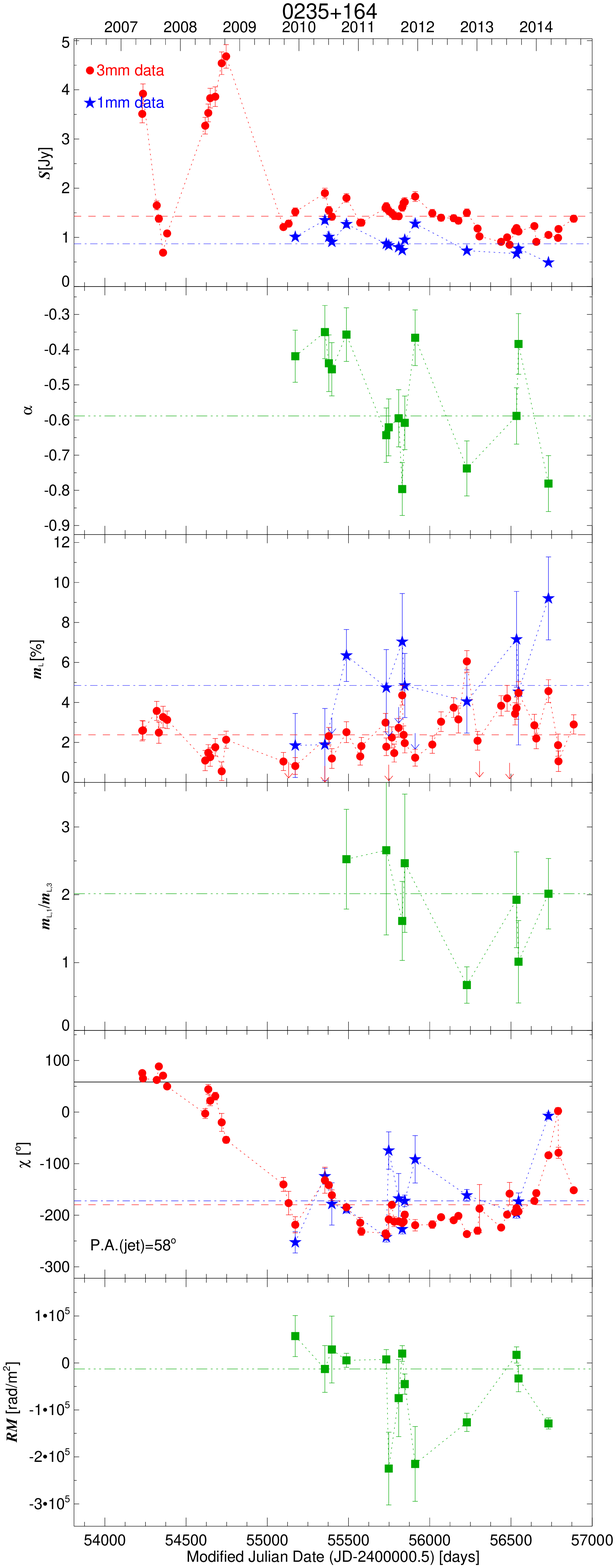}
	\end{subfigure}
	\caption{Daily averages of 3\,mm (red circles) and 1\,mm (blue stars) fully calibrated measurements of total flux ($S$), linear polarisation degree ($m_L$), and polarisation angle ($\chi$) of the 36 sources in the POLAMI sample of variable sources as a function of time. 
	              Comparison of these quantities at the two wavelengths are also shown (by green squares) for the spectral index ($\alpha$), ratio of 1\,mm to 3\,mm polarisation degree ($m_{\rm{L,1}}/m_{\rm{L,3}}$), and rotation measure ($RM$). 
	              Arrows are $2\sigma$ upper limits. 
	              The red dashed and blue dash--dotted lines indicate medians of 3 and 1\,mm values on every one of the plots, respectively, whereas dash--dotted--dotted--dotted lies represent medias of $\alpha$, $m_{\rm{L,1}}/m_{\rm{L,3}}$, and $RM$.
	              The black continuous line on the $\chi$ plots symbolises the jet position angle, i.e. P.A. (jet), according to \citet{Jorstad:2017VLBApaper,Molina:2014p22482}, see text, but is only plotted for sources for which the values where $\chi$ ranges include the P.A. (jet) value.}
	  \label{timevol}
\end{figure*}

\setcounter{figure}{9}
\begin{figure*}
	\begin{subfigure}[c]{0.49\textwidth}
		\includegraphics[width=\textwidth]{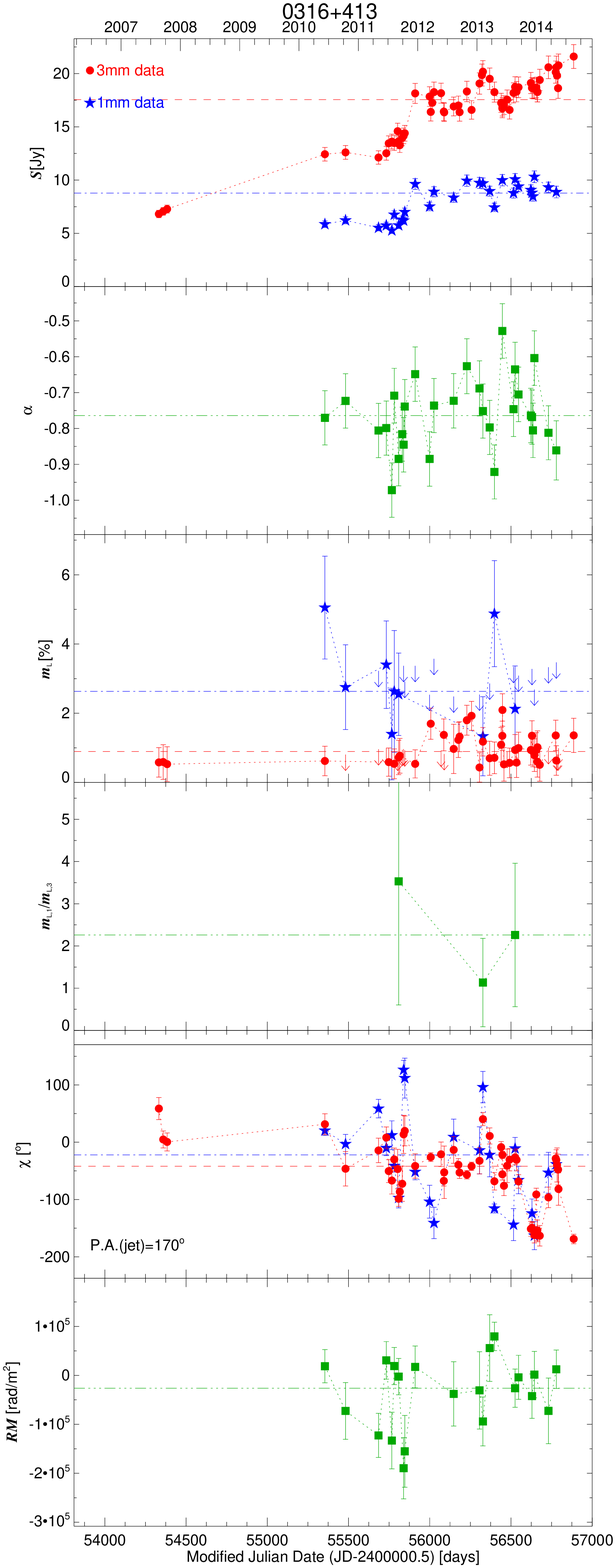}
	\end{subfigure}
	\begin{subfigure}[c]{0.49\textwidth}
		\includegraphics[width=\textwidth]{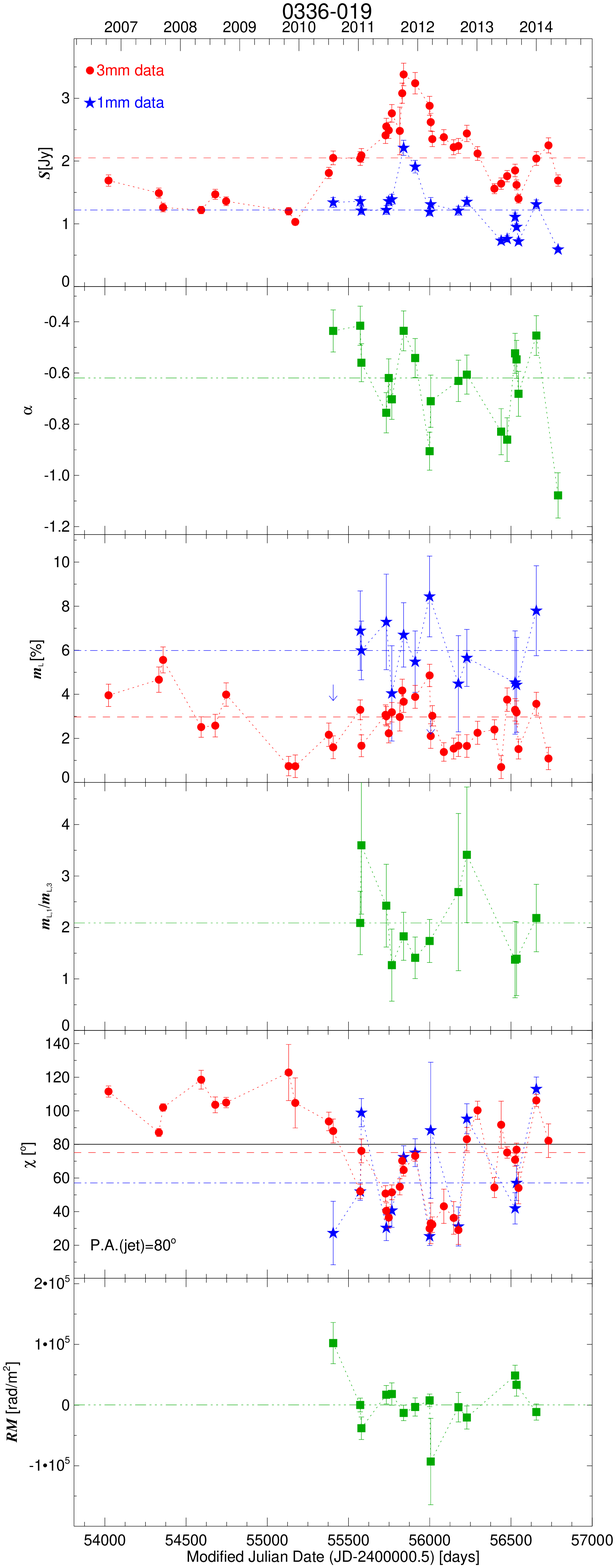}
	\end{subfigure}
   \caption{Continued.}
\end{figure*}

\setcounter{figure}{9}
\begin{figure*}
	\begin{subfigure}[c]{0.49\textwidth}
		\includegraphics[width=\textwidth]{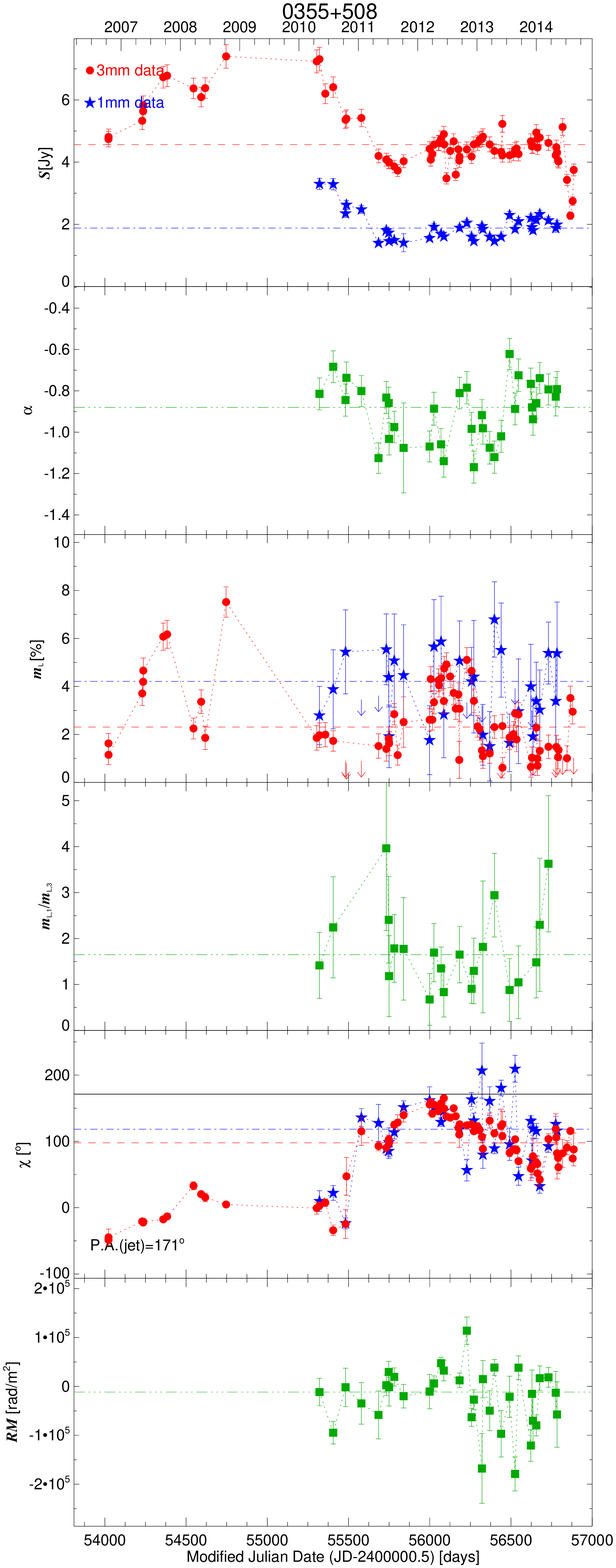}
	\end{subfigure}
	\begin{subfigure}[c]{0.49\textwidth}
		\includegraphics[width=\textwidth]{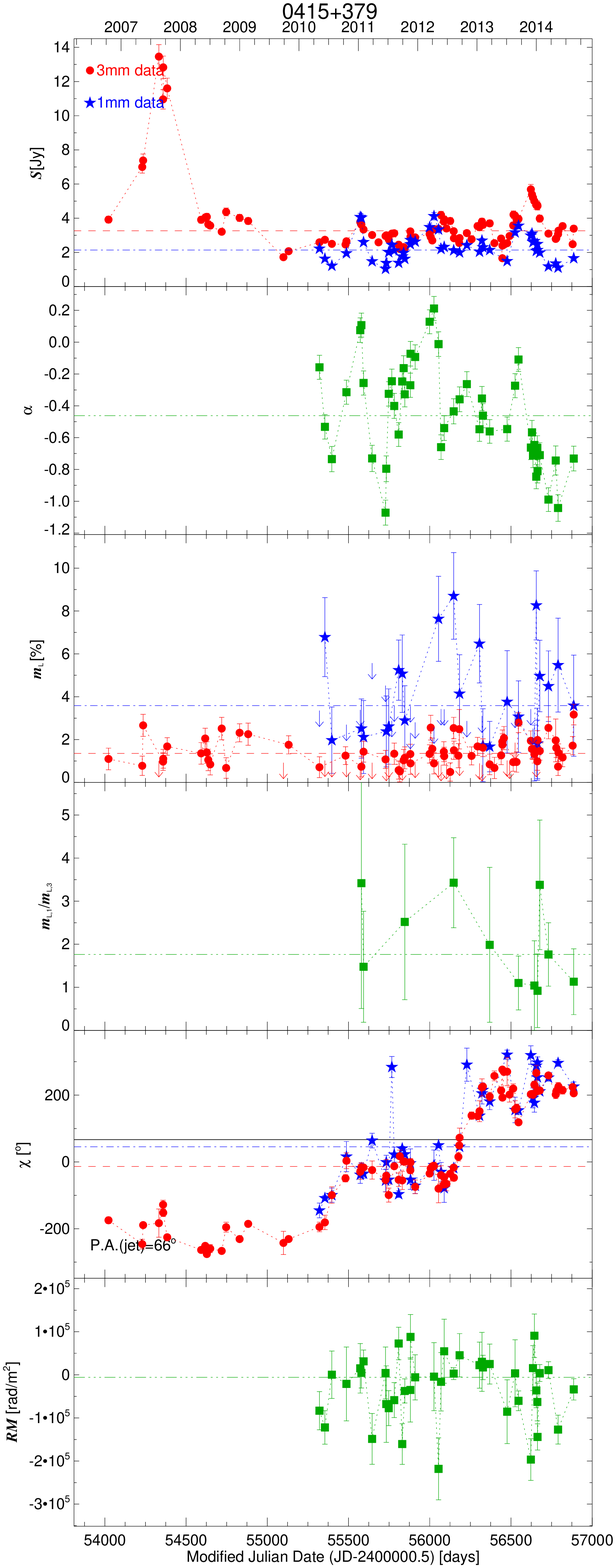}
	\end{subfigure}
   \caption{Continued.}
\end{figure*}

\setcounter{figure}{9}
\begin{figure*}
	\begin{subfigure}[c]{0.49\textwidth}
		\includegraphics[width=\textwidth]{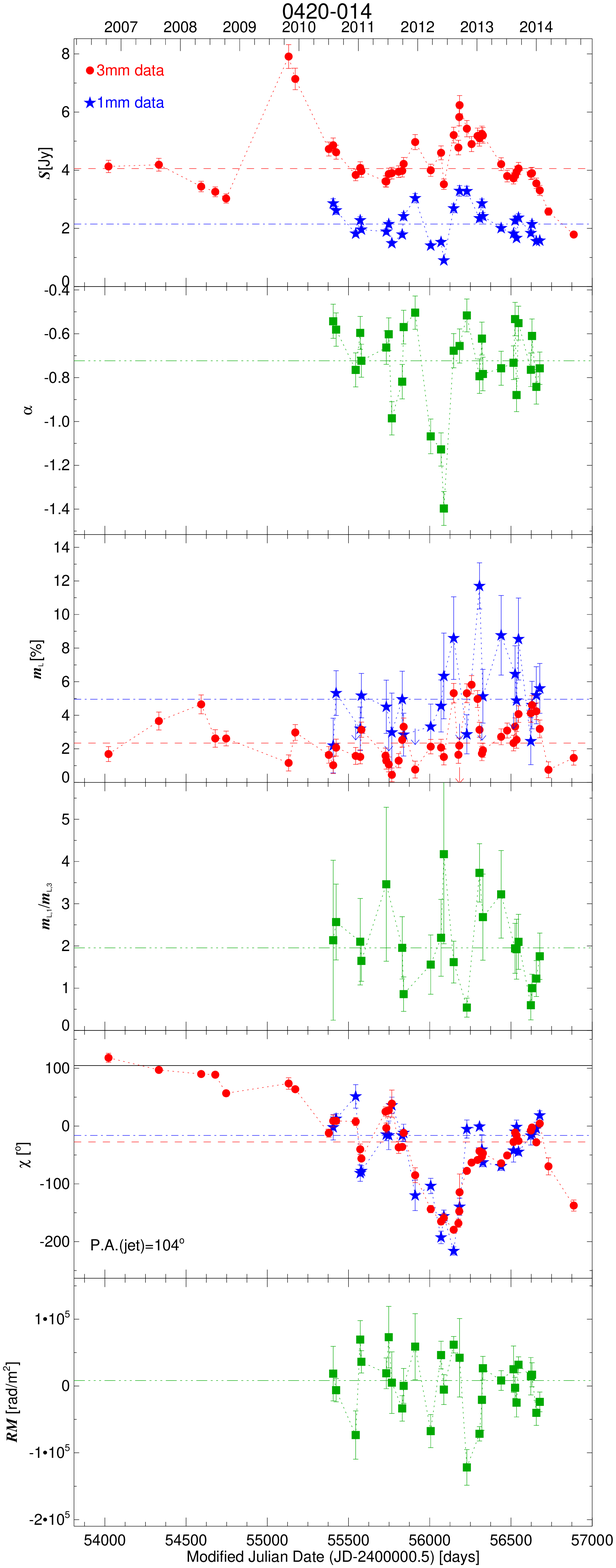}
	\end{subfigure}
	\begin{subfigure}[c]{0.49\textwidth}
		\includegraphics[width=\textwidth]{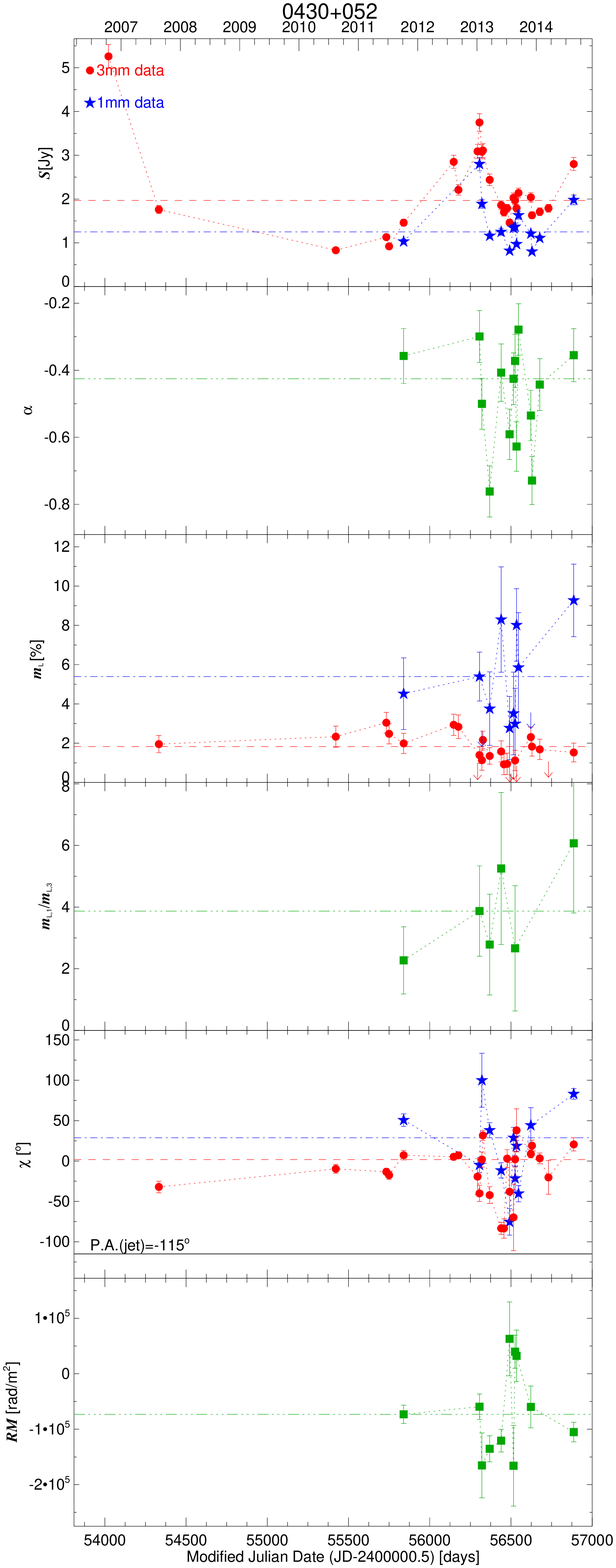}
	\end{subfigure}
   \caption{Continued.}
\end{figure*}

\setcounter{figure}{9}
\begin{figure*}
	\begin{subfigure}[c]{0.49\textwidth}
		\includegraphics[width=\textwidth]{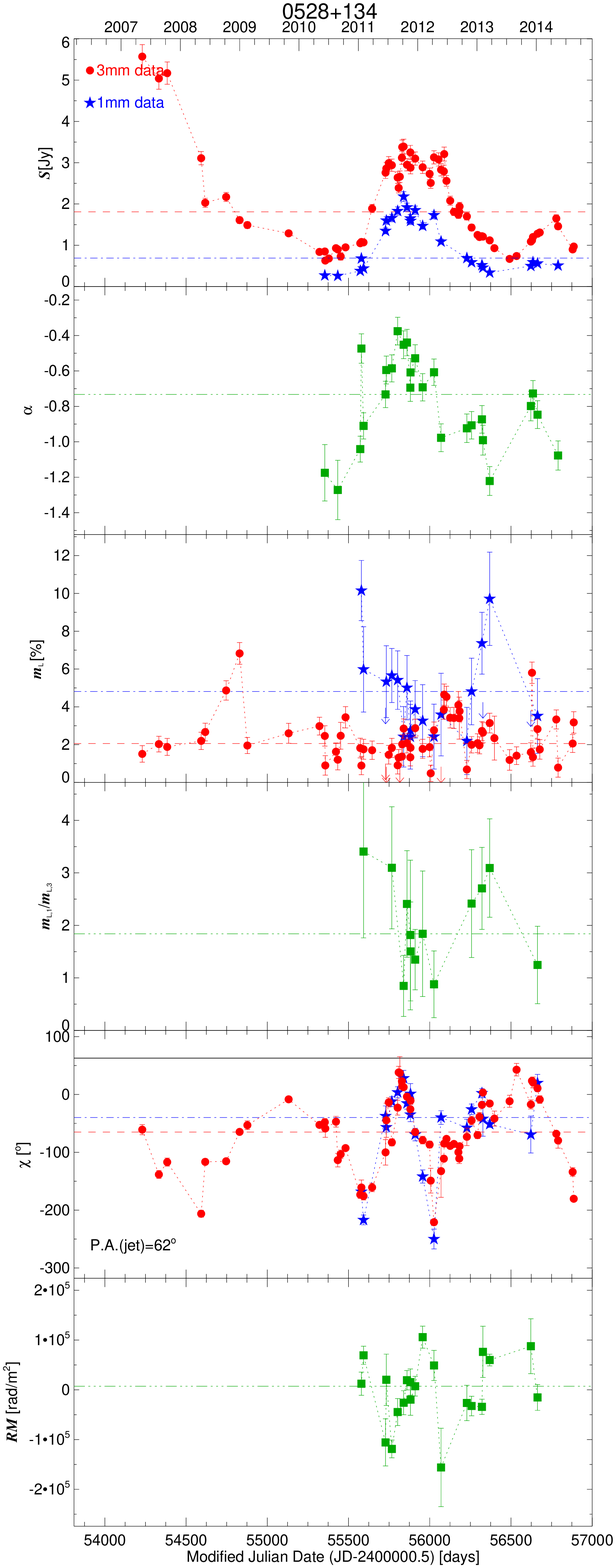}
	\end{subfigure}
	\begin{subfigure}[c]{0.49\textwidth}
		\includegraphics[width=\textwidth]{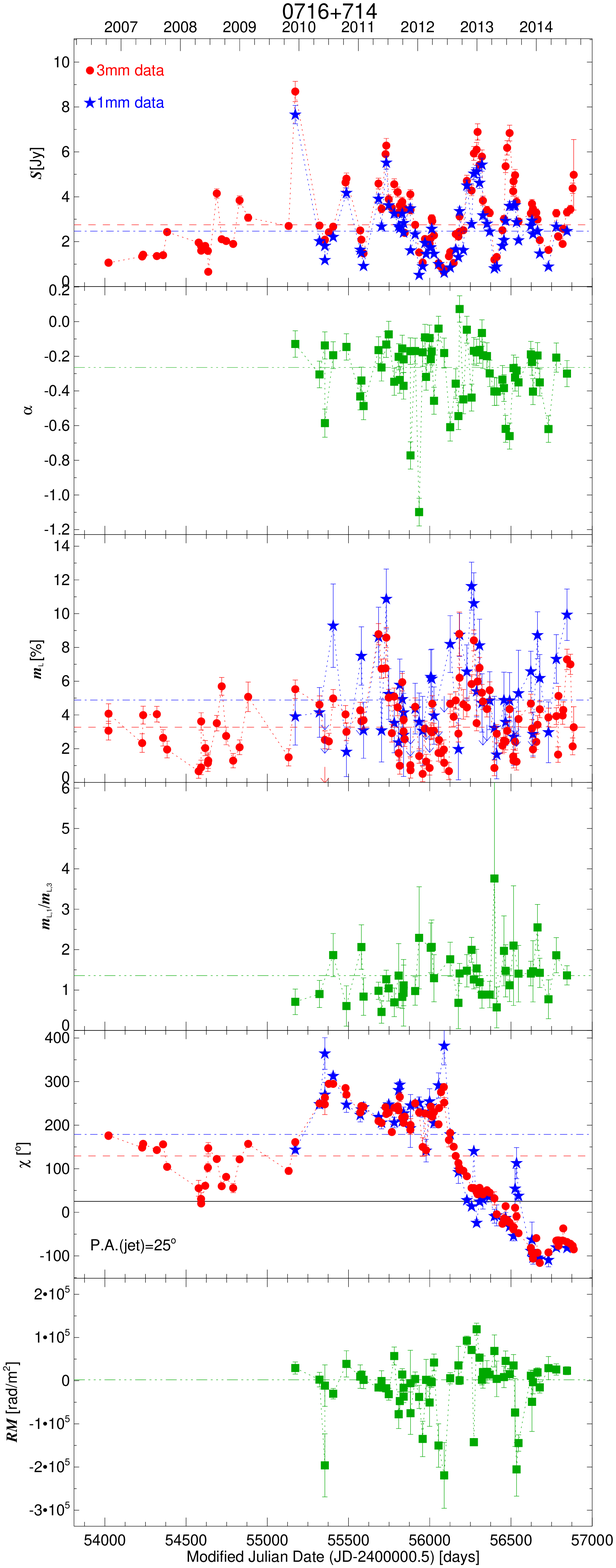}
	\end{subfigure}
   \caption{Continued.}
\end{figure*}

\setcounter{figure}{9}
\begin{figure*}
	\begin{subfigure}[c]{0.49\textwidth}
		\includegraphics[width=\textwidth]{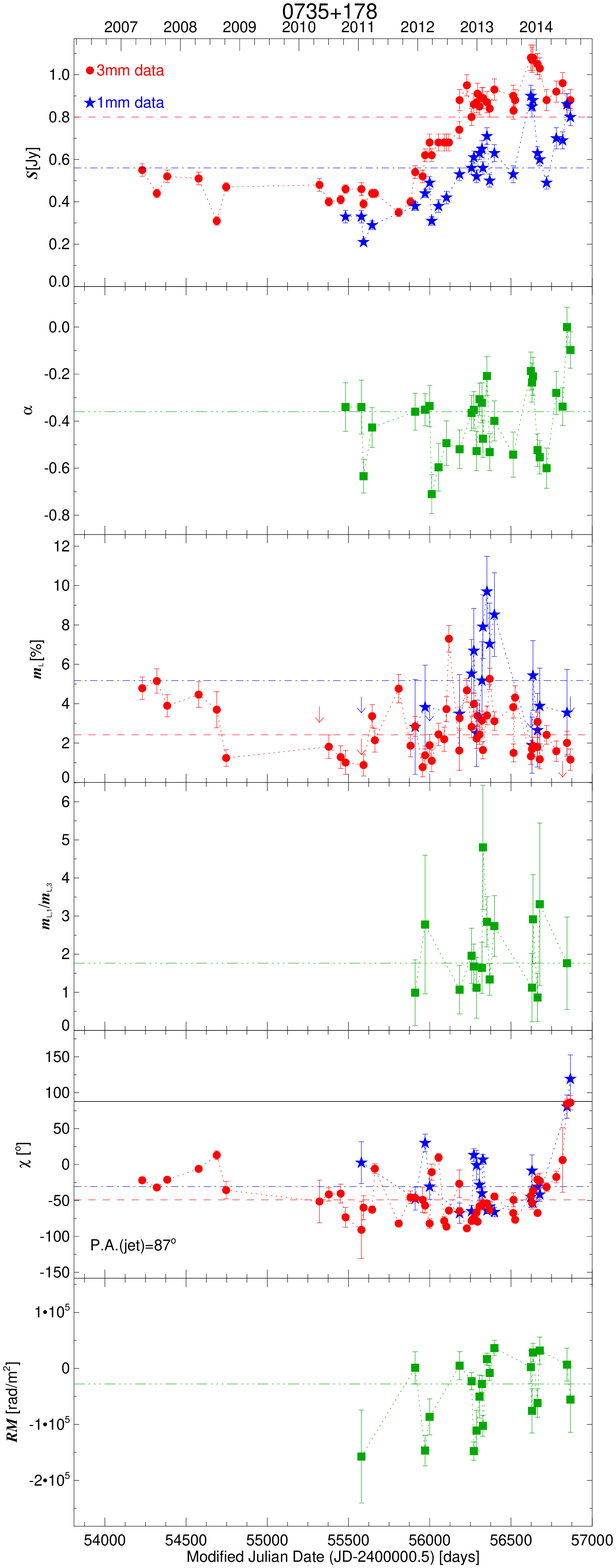}
	\end{subfigure}
	\begin{subfigure}[c]{0.49\textwidth}
		\includegraphics[width=\textwidth]{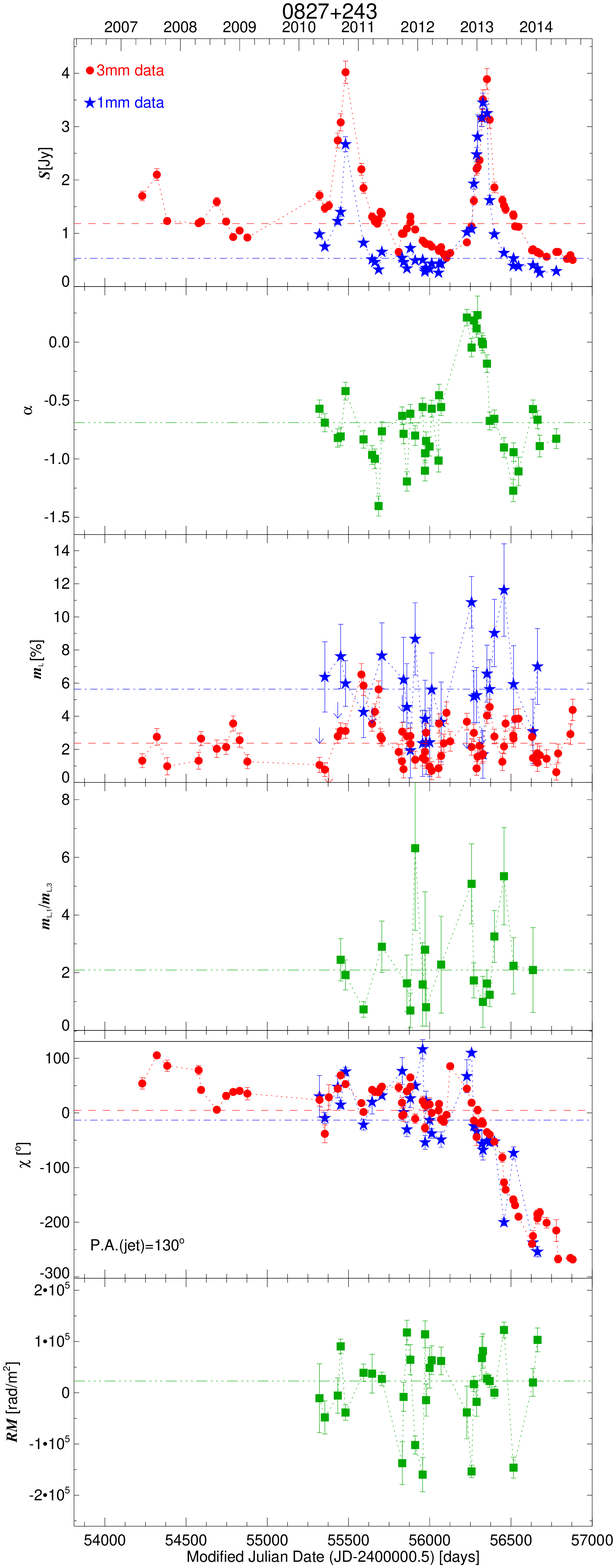}
	\end{subfigure}
   \caption{Continued.}
\end{figure*}

\setcounter{figure}{9}
\begin{figure*}
	\begin{subfigure}[c]{0.49\textwidth}
		\includegraphics[width=\textwidth]{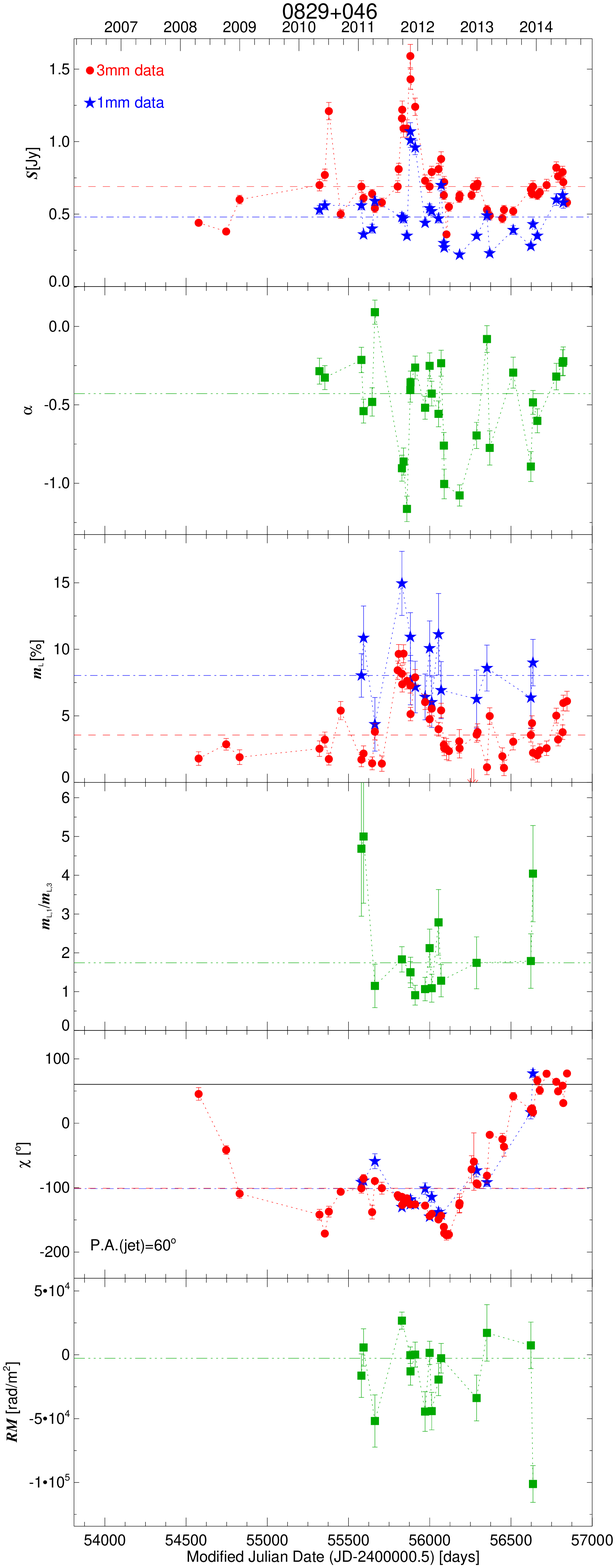}
	\end{subfigure}
	\begin{subfigure}[c]{0.49\textwidth}
		\includegraphics[width=\textwidth]{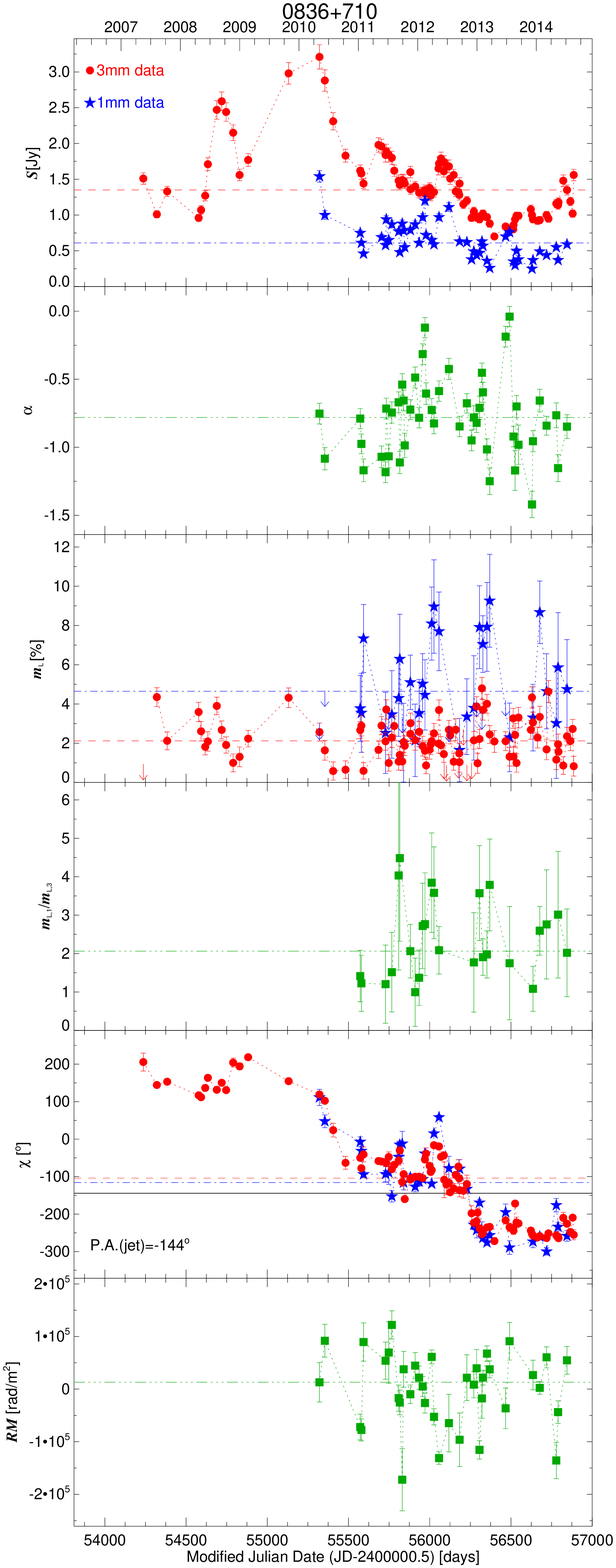}
	\end{subfigure}
   \caption{Continued.}
\end{figure*}

\setcounter{figure}{9}
\begin{figure*}
	\begin{subfigure}[c]{0.49\textwidth}
		\includegraphics[width=\textwidth]{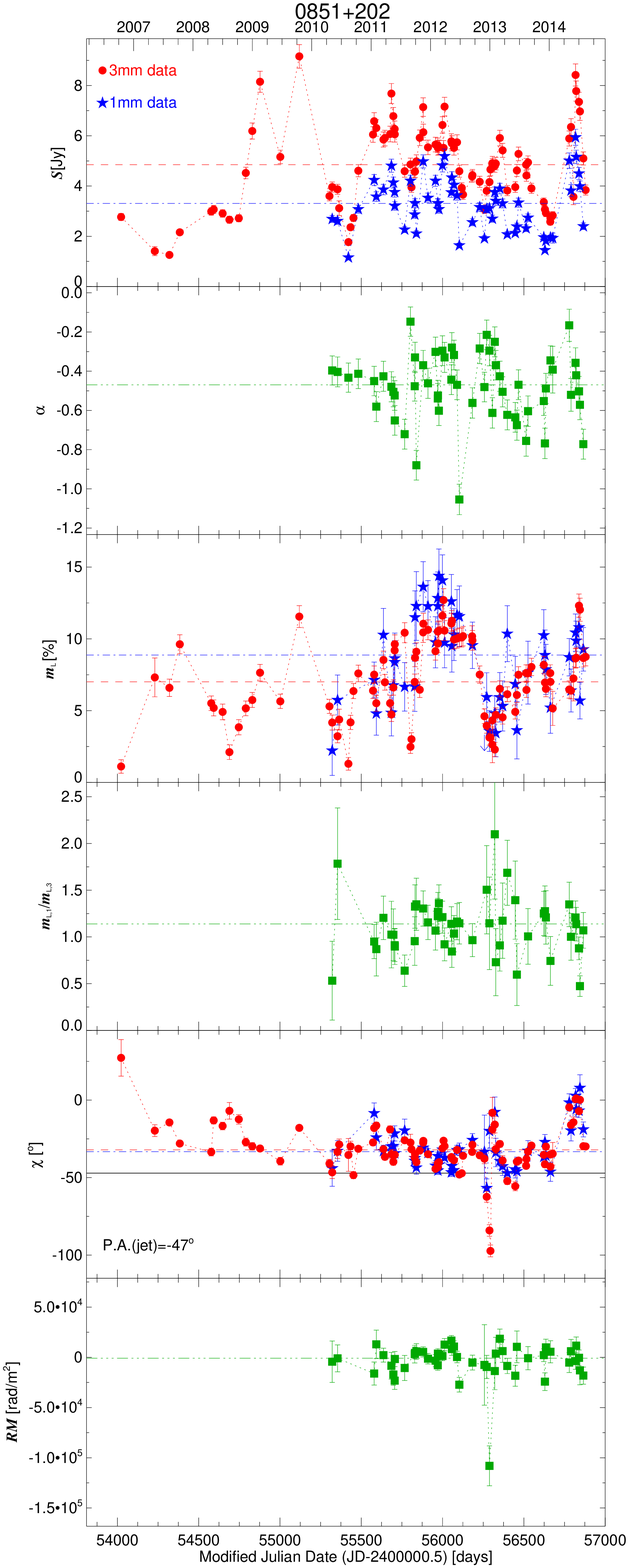}
	\end{subfigure}
	\begin{subfigure}[c]{0.49\textwidth}
		\includegraphics[width=\textwidth]{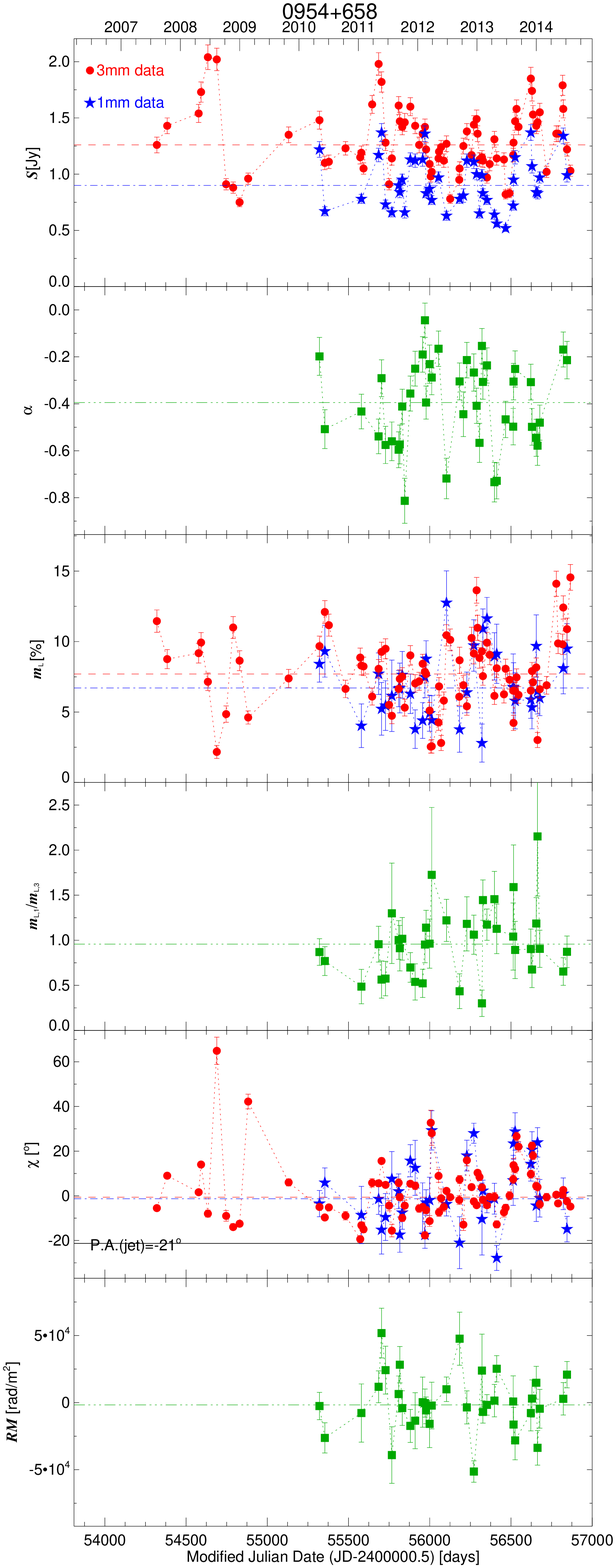}
	\end{subfigure}
   \caption{Continued.}
\end{figure*}

\setcounter{figure}{9}
\begin{figure*}
	\begin{subfigure}[c]{0.49\textwidth}
		\includegraphics[width=\textwidth]{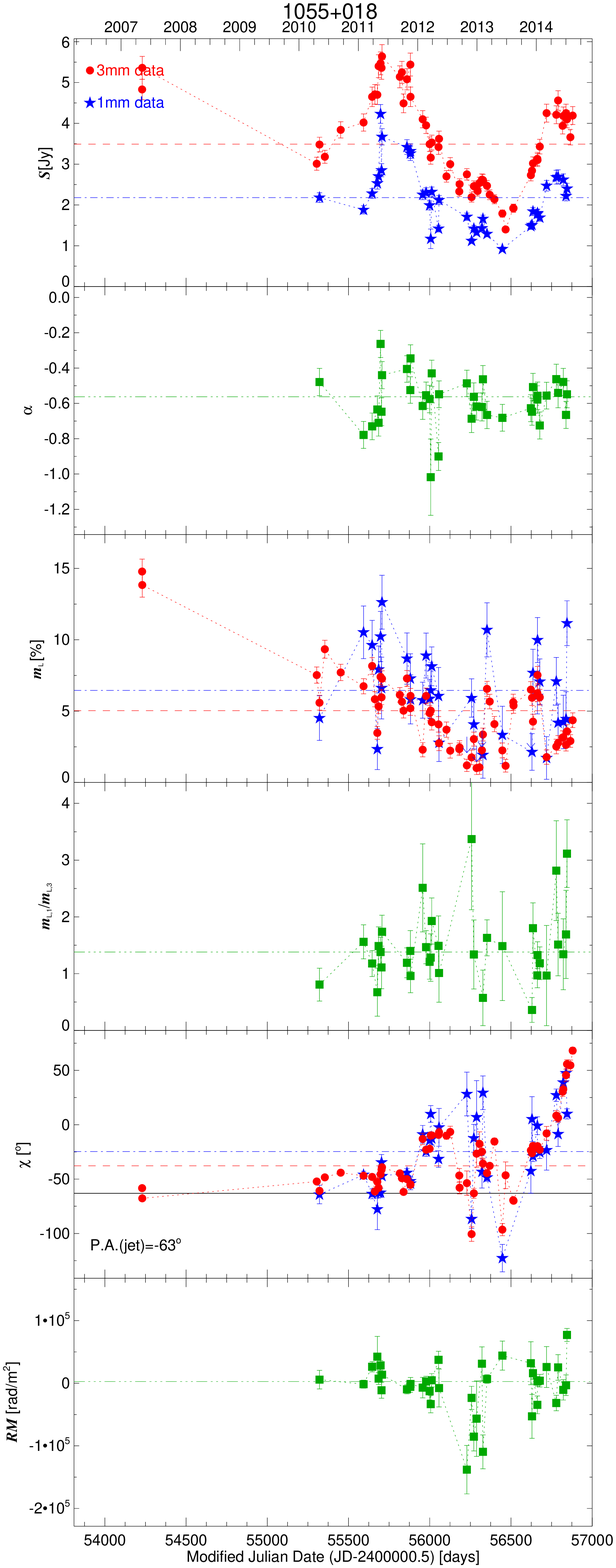}
	\end{subfigure}
	\begin{subfigure}[c]{0.49\textwidth}
		\includegraphics[width=\textwidth]{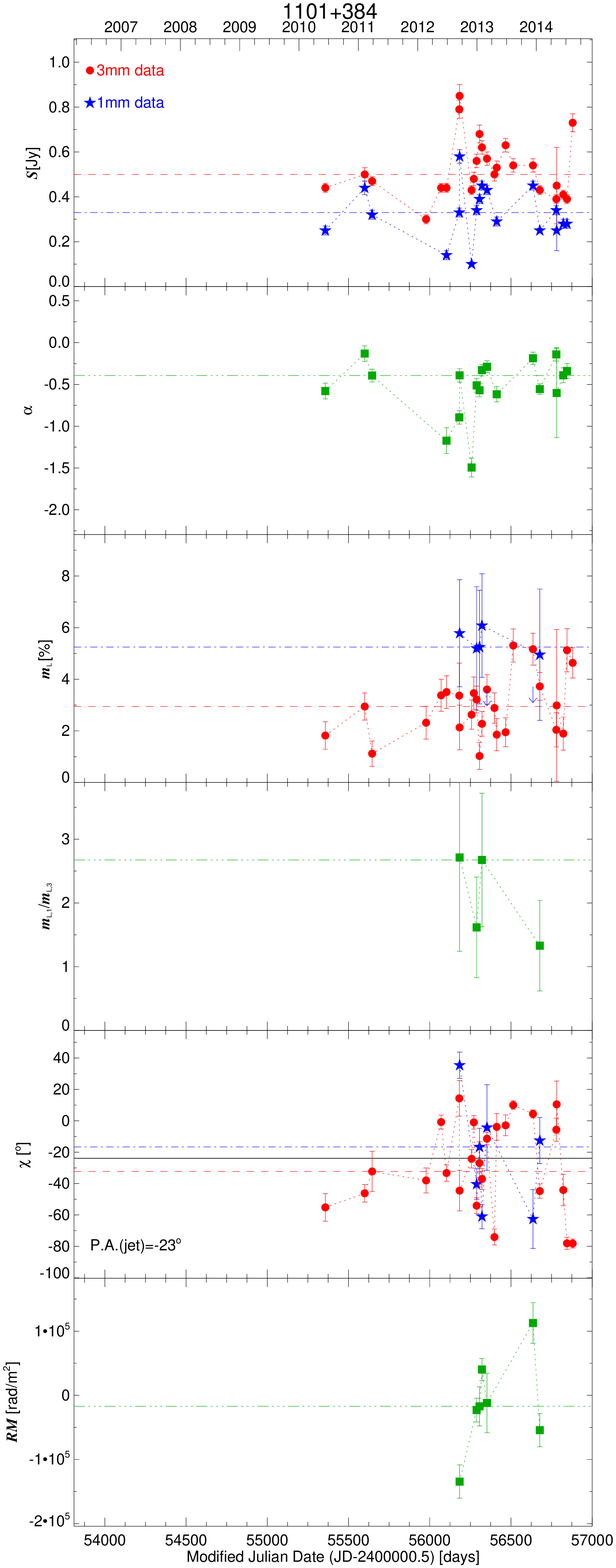}
	\end{subfigure}
   \caption{Continued.}
\end{figure*}

\setcounter{figure}{9}
\begin{figure*}
	\begin{subfigure}[c]{0.49\textwidth}
		\includegraphics[width=\textwidth]{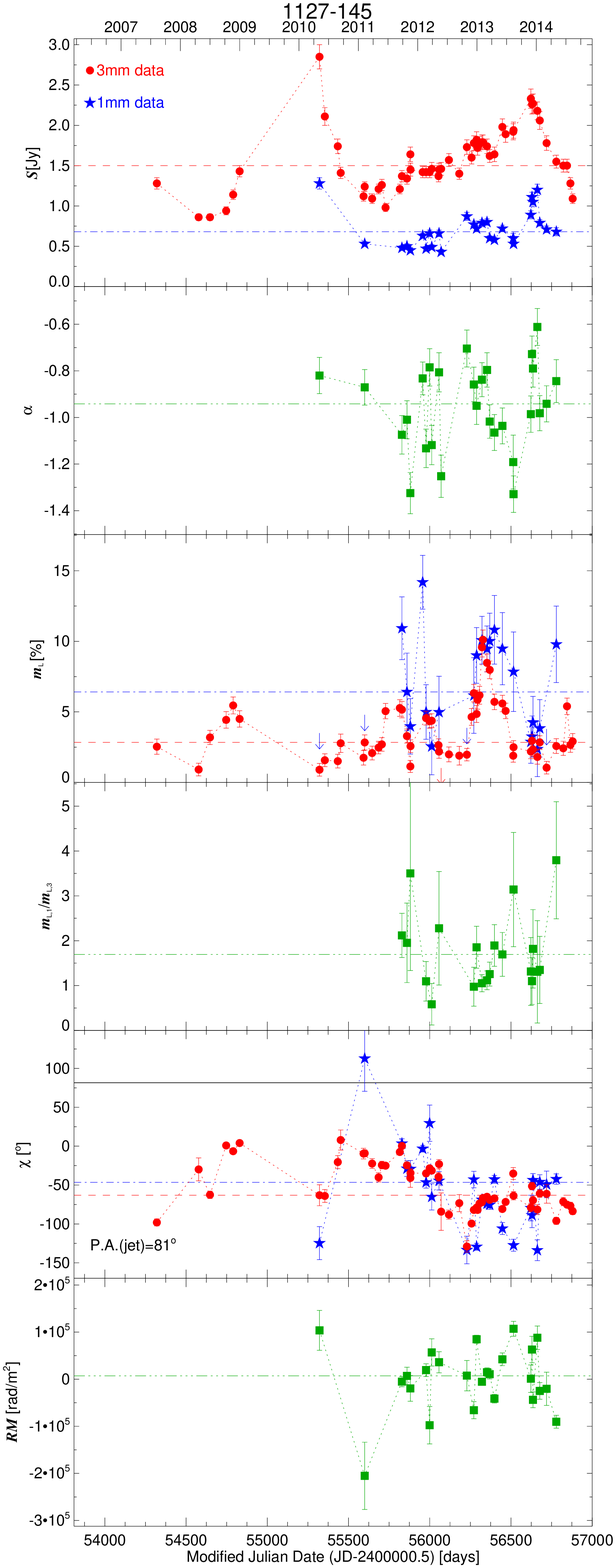}
	\end{subfigure}
	\begin{subfigure}[c]{0.49\textwidth}
		\includegraphics[width=\textwidth]{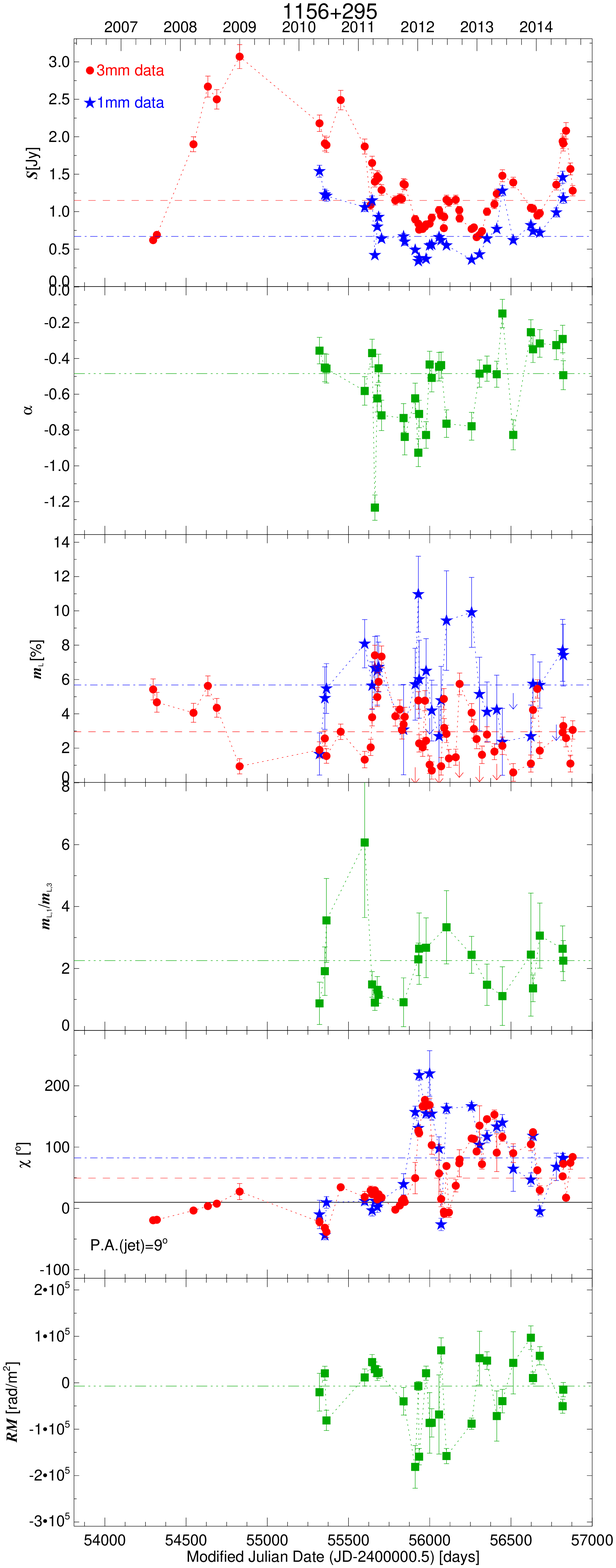}
	\end{subfigure}
   \caption{Continued.}
\end{figure*}

\setcounter{figure}{9}
\begin{figure*}
	\begin{subfigure}[c]{0.49\textwidth}
		\includegraphics[width=\textwidth]{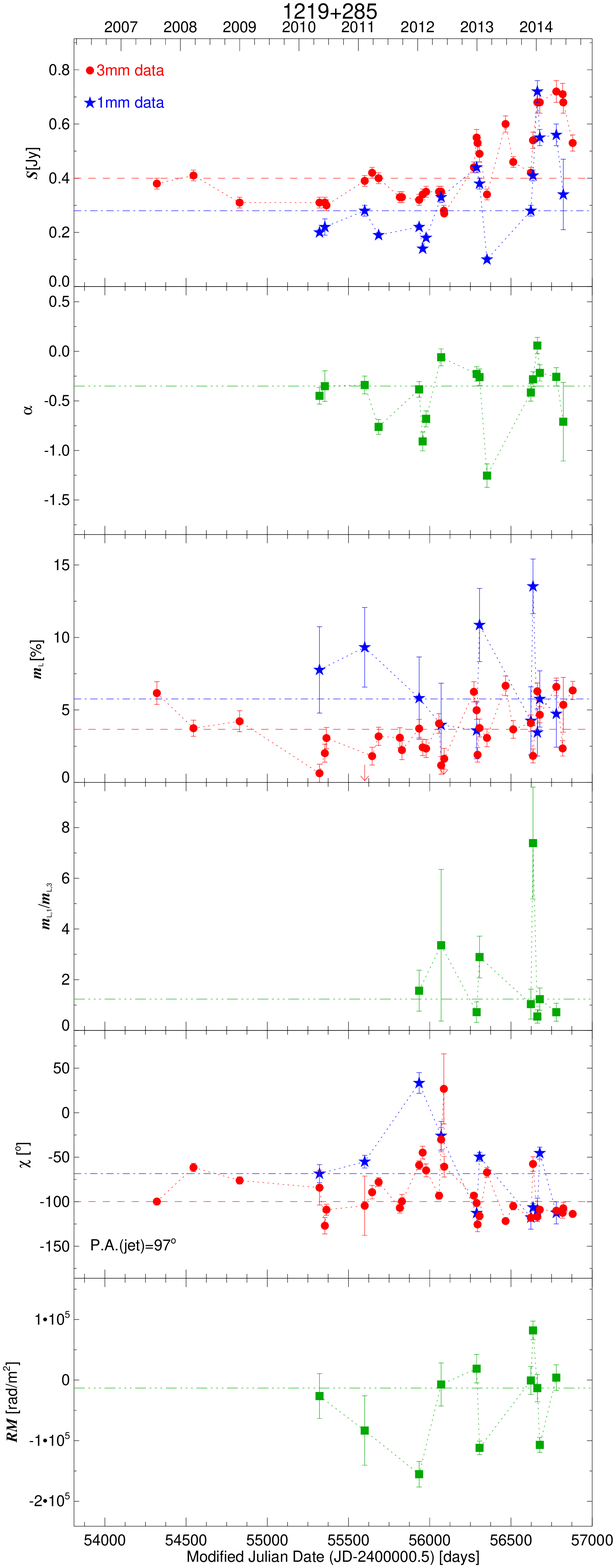}
	\end{subfigure}
	\begin{subfigure}[c]{0.49\textwidth}
		\includegraphics[width=\textwidth]{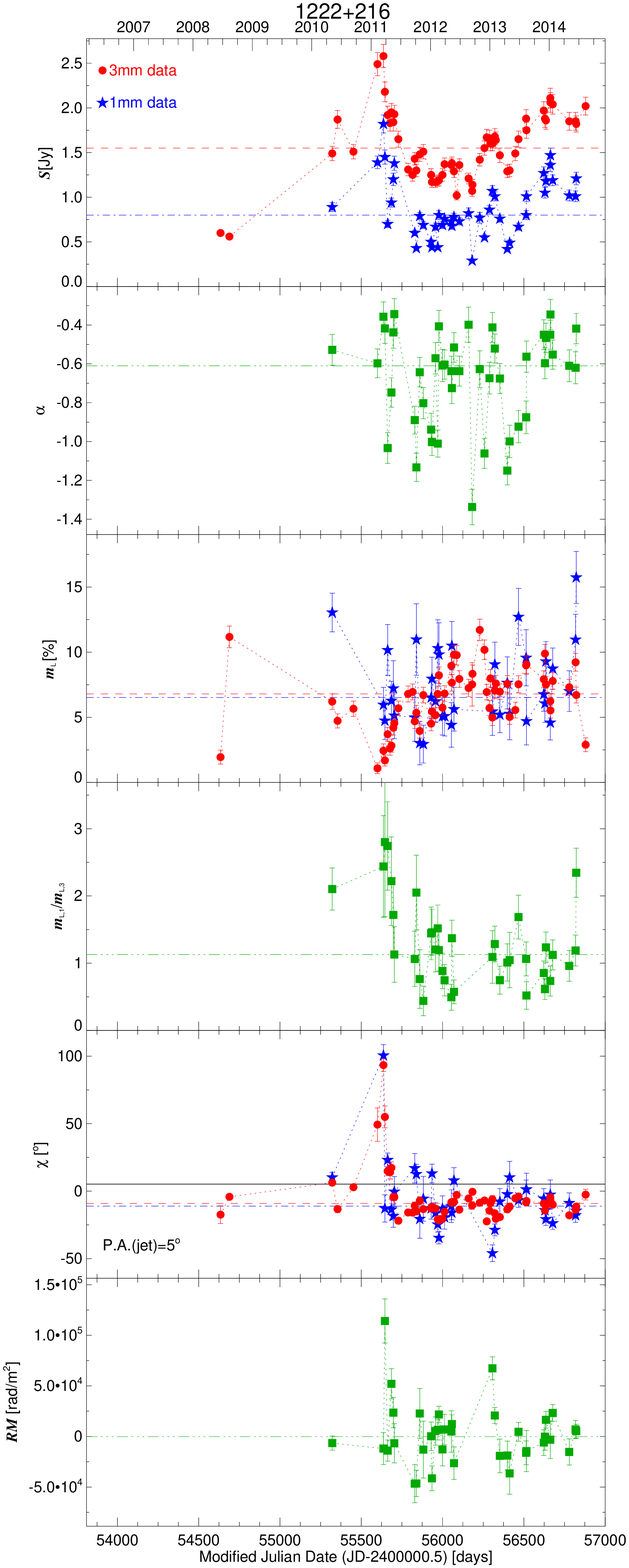}
	\end{subfigure}
   \caption{Continued.}
\end{figure*}

\setcounter{figure}{9}
\begin{figure*}
	\begin{subfigure}[c]{0.49\textwidth}
		\includegraphics[width=\textwidth]{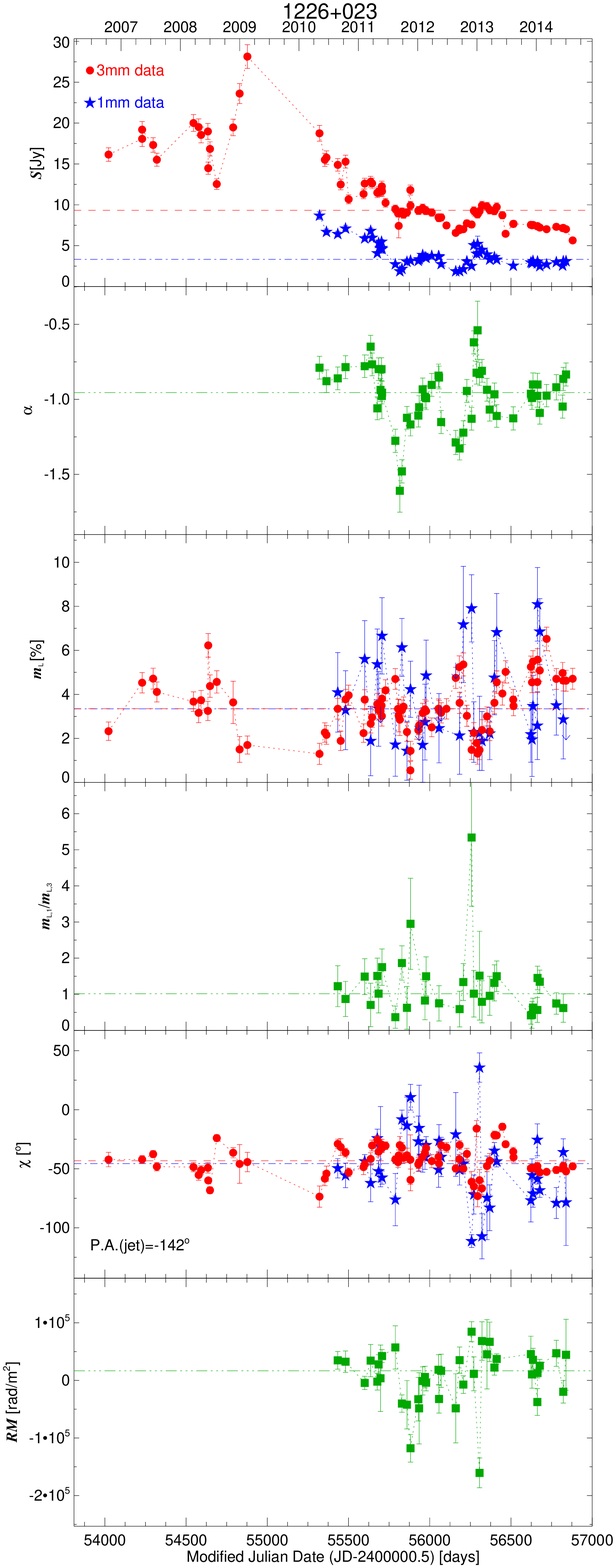}
	\end{subfigure}
	\begin{subfigure}[c]{0.49\textwidth}
		\includegraphics[width=\textwidth]{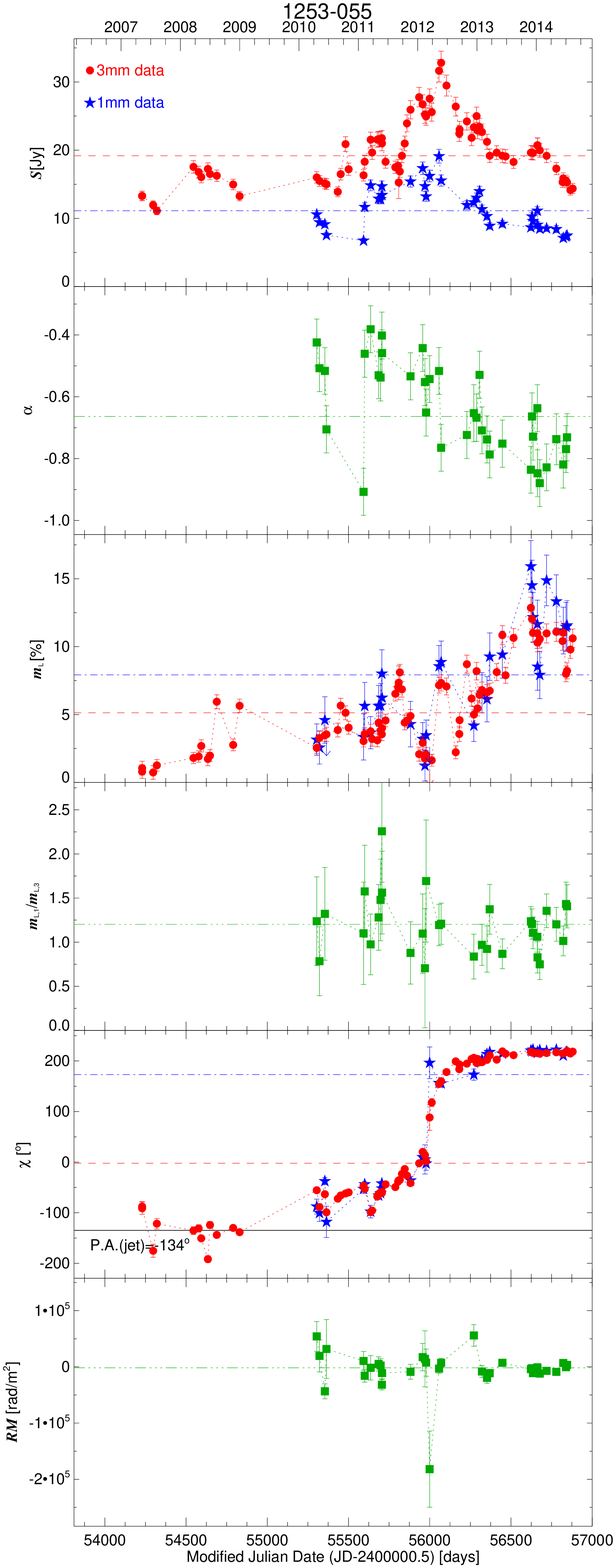}
	\end{subfigure}
   \caption{Continued.}
\end{figure*}

\setcounter{figure}{9}
\begin{figure*}
	\begin{subfigure}[c]{0.49\textwidth}
		\includegraphics[width=\textwidth]{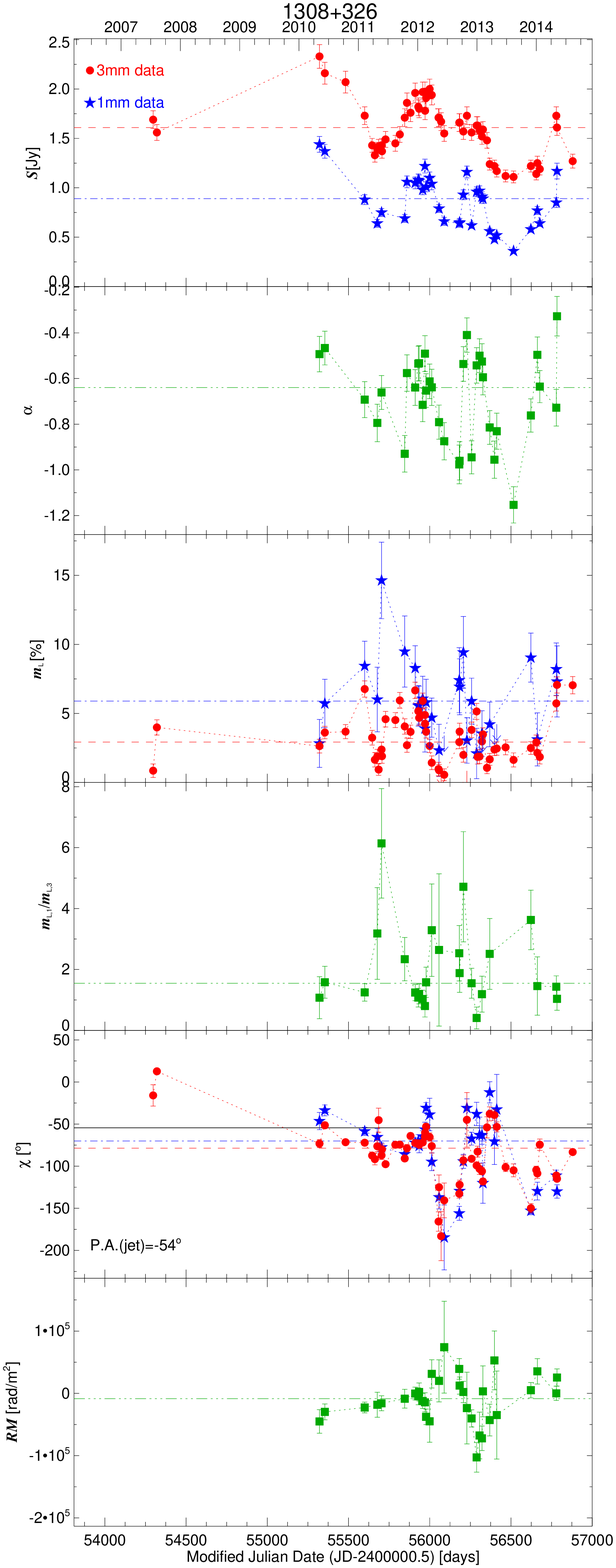}
	\end{subfigure}
	\begin{subfigure}[c]{0.49\textwidth}
		\includegraphics[width=\textwidth]{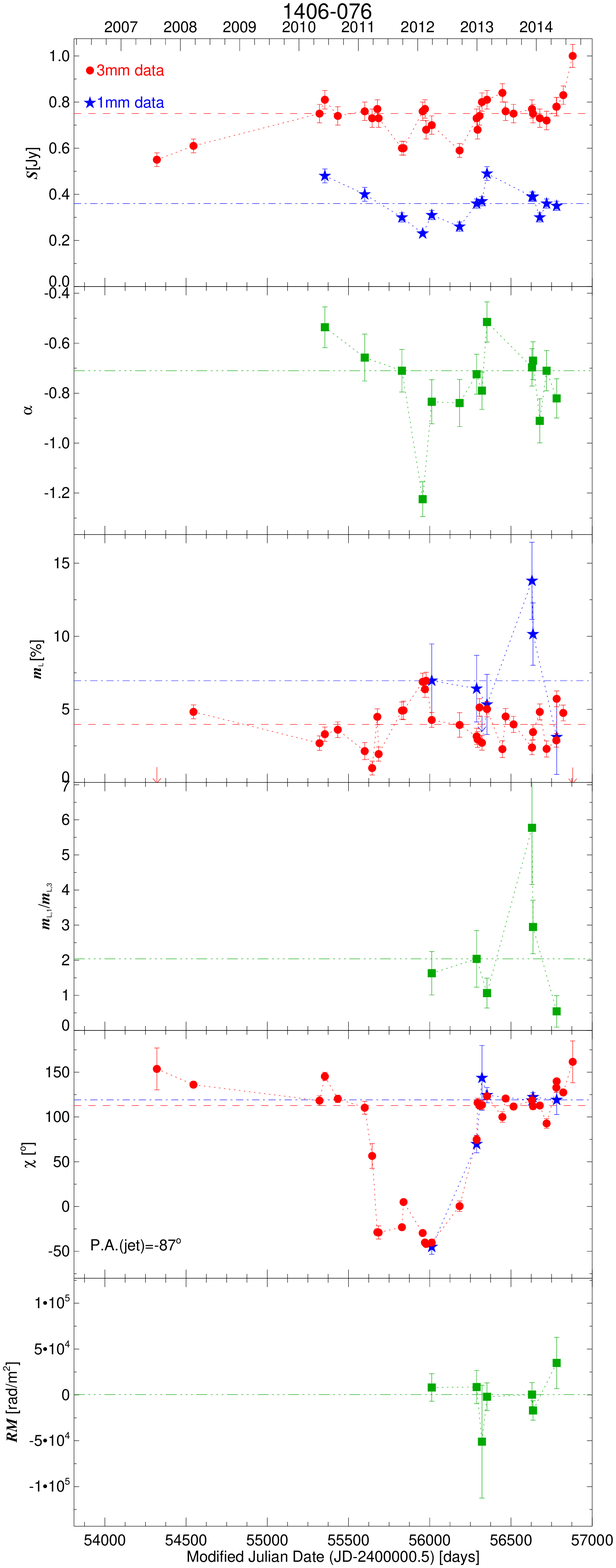}
	\end{subfigure}
   \caption{Continued.}
\end{figure*}

\setcounter{figure}{9}
\begin{figure*}
	\begin{subfigure}[c]{0.49\textwidth}
		\includegraphics[width=\textwidth]{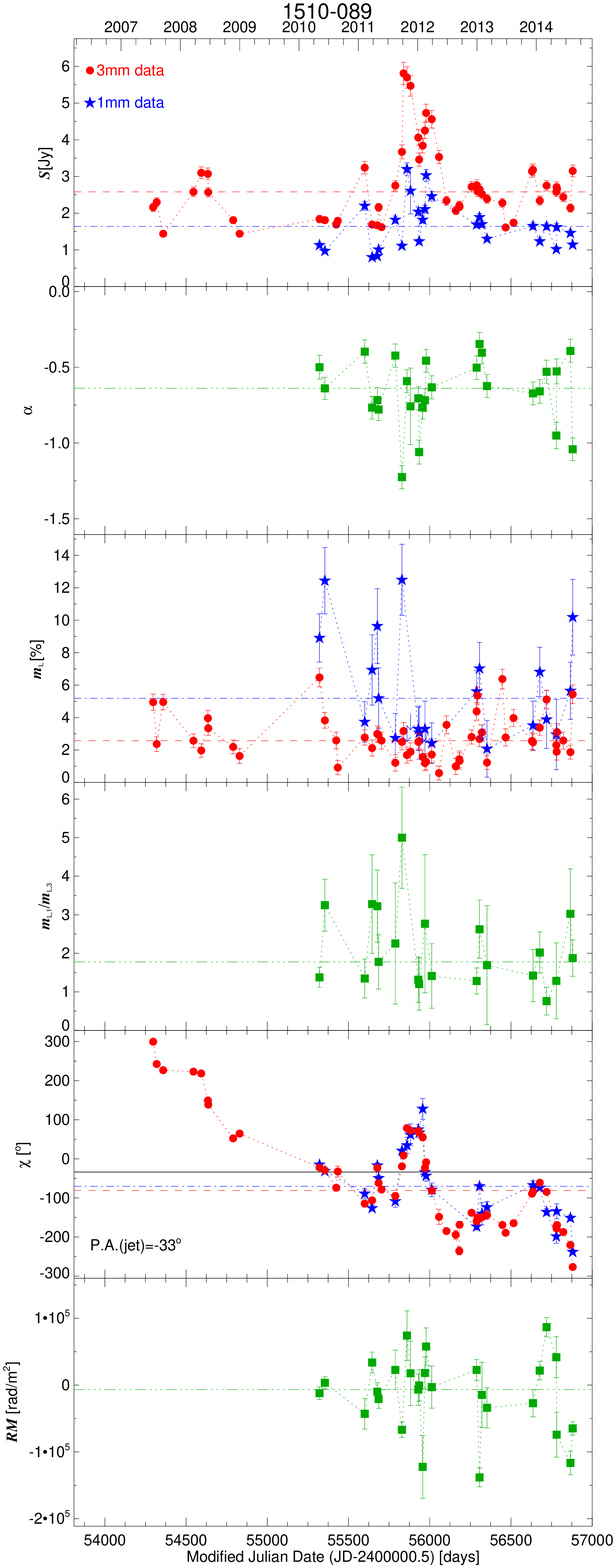}
	\end{subfigure}
	\begin{subfigure}[c]{0.49\textwidth}
		\includegraphics[width=\textwidth]{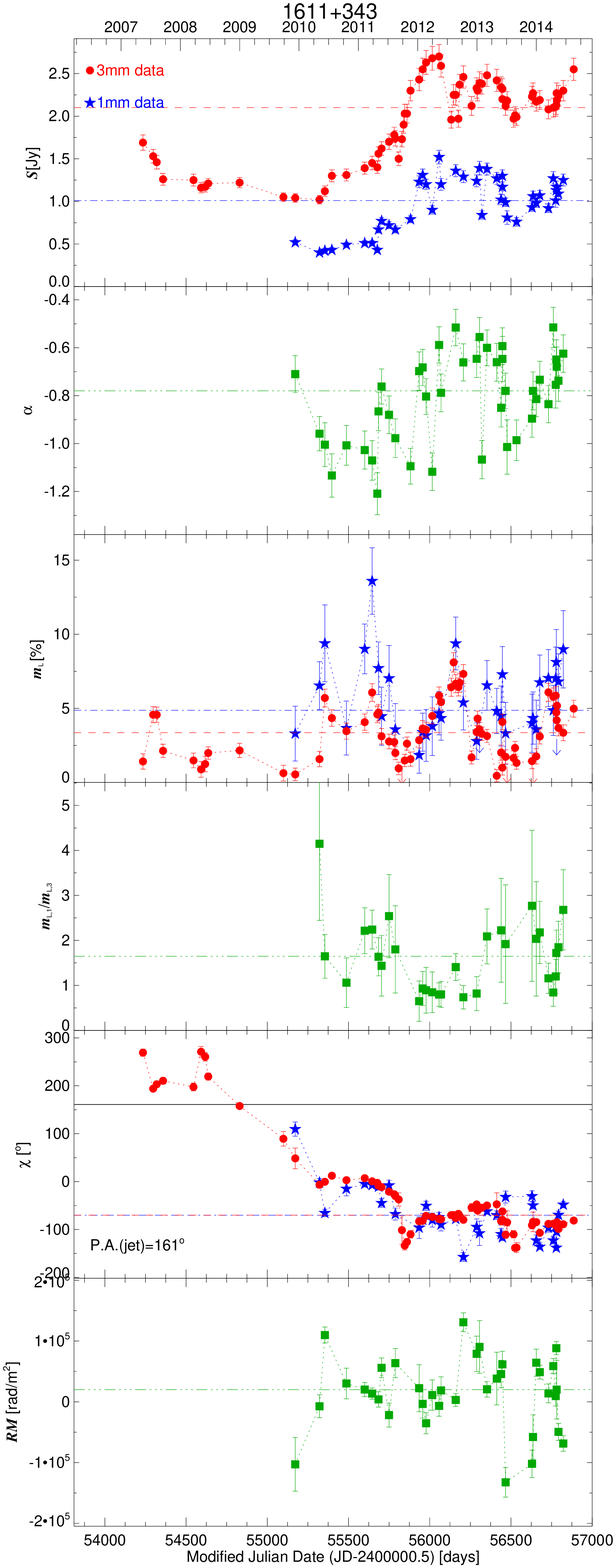}
	\end{subfigure}
   \caption{Continued.}
\end{figure*}

\setcounter{figure}{9}
\begin{figure*}
	\begin{subfigure}[c]{0.49\textwidth}
		\includegraphics[width=\textwidth]{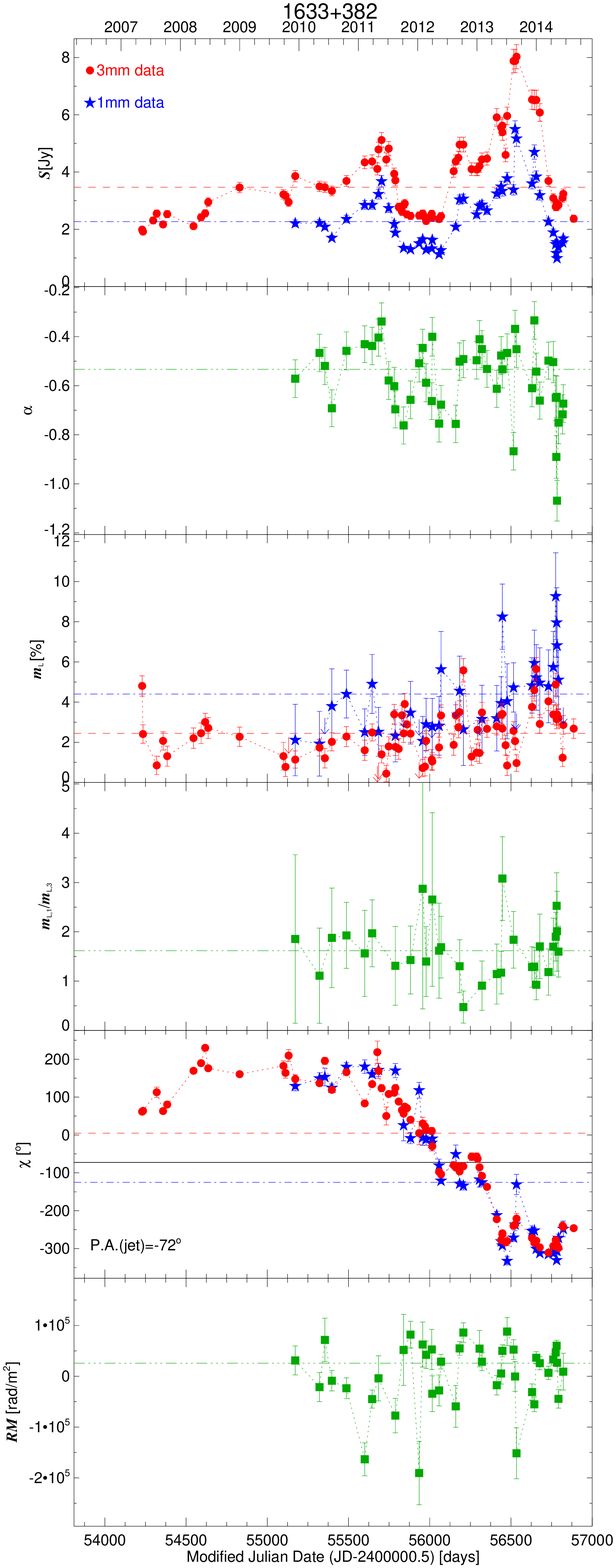}
	\end{subfigure}
	\begin{subfigure}[c]{0.49\textwidth}
		\includegraphics[width=\textwidth]{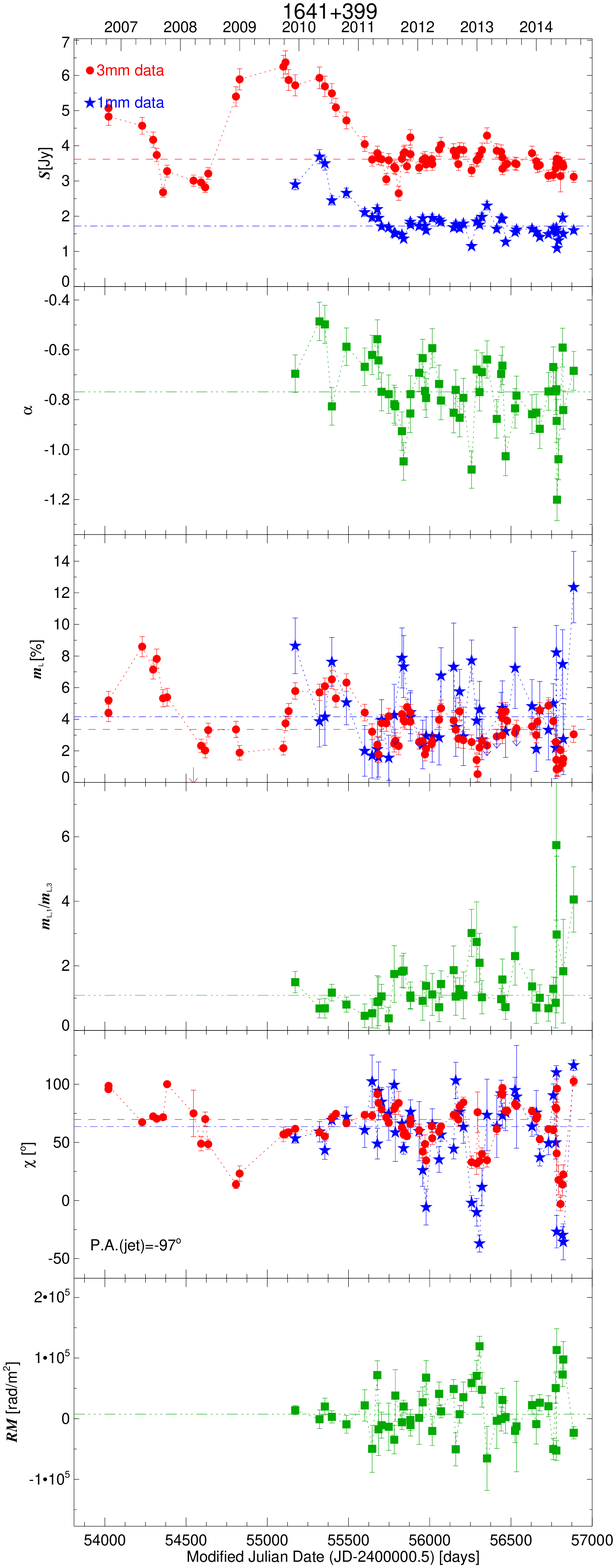}
	\end{subfigure}
   \caption{Continued.}
\end{figure*}

\setcounter{figure}{9}
\begin{figure*}
	\begin{subfigure}[c]{0.49\textwidth}
		\includegraphics[width=\textwidth]{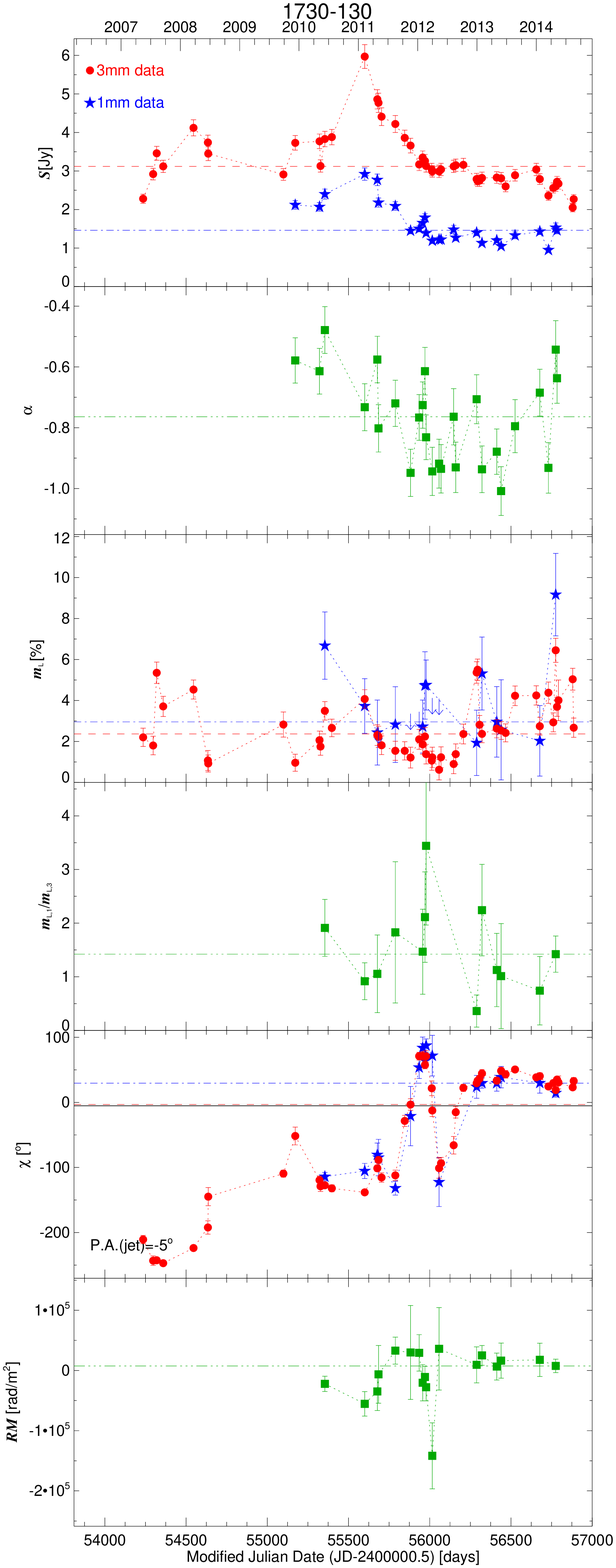}
	\end{subfigure}
	\begin{subfigure}[c]{0.49\textwidth}
		\includegraphics[width=\textwidth]{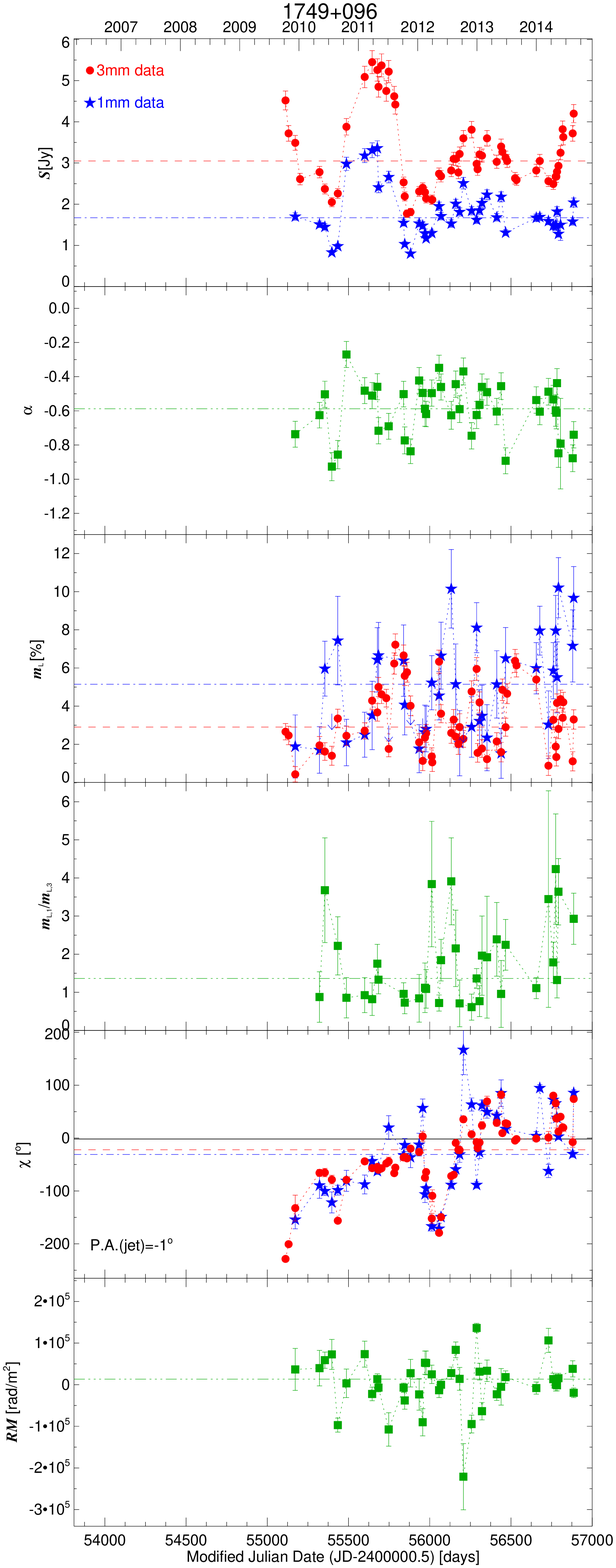}
	\end{subfigure}
   \caption{Continued.}
\end{figure*}

\setcounter{figure}{9}
\begin{figure*}
	\begin{subfigure}[c]{0.49\textwidth}
		\includegraphics[width=\textwidth]{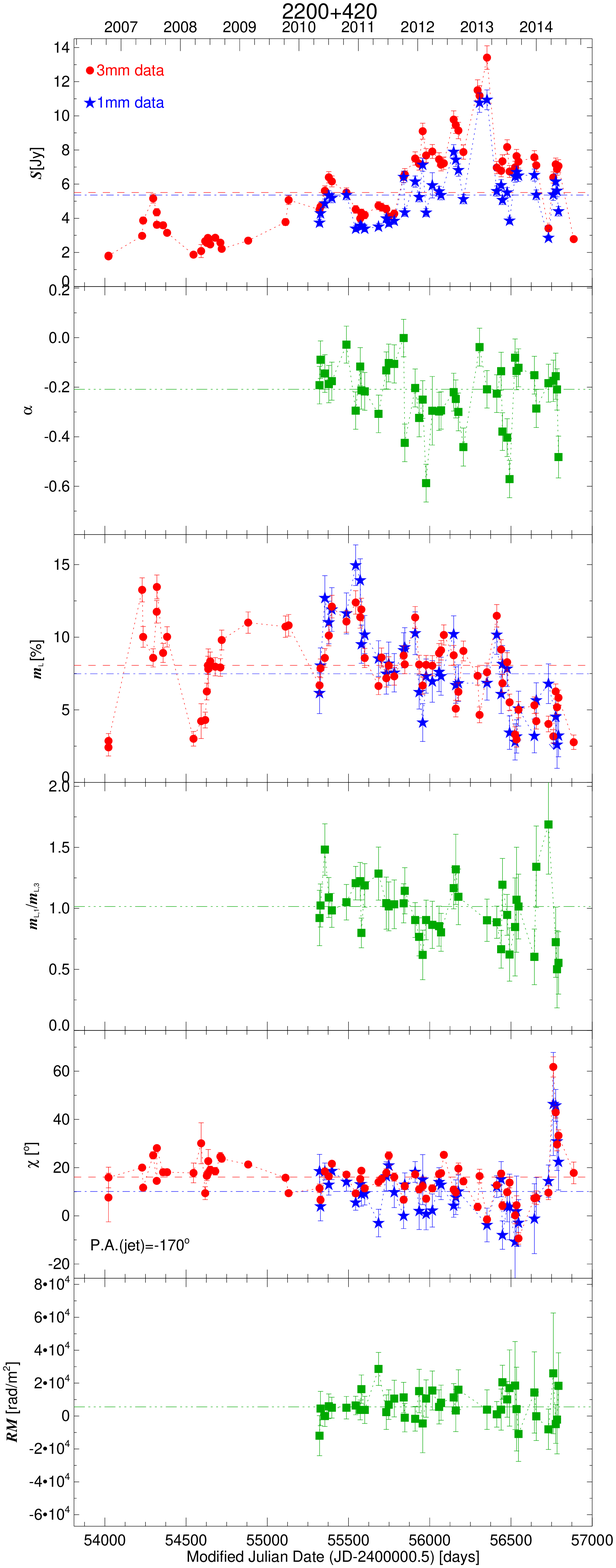}
	\end{subfigure}
	\begin{subfigure}[c]{0.49\textwidth}
		\includegraphics[width=\textwidth]{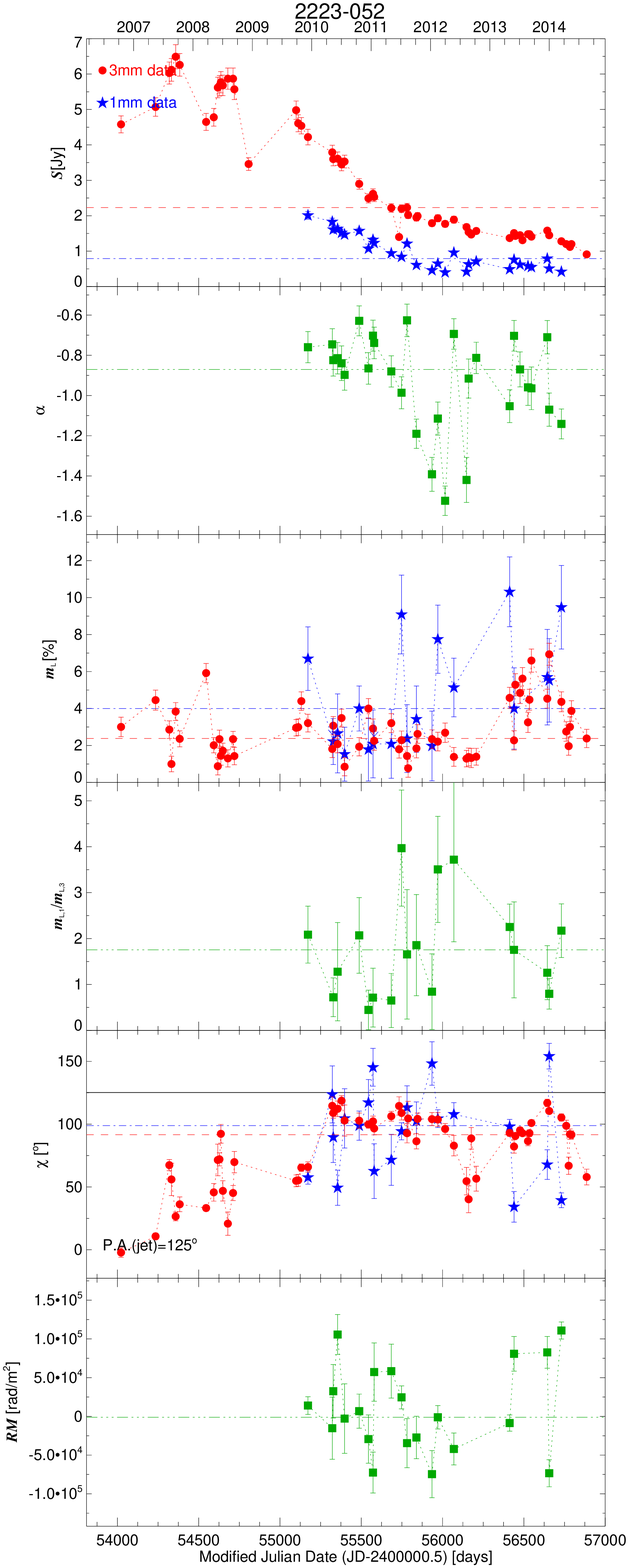}
	\end{subfigure}
   \caption{Continued.}
\end{figure*}

\setcounter{figure}{9}
\begin{figure*}
	\begin{subfigure}[c]{0.49\textwidth}
		\includegraphics[width=\textwidth]{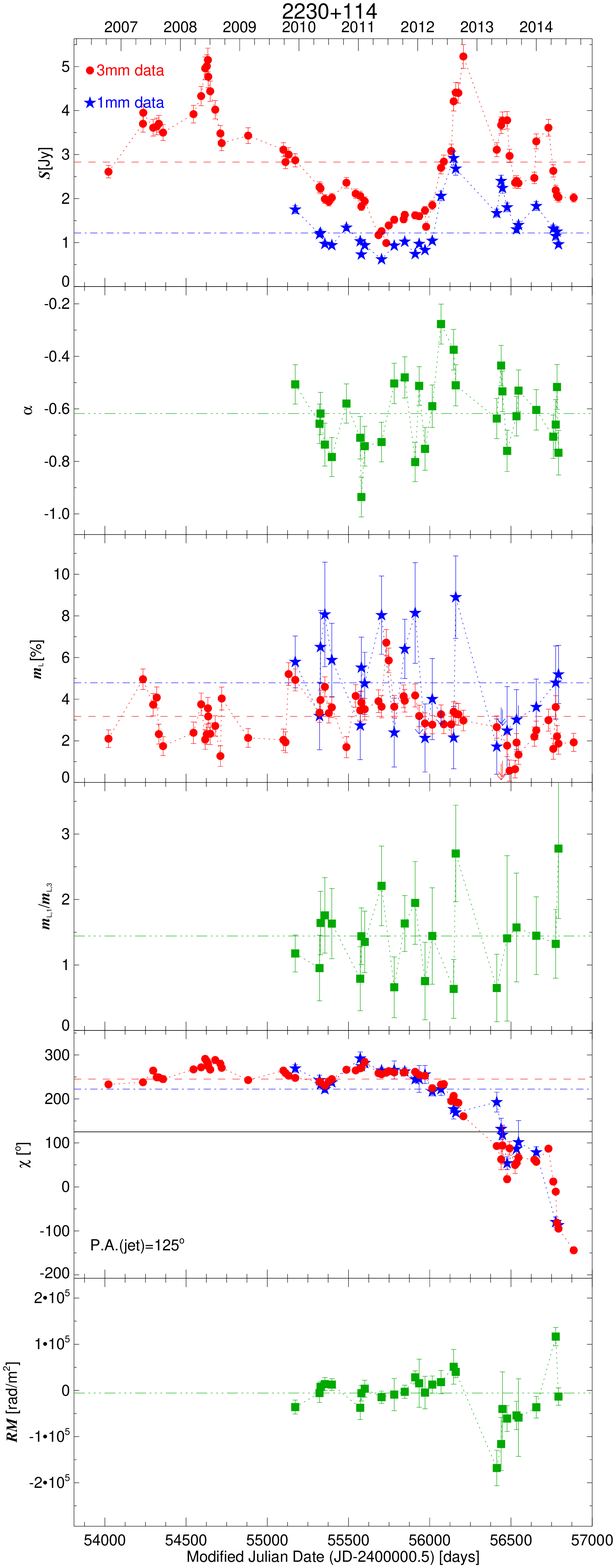}
	\end{subfigure}
	\begin{subfigure}[c]{0.49\textwidth}
	        \includegraphics[width=\textwidth]{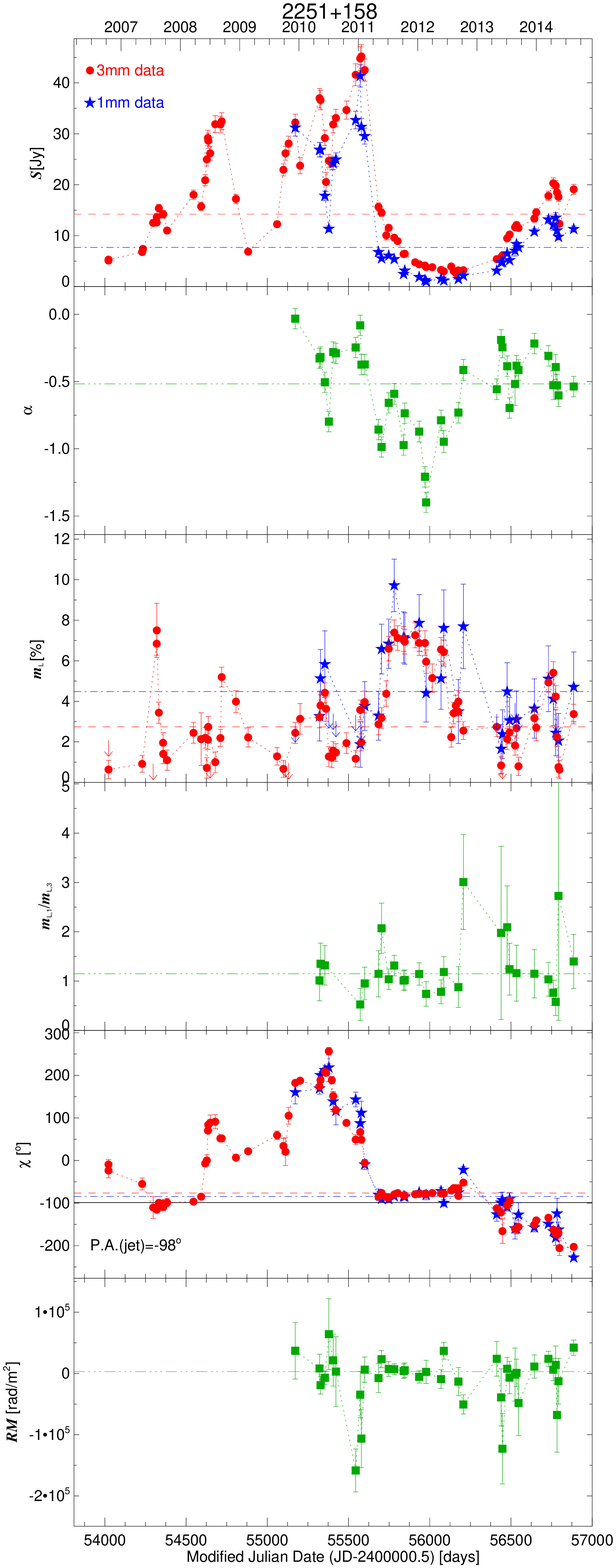}
	\end{subfigure}
   \caption{Continued.}
\end{figure*}

\appendix
\section{Analysis tools}
\label{appendix}

The statistical methods and definitions used in this paper as analysis tools for the interpretation of our data are presented in this Appendix.
These include: 

\begin{itemize}

\item[(i)] \textit{$\chi^2$ test for variability}. We tested the variability of the data from the $\chi^2$ obtained from fitting a constant against the distributions built from our time-dependent data sets. The resulting probability along with the $\chi^2$ and the degrees of freedom for every analysed variable are given in the Tables shown in this paper (e.g. Table~\ref{tab:I} for the analysis of total flux results).\\
	
\item[(ii)] \textit{Power spectral density}. We estimated the power spectral density (PSD) slopes under the assumption that the underlying model resembles a power-law (PSD $\propto f^{-\beta}$). 
A variant of the Power Spectral Response method \citep{Uttley:2002p12154} as detailed in \citet{Ramakrishnan:2015p25359} was used for estimating the slopes. 
This method involves the comparison of the observed periodogram against those obtained using the simulated time-dependent data sets. 
The data trains were simulated following the method proposed by \citet{Emmanoulopoulos:2013p25298}, which accounts for the PSD slope and the distribution of the data. 
To circumvent the problems imposed by the finite length and uneven sampling of the observed data on the estimation of the PSD slope, we simulated data trains with increased time resolution and 100 times longer than the observed ones, and through the convolution of Hanning window function with the observed data following \citet{MaxMoerbeck:2014p25302}. 
We also added Gaussian noise with variance matching those of the observations for the simulated data 
before resampling the simulated light curves. 
We ran this simulation for the PSD slopes in the range, 0.5--3, in steps of 0.05. 
The probability at every slope was obtained from a goodness of fit statistic that compares the average and standard deviation of the simulated periodograms with those of the observed ones. 
The best-fit PSD slope was taken as the one with the highest probability. 
PSD slopes were estimated only for those cases where more than 10 actual measurements were available in a data train.\\

\item[(iii)] \textit{Fractional variability amplitude}. To quantify the relative amount of variability of every variable measured on every source we estimated the fractional root mean squared variability amplitude (or fractional variability amplitude, $F$) from the relation \citep{Vaughan:2003p17038}:

\begin{equation}
	F = \sqrt{\frac{S^2 - \overline{\sigma^2_{\rm err}}}{\overline{x}^2}},
\end{equation}
where $S^2, \overline{\sigma^2_{\rm err}}$ and $\overline{x}^2$ are the variance, the mean squared error, and the squared arithmetic mean of every data train, respectively. 
The uncertainty on $F$, was computed and discussed by \citet{Vaughan:2003p17038}, and is expressed as:

\begin{equation}
	{\rm err}(F) = \sqrt{ \left\{ \sqrt{\frac{1}{2N}} \frac{ \overline{\sigma_{\rm err}^{2}}}{ \bar{x}^{2}F }  \right\}^{2} + 
		\left\{ \sqrt{\frac{\overline{\sigma_{\rm{err}}^{2}}}{N}}
			\frac{1}{\bar{x}}  \right\}^{2} },
\end{equation}
where $N$ is the number of points in every data train.\\

\item[(iv)] \textit{Correlation analysis}. The correlated variability between different variables were studied using the discrete correlation function (DCF) of \citet{Edelson:1988p12129}. 
This method is efficient when the sampling of the observed data is uneven. 
The DCF is defined as:

\begin{equation}
	{\rm DCF}_{ij} = \frac{(a_i-\bar{a})(b_j-\bar{b}) }{\sigma_a \sigma_b},
	\label{eq_DCF}
\end{equation}
where $a_i, b_j$ are the observed data at times $t_i$ and $t_j$ and $\bar{a}, \bar{b}, \sigma_a$ and $\sigma_b$ are the means and standard deviations of the entire data trains of variables $a$ and $b$, respectively. 

Unlike most other work that bin the data using equal widths, here we bin the associated time-lag such that it mitigates the chances of overlapping with the adjacent bins. 
This condition is ensured by the resolution parameter ($\epsilon$), that is defined such that $\tau_{i+1} - \tau_{i} < \epsilon$. 
In this process, we also ensure that there are at least 10 points within each bin, see \citet{Alexander:2013p25293} for more details on the binning method used here.

The average of the bins yields the DCF($\tau$). 
Following \citet{Welsh:1999p13803}, we applied the local normalisation to the DCF, which restricts the peak within [$-$1, +1] interval. 
The uncertainties on the DCF were estimated from a model-independent Monte Carlo method \citep{Peterson:1998p25344}.

We tested the significance of the DCF peak by cross-correlating 5000 data trains simulated using the best-fit PSD slopes. 
Based on this simulation, at every time-lag we constructed a cross-correlation distribution from which the 99.73\,\% significance levels were determined. 

Using the DCF relation shown in equation~\ref{eq_DCF} and the procedure employed to estimate the cross-correlation, we also computed the autocorrelation functions (ACF) of every variable analysed in this work. 
The zero-crossing time of the main lobe of the ACF ($\tau_0$) provides an estimate of the time scale of variability, that we show for every variable and every source in the tables presented in this work.

\end{itemize}

For all methods and definitions discussed above, only those measurements with modulus larger than the measurement uncertainty were considered, except for the case of the variability study of the linear polarisation angle. 

\bsp	
\label{lastpage}
\end{document}